\documentclass[a4paper,fleqn]{cas-dc}
 \usepackage[numbers]{natbib}

\def\tsc#1{\csdef{#1}{\textsc{\lowercase{#1}}\xspace}}
\tsc{WGM}
\tsc{QE}
\tsc{EP}
\tsc{PMS}
\tsc{BEC}
\tsc{DE}
\usepackage{enumitem}
\usepackage{multirow}
\begin{document}
\let\WriteBookmarks\relax
\def\floatpagepagefraction{1}
\def\textpagefraction{.001}
\shorttitle{SixGMAN}
\shortauthors{Matin Rafiei et~al.}

\title [mode = title]{SixGman: An Open‑Source Planner for Fixed 6G Hierarchical Optical Access–Core Networks}                      

\author[1]{{Matin Rafiei Forooshani}}[type=editor]
\ead{matinrafiei@aut.ac.ir}
\credit{Writing – original draft, Software, Methodology, Investigation, Formal analysis, Conceptualization}

\author[2]{{Farhad Arpanaei}}[type=editor,
                        orcid=0000-0003-1061-0614]
\cormark[1]
\ead{farpanae@it.uc3m.es}
\credit{Supervision, Writing – review \& editing, Software, Methodology, Investigation, Formal analysis, Conceptualization}

\author[1]{{Hamzeh Beyranvand}}[type=editor,
                            orcid=0000-0001-7013-5865]
\ead{beyranvand@aut.ac.ir}
\credit{Supervision, Writing –
review \& editing, Validation}
\author[2]{{Alfonso~S\'anchez-Maci\'an}}[type=editor,
                                    orcid=000-0002-2220-0594]
\ead{alfonsan@it.uc3m.es}
\credit{Writing –
review \& editing, Validation}

\author[3]{{Juan Pedro Fernández-Palacios}}[type=editor,
                            orcid=]
\ead{juanpedro.fernandez-palaciosgimenez@telefonica.com }

\credit{Writing –
review \& editing, Validation}

\author[2]{{Jos\'{e} Alberto Hern\'{a}ndez}}[type=editor,
                                        orcid=0000-0002-9551-4308]
\ead{jahgutie@it.uc3m.es }
\credit{Writing –
review \& editing, Validation}
\author[2]{{David Larrabeiti}}[type=editor,
                            orcid=0000-0003-4983-0243]
\ead{dlarra@it.uc3m.es}

\credit{Writing –
review \& editing, Validation}

\affiliation[1]{organization={Department of Electrical Engineering, Amirkabir University of Technology (Tehran Polytechnic)},
  	city={Tehran},
  	country={Iran}}

\affiliation[2]{organization={Department of Telematic Engineering, Universidad Carlos III de Madrid (UC3M), 
Madrid, 28911 },
  	city={Legan\'{e}s},
  	country={Spain}}

\affiliation[3]{organization={Telefónica Global CTIO, S/N, 28050},
  	city={Madrid},
  	country={Spain}}  
    
\cortext[1]{Corresponding author}

\fntext[1]{}


\begin{abstract}
This paper introduces \textit{SixGman}, an open-source optical network planning tool designed for the comprehensive evaluation of access-metro-core aggregation network architectures. To support detailed scenario specification and performance analysis, SixGman integrates a modular framework encompassing traffic generation, dual-homed routing, Quality of Transmission (QoT) estimation, and both spectrum and fiber assignment. Furthermore, the tool facilitates sustainability and viability assessments through dedicated techno-economic and energy consumption modules, complemented by integrated visualization capabilities. By utilizing clearly defined functions and standardized interfaces for each module, SixGman provides a flexible, transparent, and reproducible environment for advanced network simulation. The tool is applied to the Telefónica MAN157 metro-urban topology, consisting of 157 optical nodes, 220 links, and four hierarchical layers (HL1--HL4), to compare two network architectures: a conventional full hierarchical scenario and an HL3-bypassed scenario in which electrical aggregation at HL3 nodes is removed. Additionally, it analyzes IP router utilization, traffic flow distribution, link congestion, and latency. To assess economic and environmental impact, SixGman calculates the network's Total Cost of Ownership (TCO)—accounting for both CapEx and OpEx—alongside total energy consumption. By utilizing clearly defined functions and standardized interfaces for each module, SixGman provides a flexible and reproducible environment for advanced network simulation. Simulation results demonstrate that bypassing HL3 nodes leads to a more uniform traffic distribution, reduced optical and electrical resource usage, lower end-to-end latency, and a decrease in both capital and operational expenditures. Specifically, the HL3-bypassed scenario achieves reductions of up to 17.5\% in TCO and 29.1\% in cumulative energy consumption compared to the full hierarchical architecture. These findings highlight the potential of SixGman as a flexible planning platform and the effectiveness of HL3 bypassing as a cost- and energy-efficient network design strategy.
\end{abstract}

\begin{highlights}
\item \textbf{Introduction of SixGman:} The paper presents SixGman, a fully open-source, modular, and extensible optical network planning platform specifically designed for multi-hierarchical Metropolitan Area Network (MAN) topologies. 
\item \textbf{Comprehensive Multi-Layer Integration:} The tool uniquely integrates physical-layer modeling, service-layer provisioning, and multi-band spectrum allocation (C, SuperC, and L-bands) within a unified framework.
\item \textbf{End-to-End Techno-Economic Assessment:} Unlike prior tools, SixGman jointly models capital (CapEx) and operational (OpEx) expenditures for both the optical layer and the electrical IP routing layer.
\item \textbf{Evaluation of Hierarchical Bypassing:} Using the real-world Telefónica MAN157 topology, the study demonstrates that bypassing electrical aggregation at HL3 nodes significantly improves network performance.
\item \textbf{Significant Cost and Energy Savings:} Simulation results show that an HL3-bypassed architecture achieves a 17.5\% reduction in Total Cost of Ownership (TCO) and a 29.1\% decrease in cumulative energy consumption compared to conventional hierarchical designs.
\item \textbf{Performance Improvements: }Bypassing HL3 nodes leads to a more uniform traffic distribution, reduced link congestion, and a 19.4\% improvement in end-to-end latency by removing intermediate aggregation delays.
\end{highlights}

\begin{graphicalabstract}
\includegraphics[width=18cm]{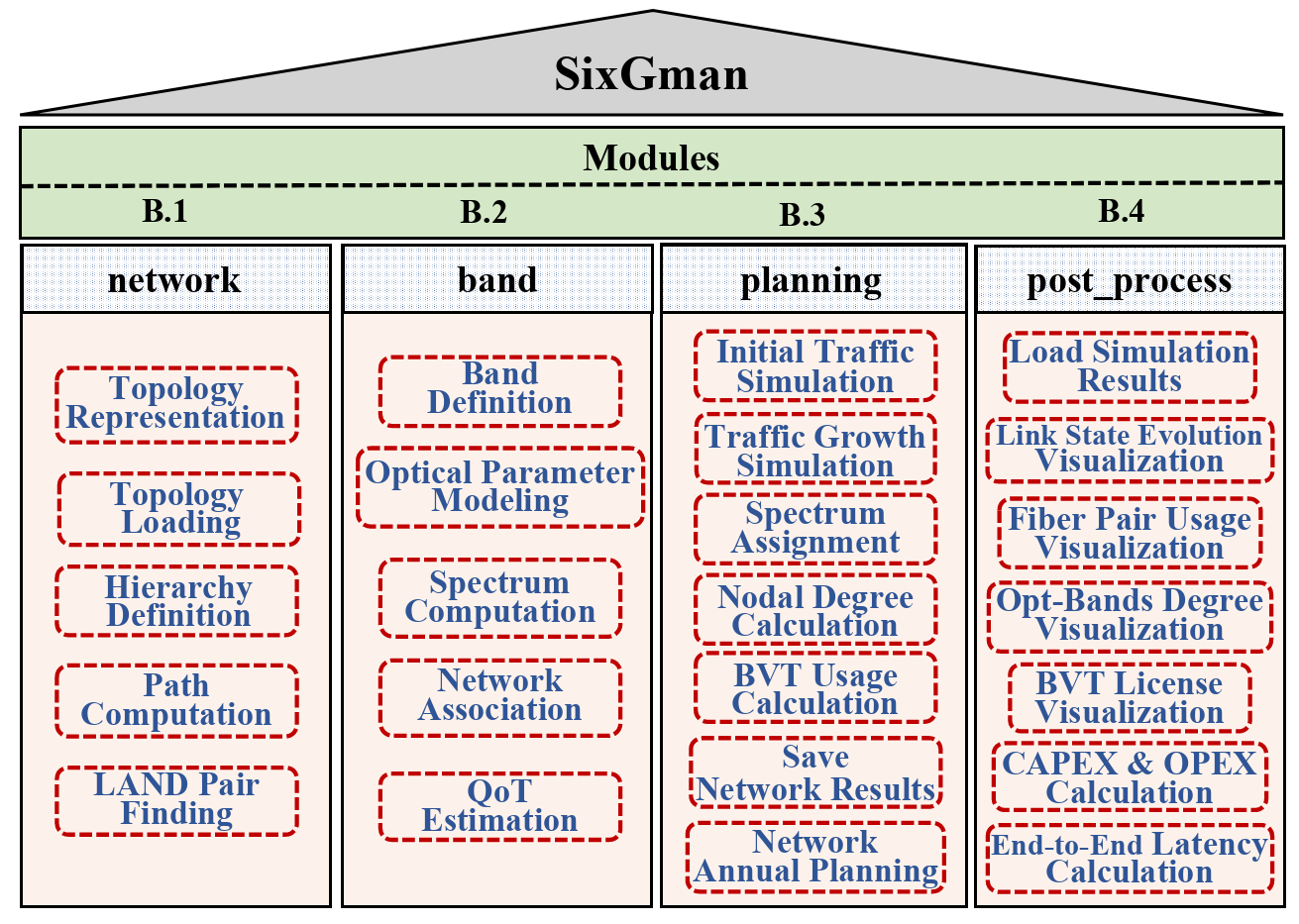}
\end{graphicalabstract}


\maketitle

\section{Introduction}

Fixed Sixth Generation (F6G) Metropolitan Area Networks (MANs) function as a critical hierarchical transport layer ~\citep{SHEN2025104330}. They aggregate high-volume wireless 6G traffic from access networks and route it toward regional and core infrastructures within densely populated urban environments ~\citep{telecom4040035}. 
Contemporary MANs are increasingly structured in multi-layer or hierarchical architectures, separating access, aggregation, and metro-core domains to improve scalability, fault isolation, and resource utilization ~\citep{kretsis2020armonia,Arpanaei23ECOC}.
In these designs, access-layer nodes handle local flows, mid-tier nodes perform aggregation and grooming, and higher-layer switching nodes provide gateway functions toward backbone networks ~\citep{TEFNET24}. However, this layered structure can impose substantial inefficiencies: frequent optical-to-electrical conversions, redundant aggregation, and suboptimal traffic routing can significantly increase both cost and energy consumption ~\citep{JUAN2023ICTON}. These concerns are particularly pronounced given the growing demands of 6G which requires ultra-low latency, energy-efficient operation, and support for distributed computing and massive connectivity ~\citep{Soleymani6GWhite}.
Moreover, emerging metro networks must meet F6G and 6G sustainability objectives while maintaining high performance. 
Addressing these challenges requires transparent and flexible planning tools capable of modeling hierarchical variations-including aggregation bypassing-while providing detailed evaluation of network performance, cost, energy consumption, latency, and Quality of Transmission (QoT) ~\citep{arpanaei2024enabling}. Motivated by these needs, this paper introduces SixGman, an open-source optical network planner designed for real-world MANs with multi-layer, multi-band, and multi-hierarchical architectures. Using SixGman, we analyze two representative network configurations: a conventional full-hierarchical design and an HL3-bypassed architecture where the electrical aggregation stage at HL3 nodes is removed. A comprehensive comparison is performed across multiple metrics, including fiber-pair consumption, spectrum usage, BVT and 100G license activation, traffic flow on links, IP and optical layer costs, energy consumption, latency, and signal quality, providing a detailed foundation for MAN modernization and 6G readiness.

MANs have received extensive attention in recent years due to their critical role in aggregating heterogeneous traffic sources and supporting emerging latency-sensitive, cloud-centric, and mobile-driven services. Early studies on metro optical systems investigated flat or two-layer architectures and wavelength-routed WDM designs~\citep{shen2009energy}. However, the rapid proliferation of broadband access, 5G, edge computing, and the forthcoming 6G era has underscored the limitations of traditional MAN architectures, emphasizing the need for scalable, flexible, and energy-efficient transport solutions~\citep{fayad2024toward,strinati20196g}.

To address these challenges, hierarchical MAN architectures have been widely studied as a means to improve scalability and operational efficiency \citep{BHANDARI2024110459,SHAKERI2024110268}. Multi-layer and multi-tier structures-typically organized into access, aggregation, and metro-core segments-provide enhanced traffic grooming, logical abstraction, and reduced control-plane complexity~\citep{ramirez2020multilayer}.

Parallel to architectural studies, the research community has increasingly recognized the importance of open and reproducible planning tools for optical transport networks. Tools such as GNPy provide accurate physical-layer modeling but mainly target long-haul or backbone domains~\citep{ferrari2020gnpy}. Similarly, open testbeds and ORAN-aligned platforms have enabled experimentation but often lack detailed support for metro hierarchies~\citep{ajayi2019openroad}. A particularly relevant contribution is the recent work by Arpanaei~\citep{arpanaei2024enabling}, which proposes a multi-band ``6D'' planning framework aimed at enabling the migration of metro-urban networks toward next-generation transmission technologies. While this study introduces several valuable concepts-such as multi-band feasibility analysis and impairment-aware planning-it remains outside the scope of hierarchical metro architectures. Specifically, the work does not analyze multi-layer or multi-hierarchical MAN structures, does not include IP-layer modeling or any assessment of  router resources, and does not evaluate energy consumption for optical or electrical-layer (or IP-layer) components. Furthermore, it does not consider latency estimation, nor does it examine the architectural consequences of bypassing intermediate aggregation layers, such as HL3 nodes. Importantly, the tool presented in~\citep{arpanaei2024enabling} is not open source, limiting its extensibility and reproducibility for the broader research community. These limitations further emphasize the need for an openly available, modular, and hierarchy-aware metro planning platform-an objective addressed by the SixGman framework introduced in this paper. Despite substantial progress, the literature lacks:
\begin{enumerate}
	\vspace{-0.3em}
	\item 
	An open-source planning platform tailored to metro, multi-layer, and multi-hierarchical architectures;
	\vspace{-0.3em}
	\item 
	A detailed evaluation of hierarchical bypass strategies such as HL3 elimination; and
	\vspace{-0.3em}
	\item 
	A comprehensive comparative study over a real, large-scale MAN topology (e.g., Telefónica MAN157) that jointly considers physical-layer impairments, traffic flow, energy consumption, latency, TCO, and resource usage.
\end{enumerate}

Therefore, this gap motivates the present work, which introduces SixGman, an open-source metro planning tool designed to support hierarchical architectures and detailed cross-layer evaluation. As a case study, we apply SixGman to compare full-hierarchical and HL3-bypassed architectures in a real-world metro topology.

This paper provides several significant contributions to the modeling, planning, and performance assessment of metro-area optical networks:

\begin{itemize}[noitemsep, topsep=0pt, parsep=0pt, partopsep=0pt, leftmargin=*]
	\setlength{\itemsep}{0.5pt}  
	\item
	\textbf{An open-source, operator-grade MAN planning platform.} We introduce SixGman, a fully open-source, modular, and extensible optical network planning tool specifically designed to support real-world multi-hierarchical MAN topologies. Unlike existing closed or research-prototype tools, SixGman integrates physical-layer modeling, service-layer provisioning, multi-band spectrum allocation, and multi-layer routing within a unified and transparent framework, facilitating reproducible research and practical deployment.
	\item
	\textbf{A comprehensive evaluation of hierarchical bypassing in MAN architectures.} Using the Telefónica MAN157 topology, we conduct an in-depth comparative analysis of full-hierarchical and HL3-bypassed architectures. The study uniquely evaluates the impact of bypassing HL3 nodes across a wide range of metrics, including fiber-pair usage, spectrum utilization, traffic flow changes, traffic congestion distribution, optical cost, IP-layer cost, optical- and IP-layer  energy consumption, latency, and QoT. This breadth of metrics provides the most complete characterization to date of hierarchical bypassing in a real metropolitan scenario.
	\item
	\textbf{Incorporation of electrical-layer (IP) cost and energy into metro network planning.} Unlike prior works that focus almost exclusively on the optical domain, our analysis jointly models the capital and operational expenditures of both the optical layer and the IP routing layer, thereby enabling true end-to-end techno-economic assessment of hierarchical architectural decisions.
	\item
	\textbf{Unified reporting of optical and IP energy consumption.} We provide a detailed quantification of energy usage across optical components (BVTs) and electrical components (IP routers). This dual-layer energy analysis reveals the cross-domain energy implications of architectural decisions and aligns MAN planning with sustainability objectives relevant for the 6G era.
	\item
	\textbf{Complete end-to-end latency evaluation.} The study computes the end-to-end latency for all demands by jointly accounting for propagation delay (per-kilometer optical latency) and electrical aggregation delay at hierarchical nodes. This allows us to isolate the effect of bypassing HL3 nodes on latency and quantify the trade-off between longer optical paths and reduced electrical processing.
\end{itemize}

The remainder of this paper is organized as follows. Section~\ref{sec:structure} introduces the overall architecture of the proposed open-source planning tool, SixGman, including the repository structure, data resources, and a high-level overview of its functional modules. Section~\ref{sec:network_functions} provides a detailed description of the algorithms and functions implemented within the network module, which is responsible for topology handling, multi-layer connectivity construction, and physical-layer modeling. Section~\ref{sec:band_functions} presents the functionalities of the band module, focusing on spectrum management, multi-band resource allocation, and channel-level parameter computation. Section~\ref{sec:planning_functions} describes the planning module, which integrates traffic provisioning, routing, protection schemes, and equipment dimensioning across the optical and electrical layers. Section~\ref{sec:post_functions} explains the features of the post\_process module, covering performance analysis, techno-economic evaluation, and energy assessment procedures. Section~\ref{sec:results} details the simulation setup and presents a comprehensive comparison of the full-hierarchical and HL3-bypassed scenarios under realistic MAN157 conditions. Finally, Section~\ref{sec:conclusion} concludes the paper by summarizing the main findings and outlining potential directions for future research.

\vspace{-1em}
\section{Structure of SixGman}
\label{sec:structure}
\subsection{Repository Layout and Data Sources}
\vspace{0.5em}

The SixGman is distributed as a Python-based open-source software package on GitHub \\ (\href{http://github.com/OS-ONDT/SixGman}{github.com/OS-ONDT/SixGman}). The repository is organized to promote modularity, reproducibility, and ease of use, as outlined below:

\vspace{-0.5em}
\begin{footnotesize}
\begin{center}
\begin{verbatim}
SixGman/
  src/sixgman/core    # Core modules
  src/sixgman/utils   # Utility functions and path handling
  tests/              # Unit tests for each module
  examples/           # Jupyter notebooks demonstrating use cases
  data/               # Example data files (e.g., .mat, .npz)
  docs/               # Documentation sources and user guides
  results/            # Simulation outputs and generated plots
\end{verbatim}
\end{center}
\end{footnotesize}
\vspace{-0.9em}

\subsubsection{Core Modules}
The directory \texttt{src/sixgman/core} contains the principal software components, including the \texttt{Network}, \texttt{Band}, \\ \texttt{PlanningTool} and \texttt{post\_process} classes. These classes provide the data structures and planning workflows required to represent optical topologies, define spectral bands, and perform routing and spectrum assignment.

\vspace{-0.6em}
\subsubsection{Utility Functions}
The \texttt{src/sixgman/utils} directory contains auxiliary functions for path management, file handling, and shared small-scale operations that support the main modules.

\vspace{-0.6em}
\subsubsection{Testing Suite}
The \texttt{tests} directory includes automated unit tests written using \texttt{pytest}. These tests verify the correctness of functions across modules, ensuring reproducibility and reducing regression risks as the planner evolves.

\vspace{-0.6em}
\subsubsection{Example Workflows}
Several Jupyter notebooks are provided in the \texttt{examples} directory. These notebooks illustrate typical usage patterns, such as loading a topology, configuring optical bands, executing network planning, and analyzing network performance.

\vspace{-0.6em}
\subsubsection{Topology Data}
The \texttt{data} directory stores input datasets used in planning and evaluation. In this work, the MAN-157 optical transport topology publicly released by Telefónica is included as a MATLAB \texttt{.mat} file. The file contains the adjacency matrix of the network, which can be imported directly into the \texttt{Network} class through SixGman's data-loading utilities.

\vspace{-0.6em}
\subsubsection{Simulation Output}
All generated network performance metrics and visualization files are saved in the \texttt{results} directory. Storing outputs separately from source code facilitates experiment traceability and post-processing.

\vspace{-0.2em}
\subsubsection{Documentation}
The \texttt{docs} directory contains the documentation build scripts and content, rendered online via Read the Docs. It provides detailed module explanations, API references, and usage examples.

\subsection{Modules Overview}

\begin{figure}
	\centering
	\includegraphics[width=8cm]{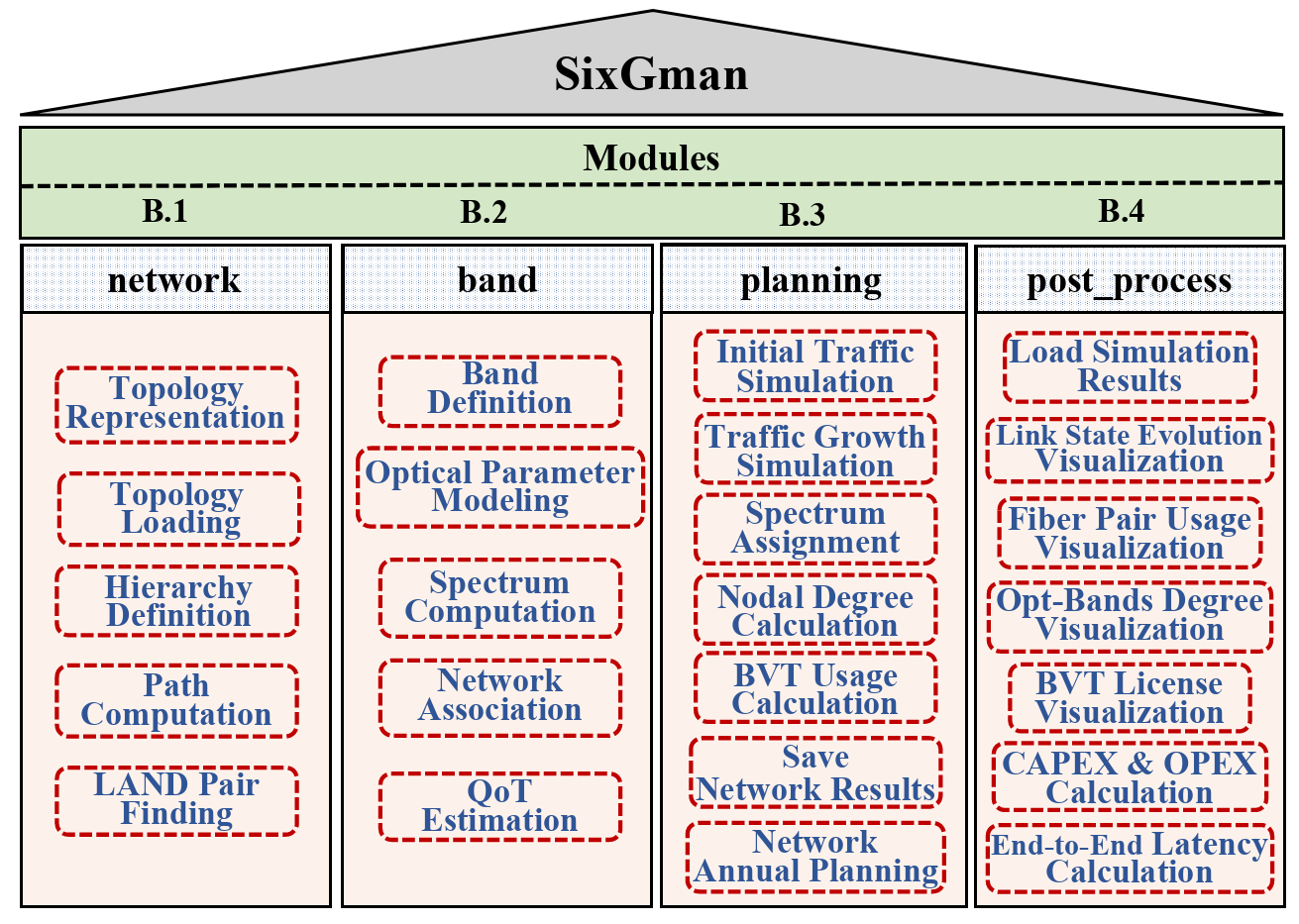}
	\caption{Overview of SixGman framework.}
	\label{fig:Arch}
\end{figure}

This section describes the functionality of the main software modules that constitute the SixGman framework. Each module encapsulates specific methods and data structures that collectively enable network representation, spectral band definition, routing and spectrum assignment, and performance evaluation. An overview of the SixGman architecture is illustrated in Fig.~\ref{fig:Arch}. For clarity, each module is presented separately, with a summary of its primary functions.

\subsubsection{Network Module}

The network module provides the foundational tools for modeling optical network topologies, importing them from various data formats, and preparing the data for planning and simulation within the SixGman framework. It serves as the structural backbone of the toolkit, enabling graph-based network representation, hierarchy definition, and path computation. The primary functionalities of the Network module as shown in Fig.~\ref{fig:Arch}.(B.1) include:
\begin{itemize}[noitemsep, topsep=0pt, parsep=0pt, partopsep=0pt, leftmargin=*]
	\setlength{\itemsep}{0.1pt}    
	\item \textbf{Topology Representation:} Models the optical network using \texttt{NetworkX} graph structures, where nodes correspond to network elements and edges represent optical links.
	\item \textbf{Topology Loading:} Supports importing adjacency matrices from \texttt{.mat}, \texttt{.npz}, and \texttt{.npy} file formats.
	\item \textbf{Hierarchy Definition:} Allows the assignment of hierarchical levels to network nodes, enabling multi-layer or regional network analysis.
	\item \textbf{Path Computation:} Implements the k-shortest path algorithm (Yen's algorithm) for computing candidate paths between source-destination pairs.
	\item \textbf{LAND Pair Finding:} Identifies link- and node-disjoint path pairs (LAND pairs) based on~\citep{arpanaei2024enabling} to support fault-tolerant and survivable network design. 
\end{itemize}

\noindent The \textbf{key class} defined in this module is as follows:

\vspace{0.5em}
\noindent\textbf{Network Class}\\
\texttt{sixgman.core.network.Network(topology\_name)}
\vspace{0.3em}

\noindent The \texttt{Network} class represents an optical network topology and its associated properties. It manages network initialization, hierarchical configuration, and path computation functionalities within the SixGman planning environment. This class provides the essential data structures required for higher-level modules, including the \texttt{Band} and \texttt{Planning} modules.

\vspace{-0.6em}
\paragraph{Attributes}
\begin{itemize}[noitemsep, topsep=0pt, parsep=0pt, partopsep=0pt, leftmargin=*]
	\setlength{\itemsep}{0.1pt}    
	\item \textbf{graph (nx.Graph):} A \texttt{NetworkX} graph object representing the network topology, where nodes correspond to optical nodes and edges represent bidirectional fiber links.
	\item \textbf{hierarchical\_levels (Dict[str, Dict[str, List[str]]]):} A dictionary that organizes nodes by their hierarchical levels, enabling structured multi-layer network modeling.
	\item \textbf{topology\_name (str):} A string specifying the name of the network topology (e.g., MAN157), used for identification and data management across modules.
\end{itemize}

\noindent A detailed explanation of the remaining functions and methods within the \texttt{Network} class will be provided in the section 3.

\vspace{-0.3em}
\subsubsection{Band Module}

The \texttt{band} module provides the necessary classes and methods to model optical transmission bands and their associated parameters, which are fundamental for multi-band optical network planning and performance evaluation. It defines the frequency grids, physical layer parameters, and spectral characteristics for each optical band, such as the C- and L-bands. The main responsibilities of the \texttt{Band} module as shown in Fig.~\ref{fig:Arch}.(B.2) include:
\begin{itemize}[noitemsep, topsep=0pt, parsep=0pt, partopsep=0pt, leftmargin=*]
	\setlength{\itemsep}{0.1pt}    
	\item \textbf{Band Definition:} Defines optical transmission bands (e.g., C-band, L-band) used in multi-band optical network design.
	\item \textbf{Optical Parameter Modeling:} Stores and computes fiber and system parameters through the \texttt{OpticalParameters} class.
	\item \textbf{Spectrum Computation:} Generates frequency grids based on the start frequency, end frequency, and channel spacing.
	\item \textbf{Network Association:} Links each band definition to a given network topology for end-to-end modeling.
	\item \textbf{QoT Estimation:} Implements the numerical structures and algorithms required for GSNR computation and nonlinear impairment-based QoT estimation in optical links.
\end{itemize}
\vspace{0.1cm}

\noindent The \textbf{key classes} defined in this module are as follows:
\vspace{0.5em}

\noindent\textbf{I) OpticalParameters Class}\\
\noindent\texttt{sixgman.core.band.OpticalParameters(h\_plank=6.626e-34,\\ target\_ber=0.01, phi\_MFL=<factory>, epsilon=0,\\ beta\_3=1.4e-40, Cr=2.8e-17, alpha\_db=0.2, beta\_2=-2.17e-26, gama=0.00121, F\_C=<factory>, F\_L=<factory>, Rs\_mat=4e10, MFL=<factory>, rof=0.1)}\\

\vspace{-1em}
\noindent This class stores and computes the fiber and system parameters used in optical network modeling. It encapsulates both linear and nonlinear transmission characteristics, enabling accurate physical layer performance evaluation.

\vspace{-0.5em}
\paragraph{Attributes}
\begin{itemize}[noitemsep, topsep=3pt, parsep=1pt, partopsep=0pt, leftmargin=*]
	\setlength{\itemsep}{0.1pt}
	
	\item \textbf{h\_plank (float):} Planck's constant (\(6.626\times10^{-34}\,\mathrm{J\,s}\)).
	\item \textbf{target\_ber (float):} Target bit error rate.
	\item \textbf{phi\_MFL (np.ndarray):} Modulation format penalty factors.
	\item \textbf{epsilon (int):} Auxiliary modeling variable.
	\item \textbf{beta\_3 (float):} Third-order dispersion coefficient (\(\mathrm{s}^3/\mathrm{m}\)).
	\item \textbf{Cr (float):} Chromatic dispersion coefficient \(\left( \frac{1}{\mathrm{m}\,\mathrm{Hz}^2} \right)\).
	\item \textbf{alpha\_db (float):} Fiber attenuation (\(\mathrm{dB}/\mathrm{km}\)).
	\item \textbf{beta\_2 (float):} Second-order dispersion coefficient (\(\mathrm{s}^2/\mathrm{m}\)).
	\item \textbf{gama (float):} Nonlinear coefficient \(\left( \frac{1}{\mathrm{W\,m}} \right)\).
	\item \textbf{F\_C, F\_L (float):} Noise figures in linear scale for the C-band and L-band, respectively.
	\item \textbf{Rs\_mat (float):} Symbol rate (Baud).
	\item \textbf{MFL (np.ndarray):} Available modulation format levels.
	\item \textbf{rof (float):} Roll-off factor of the signal spectrum.
	\item \textbf{alpha\_norm (float):} Normalized fiber attenuation (\(1/\mathrm{m}\)).
	\item \textbf{L\_eff\_a (float):} Effective fiber length (m).
	\item \textbf{B\_ch (float):} Channel bandwidth (Hz).
	\item \textbf{target\_SNR\_dB (list):} Target SNR of modulation formats (PM-64QAM, PM-32QAM, PM-16QAM, PM-8QAM, PM-QPSK, PM-BPSK) to reach the target BER (Bit Error Rate)

\end{itemize}

\vspace{0.2em}
\noindent\textbf{II) Band Class}\\
\noindent\texttt{sixgman.core.band.Band(name, start\_freq, end\_freq,\\
	 opt\_params, network\_instance, channel\_spacing=0.05)}\\

\vspace{-1em}
\noindent This class represents an optical transmission band, capturing its spectral characteristics and channel configuration. It defines the frequency grid for each band and associates it with relevant network and optical parameters, integrating the physical-layer model to support spectrum allocation, modulation format selection, and GSNR-based performance evaluation.

\vspace{-0.6em}
\paragraph{Attributes}
\begin{itemize}[noitemsep, topsep=1pt, parsep=1pt, partopsep=0pt, leftmargin=*]
	\setlength{\itemsep}{0.1pt}    
	\item \textbf{name (str):} Identifier of the optical band (e.g., ‘C’, ‘L’).
	\item \textbf{start\_freq (float):} Start frequency (THz).
	\item \textbf{end\_freq (float):} End frequency (THz).
	\item \textbf{opt\_params (OpticalParameters):} Optical transmission parameters
	\item \textbf{channel\_spacing (float):} Channel spacing (THz), default 0.05 THz (50 GHz).
	\item \textbf{spectrum (np.ndarray):} Array of center frequencies for all channels in the band.
	\item \textbf{num\_channels (int):} Total number of channels available within the frequency range.
\end{itemize}

A detailed explanation of additional methods within these classes, such as \texttt{compute\_spectrum}, will be presented in section~\ref{sec:band_functions}.

\subsubsection{Planning Module}

The \texttt{planning} module constitutes the optimization core of the SixGman framework. It integrates network topology, spectral resources, and physical layer parameters to perform traffic- and spectrum-aware planning for metro and urban optical transport networks. This module serves as the central control layer that coordinates the operation of other modules and executes the overall network design and optimization workflow.

The main responsibilities of the \texttt{Planning} module as shown in Fig.~\ref{fig:Arch}.(B.3) are summarized below:

\begin{itemize}[noitemsep, topsep=0pt, parsep=0pt, partopsep=0pt, leftmargin=*]
	\setlength{\itemsep}{0.1pt}
	
	\item \textbf{Initial Traffic Simulation:} 
	Generates the baseline traffic matrix for the first planning year based on the modeled node hierarchy. This traffic initialization forms the foundation for subsequent annual planning iterations and determines the starting point for spectrum and bitrate variable transponders (BVTs) allocation.
	
	\item \textbf{Traffic Growth Simulation:} 
	Performs iterative traffic projection across planning periods to emulate the progressive increase in data demand. Growth factor may be applied globally, enabling modeling of metro-area traffic expansion over multiple years.
	
	\item \textbf{Spectrum Assignment:} 
	Implements wavelength and spectrum resource allocation across the available optical bands (e.g., C-, SuperC-, and L-band). The algorithm ensures conflict-free assignment under spectral continuity and contiguity constraints while maintaining the coexistence of multiple transmission bands. Spectrum assignment is directly influenced by link capacity, fiber-pair availability, and signal-quality constraints such as GSNR.
	
	\item \textbf{Nodal Degree Calculation:} 
	Quantifies the degree of connectivity for each node, including the number of active fiber pairs per node. This calculation provides a topological metric for evaluating network scalability and hierarchical interconnection structure, essential for cost modeling and resilience analysis.
	
	\item \textbf{BVT Usage Calculation:} 
	Determines the number and type of BVTs required to satisfy the planned traffic demand. The calculation accounts for modulation formats, spectral efficiency, and available 100G license capacity, thereby linking physical-layer transmission parameters to network-layer resource utilization.
	
	\item \textbf{Save Network Results:} 
	Consolidates and exports all annual planning outcomes including link utilization, fiber-pair allocation, GSNR evolution, and BVT statistics into structured NumPy (\texttt{.npz}) files. These data serve as inputs to the \texttt{post\_process} module for further visualization and cost analysis.
	
	\item \textbf{Network Annual Planning:} 
	Performs the integrated planning cycle for each simulation year. This process combines traffic updates, path computation, spectrum assignment, and transponder deployment into a cohesive optimization loop. The annual planning phase ensures temporal consistency of capacity expansion while capturing the evolution of network performance indicators over time.
	
\end{itemize}

\vspace{0.5em}

\noindent The \textbf{key class} defined in this module is as follows:
\vspace{0.5em}

\noindent\textbf{PlanningTool Class}\\
\noindent\texttt{sixgman.core.planning.PlanningTool(network\_instance,\\ bands, optical\_parameters, grid\_center,  period\_time=10)}\\

\vspace{-0.7em}
\noindent The \texttt{PlanningTool} class is the main component of the SixGman framework, responsible for end-to-end optical network planning and optimization. It integrates physical-layer modeling with hierarchical network topology to evaluate and design efficient multi-band optical metro networks.

\vspace{-0.8em}
\paragraph{Attributes}
\begin{itemize}[noitemsep, topsep=2pt, parsep=2pt, partopsep=0pt, leftmargin=*]
	\setlength{\itemsep}{0.1pt}
	\item \textbf{network (Network):} A reference to the network topology object containing all nodes, links, and hierarchical information used in the planning process.
	\item \textbf{bands (List[Band]):} A list of \texttt{Band} instances representing the optical bands (e.g., C, L, S) with their respective spectral and physical parameters.
	\item \textbf{optical\_parameters (dict):} Dictionary containing the optical transmision parameters.
	\item \textbf{grid\_center (np.ndarray):} Numpy array containing the center frequency of frequency channels.
	\item \textbf{period\_time (int):} Defines the time period over which the planning procedure is performed. The unit is year.
\end{itemize}

The \texttt{PlanningTool} class provides the interface between the physical and network layers, enabling integrated optimization across spectral and topological dimensions. It supports dynamic multi-band coexistence, topology-aware path selection, and traffic-driven spectrum assignment strategies. These capabilities collectively allow the SixGman framework to perform realistic and flexible metro-area optical network design. Detailed descriptions of additional functions within the \texttt{PlanningTool} class, such as planner initialization and annual traffic simulation, are provided in section~\ref{sec:planning_functions}.

\vspace{-0.3em}
\subsubsection{post\_process Module}
\vspace{0.2em}
The \texttt{post\_process} module provides analytical and visualization tools for managing, interpreting, and evaluating the outcomes of hierarchical optical network simulations and planning processes. It functions as the post-processing and data interpretation layer within the SixGman framework, enabling structured access to key network performance metrics. The module ensures that simulation outputs from the planning stage are efficiently organized, visualized, and analyzed for both technical assessment and cost evaluation.

The main responsibilities of the \texttt{post\_process} module as shown in Fig.~\ref{fig:Arch}.(B.4) include:

\begin{itemize}[noitemsep, topsep=0pt, parsep=0pt, partopsep=0pt, leftmargin=*]
	\setlength{\itemsep}{0.1pt}
	\item \textbf{Load Simulation Results:}
	Reads and organizes stored network data (e.g., link usage, BVT allocation, GSNR performance) from compressed \texttt{.npz} files for each hierarchy level. The data are parsed into structured dictionaries, enabling consistent access to multi-level simulation outputs.
	
	\item \textbf{Link State Evolution Visualization:}
	Produces time-evolution plots showing how the number of fiber pairs per link changes over simulation years, highlighting the evolution of spectrum usage, fiber-pair allocation, and hierarchical interconnections under increasing traffic demand.
	
	\item \textbf{Fiber Pair Usage Visualization:}
	Analyzes and plots cumulative FP utilization over time, representing both the total deployed fiber length (in kilometers) and the number of active fiber pairs. This visualization supports the assessment of physical infrastructure growth throughout the planning horizon.
	
	\item \textbf{Optical Bands Degree Visualization:}
	Depicts the evolution of band degree across different optical bands (C-band, SuperC-band, and L-band). The resulting plots illustrate the temporal variation in band utilization and highlight the progressive activation of multi-band transmission resources over the simulation period.
	
	\item \textbf{BVT License Visualization:}
	Illustrates the deployment trend of BVTs and 100G licenses over time. The plots quantify per-band BVT utilization and the corresponding licensing requirements, supporting operational cost analysis.
	
	\item \textbf{CAPEX \& OPEX Calculation:}
	Computes capital and operational expenditures for each simulation year based on network component usage (e.g., licenses, switches, and fiber infrastructure). The calculated metrics enable cost-performance comparisons across different planning scenarios and hierarchy levels.
	
	\item \textbf{End-to-End Latency Calculation:}
	Computes the cumulative end-to-end (E2E) latency from the lowest hierarchical nodes (e.g., HL4) to the top-level nodes (e.g., HL1). The function recursively traverses multi-hop hierarchical paths, integrating both per-link propagation delay and per-level processing overhead to generate comprehensive latency statistics.
	
\end{itemize}

\vspace{0.5em}
\noindent The \textbf{key class} defined in this module is as follows:
\vspace{0.5em}

\noindent\textbf{analyse\_result Class}\\
\noindent\texttt{sixgman.core.post\_process.analyse\_result(\\
	network\_instance, period\_time, processing\_level\_list, results\_directory, save\_directory)}\\

\vspace{-1em}
\noindent The \texttt{analyse\_result} class provides mechanisms for analyzing and managing the outputs of hierarchical optical network simulations. It supports the loading, organization, and retrieval of both link-level and BVT-level metrics, enabling systematic evaluation of planning outcomes within the SixGman framework.

\vspace{-0.6em}
\paragraph{Attributes}
\begin{itemize}[noitemsep, topsep=0pt, parsep=0pt, partopsep=0pt, leftmargin=*]
	\setlength{\itemsep}{0.1pt}
	\item \textbf{network (Network):} Reference to the network topology object containing node and link metadata used during simulation.
	\item \textbf{period\_time (int):} Duration of the planning or simulation period (e.g., in years or simulation units).
	\item \textbf{processing\_level\_list (List[int]):} List of hierarchy levels to be processed (e.g., [4, 3, 2]) for multi-layer analysis.
	\item \textbf{results\_directory (Path):} Directory path containing stored result files in \texttt{.npz} format for each hierarchy level.
	\item \textbf{save\_directory (Path):} Directory path where output plots will save in.
\end{itemize}

\noindent The \texttt{post\_process} module thus enables detailed inspection of network planning outcomes, providing an essential link between simulation output and performance interpretation. It ensures that results from hierarchical, multi-band planning can be efficiently accessed, visualized, and correlated with input configurations, supporting transparent and reproducible optical network research.

\vspace{-1em}
\section{Network Module Functions}
\label{sec:network_functions}
This section describes the functionality of each method implemented in the \texttt{Network} class, the only class included in this module.

\vspace{-0.5em}
\subsection{\texttt{load\_topology()} Function}

\noindent\texttt{load\_topology(filepath, matrixName=None) $\rightarrow$ nx.Graph}

\vspace{-0.6em}
\paragraph{Description}
The \texttt{load\_topology()} function loads an optical network topology from external data files and converts it into a \texttt{NetworkX} graph structure. It reads an adjacency matrix describing the network connectivity and initializes corresponding network attributes within the SixGman framework.

{\small
	\vspace{-0.6em}
	\paragraph{Supported Formats}
	\begin{itemize}[noitemsep, topsep=0pt, parsep=0pt, partopsep=0pt, leftmargin=*]
		\setlength{\itemsep}{0.1pt}    
		\item \texttt{.mat} – MATLAB file (requires \texttt{matrixName} to specify the variable).
		\item \texttt{.npz} – NumPy compressed archive (variable name required).
		\item \texttt{.npy} – NumPy single-array file.
	\end{itemize}
	
	\vspace{-0.6em}
	\paragraph{Parameters}
	\begin{itemize}[noitemsep, topsep=0pt, parsep=0pt, partopsep=0pt, leftmargin=*]
		\item \texttt{filepath (str):} Path to the file containing the network topology data.
		\item \texttt{matrixName (str):} Name of the adjacency matrix variable in \texttt{.mat} or \texttt{.npz} files.
	\end{itemize}
	
	\vspace{-0.6em}
	\paragraph{Returns}
	\begin{itemize}[noitemsep, topsep=0pt, leftmargin=*]
		\item \texttt{nx.Graph:} A \texttt{NetworkX} graph representing the loaded network topology.
	\end{itemize}
}

\vspace{-0.5em}
\subsection{\texttt{define\_hierarchy()} Function}

\noindent\texttt{define\_hierarchy(**kwargs) $\rightarrow$ Dict[str, Dict[str, List[str]]]}

\vspace{-0.6em}
\paragraph{Description}
The \texttt{define\_hierarchy()} function assigns nodes in the network to different hierarchical levels (e.g., HL1, HL2, HL3). It supports flexible configuration of node placement, distinguishing between standalone and co-located nodes across hierarchy levels. Standalone nodes are unique to their respective levels, while co-located nodes share membership with previous levels if not explicitly specified.

{\small
	\vspace{-0.5em}
	\paragraph{Behavior}
	\begin{itemize}[noitemsep, topsep=0pt, parsep=0pt, partopsep=0pt, leftmargin=*]
		\item Accepts keyword arguments such as \texttt{HL1\_standalone}, \texttt{HL2\_colocated}, etc.
		\item \texttt{standalone} nodes are unique to that hierarchy level.
		\item If a \texttt{\_colocated} list is not defined, it is automatically accumulated from all prior \texttt{standalone} nodes.
	\end{itemize}
	\vspace{-0.5em}
	\paragraph{Parameters}
	\begin{itemize}[noitemsep, topsep=0pt, parsep=0pt, partopsep=0pt, leftmargin=*]
		\item \texttt{**kwargs:} Variable keyword arguments defining hierarchy membership, such as \texttt{HLx\_standalone} and \texttt{HLx\_colocated}.
	\end{itemize}
	\vspace{-0.5em}
	\paragraph{Returns}
	\begin{itemize}[noitemsep, topsep=0pt, leftmargin=*]
		\item \texttt{Dict[str, Dict[str, List[str]]]:} Updated hierarchical levels structure of the network.
	\end{itemize}
}
\vspace{-0.5em}
\subsection{\texttt{\_calc\_all\_hierarchical\_nodes()} Function}
\noindent\texttt{\_calc\_all\_hierarchical\_nodes() $\rightarrow$ List[int]}

\vspace{-0.6em}
\paragraph{Description}
The \texttt{\_calc\_all\_hierarchical\_nodes()} function computes a complete list of hierarchical nodes in the network, aggregating both standalone and co-located nodes across all hierarchy levels. It provides a convenient way to retrieve all nodes that participate in hierarchical planning or analysis.

{\small
	\vspace{-0.5em}
	\paragraph{Behavior}
	\begin{itemize}[noitemsep, topsep=0pt, parsep=0pt, partopsep=0pt, leftmargin=*]
		\item Uses the internal attribute \texttt{self.hierarchical\_levels} to identify nodes at each hierarchy level.
		\item Aggregates nodes from both \texttt{standalone} and \texttt{colocated} categories.
		\item Returns the full list without duplicates.
	\end{itemize}
	\vspace{-0.5em}
	\paragraph{Parameters}
	\begin{itemize}[noitemsep, topsep=0pt, parsep=0pt, partopsep=0pt, leftmargin=*]
		\item None
	\end{itemize}
	\vspace{-0.5em}
	\paragraph{Returns}
	\begin{itemize}[noitemsep, topsep=0pt, leftmargin=*]
		\item \texttt{List[int]:} A list of all hierarchical nodes in the network across all levels.
	\end{itemize}
}

\vspace{-0.5em}
\subsection{\texttt{get\_standalone\_hierarchy\_level()} Function}
\noindent\texttt{get\_standalone\_hierarchy\_level(node: int) $\rightarrow$ Optional[int]}

\vspace{-0.6em}
\paragraph{Description}
The \texttt{get\_standalone\_hierarchy\_level()} function returns the standalone hierarchical level of a given network node. It checks whether the input node belongs to any standalone set of nodes. If the node is not found in any standalone level, the function returns \texttt{None}.

{\small
	\vspace{-0.5em}
	\paragraph{Parameters}
	\begin{itemize}[noitemsep, topsep=0pt, parsep=0pt, partopsep=0pt, leftmargin=*]
		\item \texttt{node (int):} The index of the node whose hierarchical level is to be determined.
	\end{itemize}
	\vspace{-0.5em}
	\paragraph{Returns}
	\begin{itemize}[noitemsep, topsep=0pt, leftmargin=*]
		\item \texttt{Optional[int]:} The standalone hierarchical level of the node (e.g., 2 for HL2). Returns \texttt{None} if the node does not belong to any standalone level.
	\end{itemize}
}
\vspace{-0.5em}
\subsection{\texttt{\_reconstruct\_yen\_path()} Function}

\noindent\texttt{\_reconstruct\_yen\_path(predecessors, path\_index, source, target) $\rightarrow$ List[int]}

\begin{figure}
	\centering
	\includegraphics[height = 7cm]{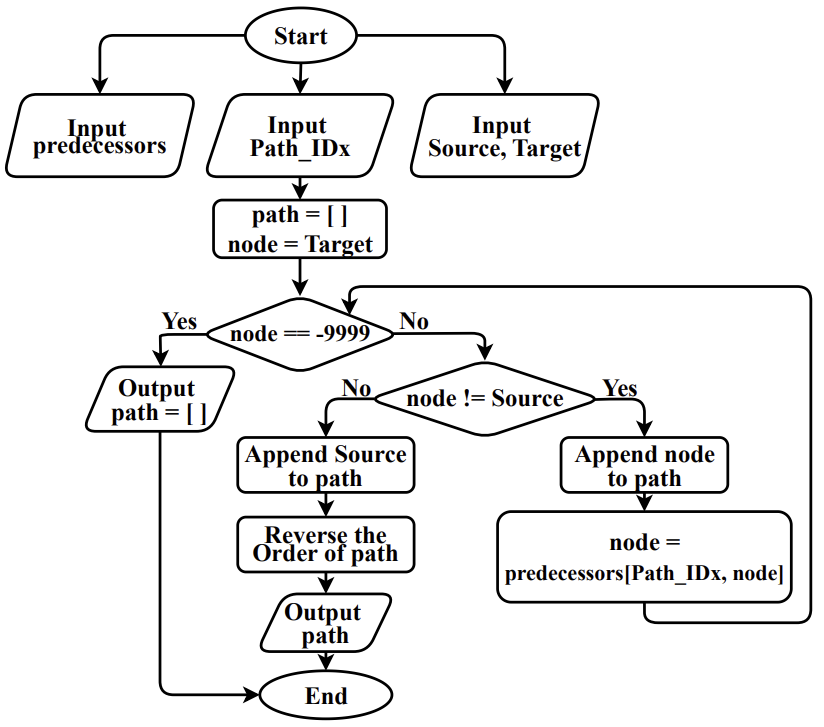}
	\caption{Flowchart of \texttt{\_reconstruct\_yen\_path()} function.}
	\label{fig:reconstruct_path_flowchart}
\end{figure}

\vspace{-0.6em}
\paragraph{Description}
The \texttt{\_reconstruct\_yen\_path()} function reconstructs a specific path from the predecessors matrix generated by Yen's algorithm. It produces an ordered list of node IDs representing the path from the source to the target. If the target is unreachable from the source, the function returns an empty list. The flowchart of this function is shown in Fig.~\ref{fig:reconstruct_path_flowchart}. This function is used internally by the \texttt{compute\_k\_shortest\_paths()} method to rebuild candidate paths after the shortest-path computation.

{\small
	\vspace{-0.5em}
	\paragraph{Parameters}
	\begin{itemize}[noitemsep, topsep=0pt, parsep=0pt, partopsep=0pt, leftmargin=*]
		\item \texttt{predecessors (np.ndarray):} Predecessors matrix obtained from Yen's algorithm with shape [k\_paths, n\_nodes].
		\item \texttt{path\_index (int):} Index of the path to reconstruct from the predecessors matrix.
		\item \texttt{source (int):} Source node ID.
		\item \texttt{target (int):} Target node ID.
	\end{itemize}
	
	\vspace{-0.5em}
	\paragraph{Returns}
	\begin{itemize}[noitemsep, topsep=0pt, leftmargin=*]
		\item \texttt{List[int]:} Ordered list of node IDs representing the reconstructed path. Returns an empty list if the target is not reachable from the source.
	\end{itemize}
}

\vspace{-0.5em}
\subsection{\texttt{\_get\_link\_indices\_in\_path()} Function}

\noindent\texttt{\_get\_link\_indices\_in\_path(path) $\rightarrow$ List[int]}

\vspace{-0.6em}
\paragraph{Description}
The \texttt{\_get\_link\_indices\_in\_path()} function returns the list of link indices corresponding to a given path in the network. Since the network graph is undirected, the function checks both directions (u$\rightarrow$v and v$\rightarrow$u) when matching node pairs to links. It is used internally by the \texttt{compute\_k\_shortest\_paths()} method to map node sequences from paths into the actual link indices stored in \texttt{self.all\_links}. 

{\small
	\vspace{-0.5em}
	\paragraph{Parameters}
	\begin{itemize}[noitemsep, topsep=0pt, parsep=0pt, partopsep=0pt, leftmargin=*]
		\item \texttt{path (List[int]):} Ordered list of node IDs representing the path.
	\end{itemize}
	
	\vspace{-0.5em}
	\paragraph{Returns}
	\begin{itemize}[noitemsep, topsep=0pt, leftmargin=*]
		\item \texttt{List[int]:} Indices of links in \texttt{self.all\_links} that form the given path.
	\end{itemize}
}

\vspace{-0.5em}
\subsection{\texttt{compute\_hierarchy\_subgraph()} Function}

\noindent\texttt{compute\_hierarchy\_subgraph(hierarchy\_level,\\ minimum\_hierarchy\_level) $\rightarrow$ Tuple[nx.Graph, np.ndarray]}

\vspace{-0.6em}
\paragraph{Description}
The \texttt{compute\_hierarchy\_subgraph()} function extracts a subgraph from the network based on hierarchical constraints. It constructs a subgraph that includes edges where at least one endpoint belongs to the specified hierarchy level (HLx), and the other endpoint is not part of any lower level between HL(x+1) and HL(minimum).

{\small
	\vspace{-0.5em}
	\paragraph{Parameters}
	\begin{itemize}[noitemsep, topsep=0pt, parsep=0pt, partopsep=0pt, leftmargin=*]
		\item \texttt{hierarchy\_level (int):} The current hierarchy level (e.g., 1 for HL1).
		\item \texttt{minimum\_hierarchy\_level (int):} The lowest hierarchy level to exclude from edge participation.
	\end{itemize}
	\vspace{-0.5em}
	\paragraph{Returns}
	\begin{itemize}[noitemsep, topsep=0pt, leftmargin=*]
		\item \texttt{Tuple[nx.Graph, np.ndarray]:} The resulting \texttt{NetworkX} subgraph and its adjacency (cost) matrix, where missing links are assigned \texttt{np.inf}.
	\end{itemize}
}

\vspace{-0.5em}
\subsection{\texttt{get\_neighbor\_nodes()} Function}

\noindent\texttt{get\_neighbor\_nodes(nodes) $\rightarrow$ List[int]}

\vspace{-0.6em}
\paragraph{Description}
The \texttt{get\_neighbor\_nodes()} function returns the unique neighbors of a given list of nodes in the network graph. The input nodes themselves are excluded, and each neighbor appears only once, regardless of how many input nodes it is connected to.

{\small
	\vspace{-0.5em}
	\paragraph{Parameters}
	\begin{itemize}[noitemsep, topsep=0pt, parsep=0pt, partopsep=0pt, leftmargin=*]
		\item \texttt{nodes (List[int]):} List of node IDs for which neighbors are to be found.
	\end{itemize}
	\vspace{-0.5em}
	\paragraph{Returns}
	\begin{itemize}[noitemsep, topsep=0pt, leftmargin=*]
		\item \texttt{List[int]:} Sorted list of unique neighbor node IDs.
	\end{itemize}
}

\vspace{-0.5em}
\subsection{\texttt{compute\_k\_shortest\_paths()} Function}
\noindent\texttt{compute\_k\_shortest\_paths(subnet\_matrix, paths, source, target, k=20) $\rightarrow$ List[Dict]}

\begin{figure}
	\centering
	\includegraphics[height = 7.5cm]{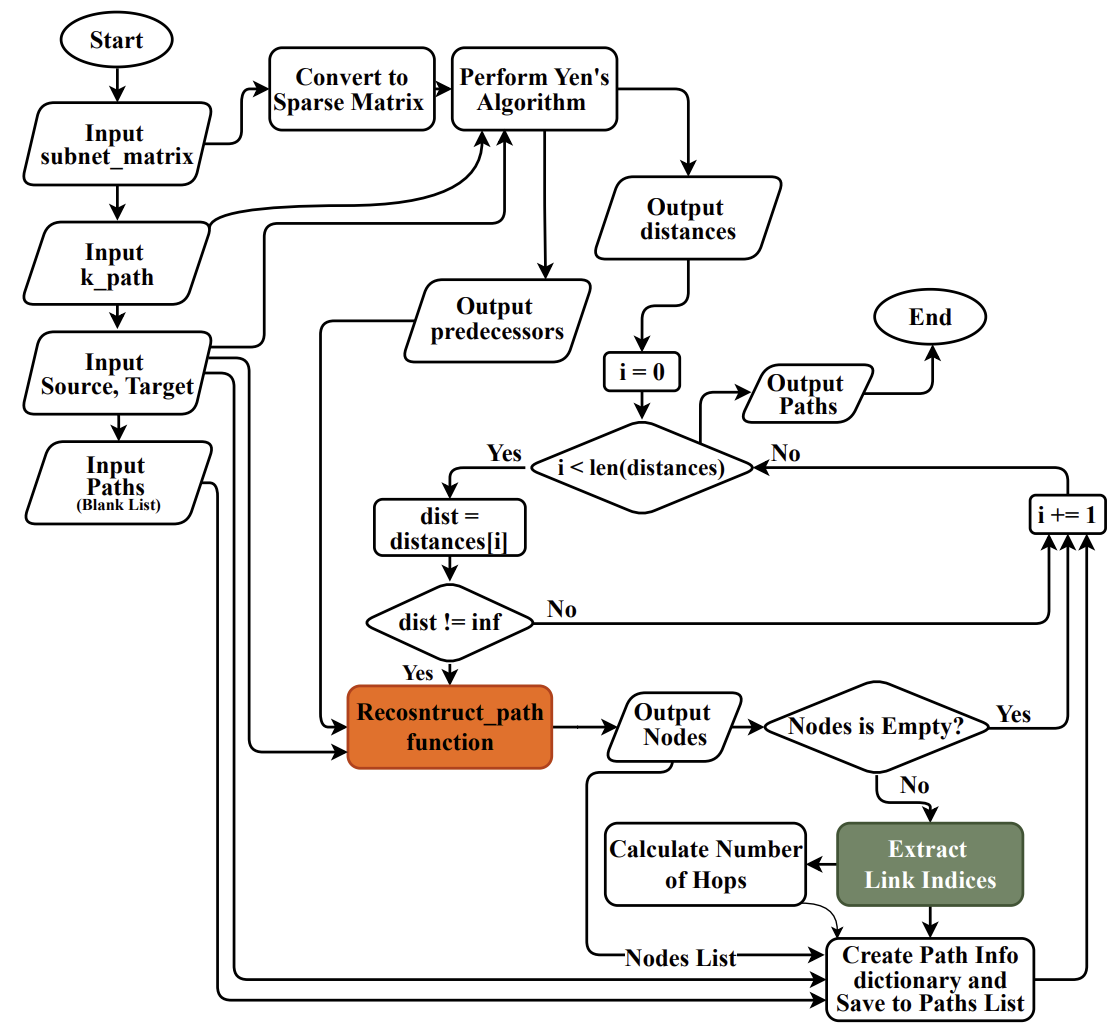}
	\caption{Flowchart of \texttt{compute\_k\_shortest\_paths()} function.}
	\label{fig:K_shortest_path_flowchart}
\end{figure}

\vspace{-0.6em}
\paragraph{Description}
The \texttt{compute\_k\_shortest\_paths()} function computes the $k$-shortest loopless paths between a source and a target node using Yen's algorithm. It generates multiple candidate routes for optical network planning, calculates their total distances, and appends the results to the provided path list with detailed topological and distance information.
The flowchart of this function is shown in Fig.~\ref{fig:K_shortest_path_flowchart}. It employs Yen's $k$-shortest paths algorithm, which incrementally generates alternative routes by perturbing the globally shortest path and recalculating minimal-cost deviations. This ensures that all resulting paths are simple (non-cyclic) and sorted in ascending order of total distance.

Initially, the adjacency matrix of the optical subnet is transformed into a compressed sparse representation to improve both memory efficiency and computational performance. Yen's algorithm is then executed for a specified number $k$ of alternative paths between each source-destination pair, returning both the total path distances and the corresponding predecessor matrix.

Using \texttt{\_reconstruct\_yen\_path()} function, each candidate path is reconstructed by backtracking through the predecessor chain from the destination to the source node. The physical link indices corresponding to the traversed node pairs are subsequently identified, taking into account bidirectional link orientation since the optical network is undirected. The corresponding \texttt{link\_indices} array is obtained using the internal helper function \texttt{\_get\_link\_indices\_in\_path()}. The resulting paths are stored as structured records containing the node sequence, link indices, total propagation distance, and hop count.

This $k$-path computation forms the analytical foundation for redundancy assessment, fiber-pair allocation, and LAND pair identification within the SixGman framework. It enables systematic evaluation of routing diversity and topological resilience, serving as a critical pre-processing stage for higher-level planning and optimization procedures.

{\small
	\vspace{-0.5em}
	\paragraph{Parameters}
	\begin{itemize}[noitemsep, topsep=0pt, parsep=0pt, partopsep=0pt, leftmargin=*]
		\item \texttt{subnet\_matrix (np.ndarray):} Adjacency matrix of the subnet representing physical connectivity.
		\item \texttt{paths (List[Dict]):} List to append path dictionaries to (can be empty initially).
		\item \texttt{source (int):} Source node ID.
		\item \texttt{target (int):} Target node ID.
		\item \texttt{k (int, optional):} Number of shortest paths to compute (default: 20).
	\end{itemize}
	\vspace{-0.5em}
	\paragraph{Returns}
	\begin{itemize}[noitemsep, topsep=0pt, leftmargin=*]
		\item \texttt{List[Dict]:} Updated list of paths, where each dictionary contains:
		\begin{itemize}[noitemsep, topsep=0pt, leftmargin=*]
			\item \texttt{src\_node (int):} Source node.
			\item \texttt{dest\_node (int):} Destination node.
			\item \texttt{nodes (List[int]):} Sequence of nodes forming the path.
			\item \texttt{links (List[int]):} Link indices corresponding to the traversed physical connections.
			\item \texttt{distance (float):} Total path distance.
			\item \texttt{num\_hops (int):} Number of hops along the path.
		\end{itemize}
	\end{itemize}
}

\vspace{-0.5em}
\subsection{\texttt{get\_node\_degrees()} Function}

\noindent\texttt{get\_node\_degrees(nodes) $\rightarrow$ np.ndarray}
\vspace{-0.6em}
\paragraph{Description}
The \texttt{get\_node\_degrees()} function returns the degree of specified nodes in the network graph. The degree represents the number of edges connected to each node.

{\small
	\vspace{-0.5em}
	\paragraph{Parameters}
	\begin{itemize}[noitemsep, topsep=0pt, parsep=0pt, partopsep=0pt, leftmargin=*]
		\item \texttt{nodes (List[int]):} List of node IDs for which the degree is to be retrieved.
	\end{itemize}
	\vspace{-0.5em}
	\paragraph{Returns}
	\begin{itemize}[noitemsep, topsep=0pt, leftmargin=*]
		\item \texttt{np.ndarray:} 2D numpy array containing node IDs and their relative degree.
	\end{itemize}
}
\vspace{-0.5em}
\subsection{\texttt{land\_pair\_finder()} Function}

\noindent\texttt{and\_pair\_finder(src\_list, candidate\_paths\_sorted,\\ num\_pairs) $\rightarrow$ pd.DataFrame}
\vspace{-0.6em}
\paragraph{Description}
The \texttt{land\_pair\_finder()} function identifies link- and node-disjoint path pairs (LAND pairs) for each source node. A LAND pair consists of a primary and secondary path that: 
\begin{itemize}[noitemsep, topsep=0pt, parsep=0pt, partopsep=0pt, leftmargin=*]
	\item Have different destination nodes (multi-homed protection),
	\item Are node-disjoint except at the source, and
	\item Are link-disjoint.
\end{itemize}

{\small
	\vspace{-0.5em}
	\paragraph{Parameters}
	\begin{itemize}[noitemsep, topsep=0pt, parsep=0pt, partopsep=0pt, leftmargin=*]
		\item \texttt{src\_list (List[int]):} List of source node IDs to process.
		\item \texttt{candidate\_paths\_sorted (pd.DataFrame):} DataFrame of candidate paths. Must include columns: \texttt{['src\_node', 'dest\_node', 'nodes', 'links', 'index', 'distance', 'num\_hops']}.
		\item \texttt{num\_pairs (int):} Maximum number of disjoint pairs to return per source node.
	\end{itemize}
	\vspace{-0.5em}
	\paragraph{Returns}
	\begin{itemize}[noitemsep, topsep=0pt, leftmargin=*]
		\item \texttt{pd.DataFrame:} DataFrame containing columns:
		
		\texttt{['primary\_path\_IDx', 'secondary\_path\_IDx', 'numHops\_secondary', 'distance\_secondary', 'src\_node']}.
	\end{itemize}
}

\vspace{-0.5em}
\subsection{\texttt{calc\_num\_pair()} Function}

\noindent\texttt{calc\_num\_pair(pairs\_disjoint\_df) $\rightarrow$ np.ndarray}
\vspace{-0.6em}
\paragraph{Description}
The \texttt{calc\_num\_pair()} function calculates the total number of candidate LAND pairs in the network based on the provided DataFrame.

{\small
	\vspace{-0.5em}
	\paragraph{Parameters}
	\begin{itemize}[noitemsep, topsep=0pt, parsep=0pt, partopsep=0pt, leftmargin=*]
		\item \texttt{pairs\_disjoint\_df (DataFrame):} DataFrame containing disjoint path pairs.
	\end{itemize}
	\vspace{-0.5em}
	\paragraph{Returns}
	\begin{itemize}[noitemsep, topsep=0pt, leftmargin=*]
		\item A numpy array containing the number of candidate LAND pairs for each source node
	\end{itemize}
}

\vspace{-1em}
\section{Band Module Functions}
\label{sec:band_functions}
In this module, two primary classes are defined. The first class, \texttt{OpticalParameters}, is a data class designed to store and compute fiber and system parameters for optical network modeling. This class does not include any user-defined functions. The second class, \texttt{Band}, encapsulates the functionality related to optical transmission bands and their frequency grid computations. In this section, we describe the functions implemented within the \texttt{Band} class in detail.

\vspace{-0.5em}
\subsection{\texttt{calc\_spectrum()} Function}

\noindent\texttt{calc\_spectrum() $\rightarrow$ np.ndarray}

\vspace{-0.6em}
\paragraph{Description}
The \texttt{calc\_spectrum()} function computes the frequency grid (spectrum) for the optical transmission band based on the defined start frequency, end frequency, and channel spacing parameters.

{\small
	\vspace{-0.5em}
	\paragraph{Returns}
	\begin{itemize}[noitemsep, topsep=0pt, leftmargin=*]
		\item \texttt{np.ndarray:} Array of center frequencies (in THz) representing the computed spectral grid.
	\end{itemize}
}

\vspace{-0.5em}
\subsection{\texttt{process\_link\_gsnr()} Function}

\noindent\texttt{process\_link\_gsnr(f\_c\_axis, Pch\_dBm, num\_Ch\_mat,\\ spectrum\_C, Nspan\_array, hierarchy\_level,\\ minimum\_hierarchy\_level, result\_directory) $\rightarrow$ Tuple}

\vspace{-0.6em}
\paragraph{Description}
The \texttt{process\_link\_gsnr()} function computes the Generalized Signal-to-Noise Ratio (GSNR) and optimal power for all network links corresponding to a specific hierarchy level subnetwork. It evaluates optical performance metrics across different launch powers and frequency channels, caching intermediate results for efficient reuse during hierarchical network analysis.

\vspace{-0.5em}
\paragraph{GSNR Model}
In this framework, the \textit{Generalized Signal-to-Noise Ratio} (GSNR) is employed as the primary quality-of-transmission (QoT) metric to evaluate the optical performance of lightpaths. The Gaussian Noise (GN) model is utilized to account for both linear effects, such as fiber attenuation and chromatic dispersion, and nonlinear interference (NLI) phenomena, including self-phase modulation (SPM), cross-phase modulation (XPM), and inter-channel stimulated Raman scattering (ISRS). 

The GN model is particularly suitable for coherently detected systems, where signal degradation due to NLI can be effectively modeled as additive Gaussian noise. The GSNR for each channel $i$ within link $l$ is expressed as
\begin{equation}
	\mathrm{GSNR}^{l,i} \simeq 
	\frac{P_{\mathrm{tx}}^{l,i}}
	{P_{\mathrm{ASE}}^{l,i} + P_{\mathrm{NLI}}^{l,i}},
	\label{eq:GSNR_link}
\end{equation}
where $P_{\mathrm{tx}}^{l,i}$ denotes the launch power at the beginning of the link, 
$P_{\mathrm{ASE}}^{l,i}$ is the amplified spontaneous emission (ASE) noise power generated by the erbium-doped fiber amplifier (EDFA), and 
$P_{\mathrm{NLI}}^{l,i}$ is the nonlinear interference (NLI) power caused by SPM, XPM, and ISRS effects.

The ASE noise power for the EDFA can be written as
\begin{equation}
	P_{\mathrm{ASE}}^{l,i} 
	= n_{\mathrm{F}} \, h f^{i} 
	\left( G^{l,i} - 1 \right)
	B_{\mathrm{ch}}^{l,i},
	\label{eq:P_ASE}
\end{equation}
where $n_{\mathrm{F}}$ is the amplifier noise figure, 
$h$ is Planck's constant, 
$f^{i}$ is the optical frequency, 
$G^{l,i}$ is the EDFA gain, and 
$B_{\mathrm{ch}}^{l,i}$ is the channel bandwidth.

The nonlinear interference power is modeled as
\begin{equation}
	P_{\mathrm{NLI}}^{l,i} 
	= G_{\mathrm{NLI}}^{l,i} 
	B_{\mathrm{ch}}^{l,i},
	\label{eq:P_NLI}
\end{equation}
where $G_{\mathrm{NLI}}^{l,i}$ represents the nonlinear interference coefficient determined by the GN model.

The end-to-end GSNR of a lightpath (LP) across multiple links is then computed through the incoherent accumulation of noise contributions as
\begin{equation}
	\begin{split}
		\mathrm{GSNR}_{\mathrm{LP}}^{i}
		= 10 \log_{10} \Bigg( 
		\Bigg[ &
		\Bigg( 
		\sum_{l=1}^{N_{\mathrm{link}}}
		\frac{1}{\mathrm{GSNR}^{l,i}}
		+ \mathrm{SNR}_{\mathrm{TRx}}^{-1}
		\Bigg)^{-1}
		\Bigg] 
		\Bigg) \\
		& - \mathrm{SNR}_{\mathrm{Pen_{filter}}}
		- \mathrm{SNR}_{\mathrm{mrg_{Aging}}}.
	\end{split}
	\label{eq:GSNR_LP}
\end{equation}

\noindent
where $\mathrm{SNR}_{\mathrm{TRx}}$, $\mathrm{SNR}_{\mathrm{Pen_{filter}}}$ and $\mathrm{SNR}_{\mathrm{mrg_{Aging}}}$ denote the transceiver SNR, 
the SNR penalty due to WSS filtering, and the SNR margin accounting for aging effects, respectively.

Equations~\eqref{eq:GSNR_link}--\eqref{eq:GSNR_LP} together describe the GSNR-based QoT estimation model used in SixGman, incorporating ASE and NLI noise sources, as well as transceiver and system-level penalties.

{\small
	\vspace{-0.5em}
	\paragraph{Parameters}
	\begin{itemize}[noitemsep, topsep=0pt, parsep=0pt, partopsep=0pt, leftmargin=*]
		\item \texttt{f\_c\_axis (np.ndarray):} Center frequencies of the channels [THz].
		\item \texttt{Pch\_dBm (np.ndarray):} Per channel candidate launch power values [dBm].
		\item \texttt{num\_Ch\_mat (np.ndarray):} Channel indices or counts used for modulation processing.
		\item \texttt{spectrum\_C (np.ndarray):} C-band frequency set, used for band-dependent parameter calculation.
		\item \texttt{Nspan\_array (np.ndarray):} Number of spans for each link in the network.
		\item \texttt{hierarchy\_level (int):} Current hierarchy level of the network topology.
		\item \texttt{minimum\_hierarchy\_level (int):} Lowest hierarchy level included in the analysis.
		\item \texttt{result\_directory (Path):} Output directory for storing GSNR results.
	\end{itemize}
	
	\vspace{-0.5em}
	\paragraph{Returns}
	\begin{itemize}[noitemsep, topsep=0pt, leftmargin=*]
		\item \texttt{Tuple:} Contains the GSNR matrix, throughput per power, optimal throughput, and optimal power array.
	\end{itemize}
}

\vspace{-1em}
\section{Planning Module Functions}
\label{sec:planning_functions}
This section describes the functionality of each method implemented in the \texttt{PlanningTool} class, which is the sole class defined in this module. The \texttt{PlanningTool} class integrates network topology, spectral, and optical performance parameters to perform network planning and resource optimization tasks within the SixGman framework.

\vspace{-0.5em}
\subsection{\texttt{initialize\_planner()} Function}

\noindent\texttt{initialize\_planner(num\_fslots, hierarchy\_level,\\ minimum\_hierarchy\_level, rolloff=0.1, SR=40.0e9,\\ Max\_bit\_rate\_BVT=array([400, 348, 280, 200, 120, 80]), Ref\_license\_capacity=array([100, 87, 70, 50, 30, 20]), \\ FP\_max\_num = 20, band\_sepration\_idx=[96, 120])}

\vspace{-0.6em}
\paragraph{Description}
The \texttt{initialize\_planner()} function initializes key planning-related variables and data structures required for network simulation and optical resource allocation. It sets up matrices and parameters for simulating BVT assignments, GSNR evaluation, fiber pair tracking, and spectrum utilization across different optical bands and hierarchy levels.

{\small
	\vspace{-0.5em}
	\paragraph{Parameters}
	\begin{itemize}[noitemsep, topsep=0pt, parsep=0pt, partopsep=0pt, leftmargin=*]
		\item
		 \texttt{num\_fslots (int):} Number of available frequency slots in the network.
		\item
		 \texttt{hierarchy\_level (int):} Target hierarchy level for current planning iteration.
		\item
		 \texttt{minimum\_hierarchy\_level (int):} Minimum hierarchy level considered for subgraph-based planning.
		\item
		 \texttt{rolloff (float):} Rolloff factor for spectral shaping (default: 0.1).
		\item
		 \texttt{SR (float):} Symbol rate in baud (default: 40e9 baud).
		\item
		 \texttt{Max\_bit\_rate\_BVT (np.ndarray):} Array of supported BVT bit rates in Gbps.  
		\\ Example: \texttt{[400, 320, 260, 200, 120, 64]} indicates bitrate of BVTs for PM-64QAM, PM-32QAM, PM-16QAM, PM-8QAM, PM-QPSK, PM-BPSK in 75~GHz channel spacing and 64~GBaud symbol rate.
		\item 
		\texttt{Ref\_license\_capacity (np.ndarray):} Array of reference license capacities of different BVT bitrates  .
		\\ Example: \texttt{[100, 80, 65, 50, 30, 16]} indicates license capacities for different BVT Types of the previous example (consider four licenses per BVT).
		\item 
		\texttt{FP\_max\_num (int):} Maximum number of available fiber pairs per link (default: 20).
		\item 
		\texttt{band\_sepration\_idx (list):} Index of the last frequency slot of the C-band and SuperC-band.
		\\ Example: for the channel spacing of 75~GHz, C-band with 4.8~THz bandwidth contains 64 frequency slots, and SuperC-band with 1.2~THz bandwidth contains 16 frequency slots, so the band\_sepration\_idx = [64, 64 + 16], note that from the 80th frequency slot to the end of the spectrum is inside the L-band (80 frequency slot).
		
	\end{itemize}
	
	\vspace{-0.5em}
	\paragraph{Returns}
	\begin{itemize}[noitemsep, topsep=0pt, leftmargin=*]
		\item This function does not return a value. It initializes internal planning matrices and parameters used by subsequent simulation and optimization routines.
	\end{itemize}
}

\vspace{-0.5em}
\subsection{\texttt{generate\_initial\_traffic\_profile()} Function}

\noindent\texttt{generate\_initial\_traffic\_profile(num\_nodes, optical\_nodes monteCarlo\_steps, min\_rate, max\_rate, seed, result\_directory) $\rightarrow$ None}

\vspace{-0.6em}
\paragraph{Description}
The \texttt{generate\_initial\_traffic\_profile()} function generates or loads the initial traffic capacity profile for network nodes using Monte Carlo simulation. It estimates the initial traffic demand or capacity for each node by performing multiple simulations, generating random capacities uniformly distributed between the specified minimum and maximum rates. If a precomputed capacity file exists in the specified directory, it is loaded to avoid redundant computation.  

The resulting per-node average capacities are stored internally in \texttt{self.HL\_capacity\_final}.

{\small
	\vspace{-0.5em}
	\paragraph{Parameters}
	\begin{itemize}[noitemsep, topsep=0pt, parsep=0pt, partopsep=0pt, leftmargin=*]
		\item \texttt{num\_nodes (int):} Number of nodes in the network for which to simulate traffic.
		\item \texttt{optical\_nodes (list):} List of all nodes with no electrical aggregation (optically bypass nodes), these nodes not have any initial traffic and can't be the destination of any node
		\item \texttt{monteCarlo\_steps (int):} Number of Monte Carlo iterations used for averaging.
		\item \texttt{min\_rate (float):} Minimum traffic rate (Gbps) per node.
		\item \texttt{max\_rate (float):} Maximum traffic rate (Gbps) per node.
		\item \texttt{seed (int):} Initial random seed for reproducibility.
		\item \texttt{result\_directory (Path):} Directory for storing or loading simulation results.
	\end{itemize}
	
	\vspace{-0.5em}
	\paragraph{Updates}
	\begin{itemize}[noitemsep, topsep=0pt, leftmargin=*]
		\item \texttt{self.HL\_capacity\_final (np.ndarray):} Final per-node traffic capacities averaged over all Monte Carlo simulations.
	\end{itemize}
	
	\paragraph{Returns}
	\begin{itemize}[noitemsep, topsep=0pt, leftmargin=*]
		\item This method does not return a value; instead, it updates the internal attribute \texttt{self.HL\_capacity\_final} with the averaged per-node traffic capacities.
	\end{itemize}
}

\vspace{-0.5em}
\subsection{\texttt{simulate\_traffic\_annual()} Function}

\noindent\texttt{simulate\_traffic\_annual(lowest\_hierarchy\_dict, CAGR,\\ result\_directory) $\rightarrow$ None}

\vspace{-0.6em}
\paragraph{Description}
The \texttt{simulate\_traffic\_annual()} function models annual traffic evolution for the lowest hierarchy-level nodes (e.g., standalone \& co-located HL4) by applying a constant Compound Annual Growth Rate (CAGR) to each node's base capacity. It computes annual traffic, incremental added traffic, required 100G licenses, and residual capacities for both standalone and co-located nodes. If precomputed results exist in the specified directory, they are loaded to save computation time; otherwise, the full simulation is executed and results are saved.

{\small
	\vspace{-0.5em}
	\paragraph{Parameters}
	\begin{itemize}[noitemsep, topsep=0pt, parsep=0pt, partopsep=0pt, leftmargin=*]
		\item \texttt{lowest\_hierarchy\_dict (dict):} Dictionary of node IDs for the lowest hierarchy level, with keys \texttt{'standalone'} and \texttt{'colocated'}.
		\item \texttt{CAGR (float):} Compound Annual Growth Rate (e.g., 0.4 for 40\% annual increase).
		\item \texttt{result\_directory (Path):} Directory where results are read from or written to.
	\end{itemize}
	\vspace{-0.5em}
	\paragraph{Updates}
	\begin{itemize}[noitemsep, topsep=0pt, leftmargin=*]
		\item \texttt{self.lowest\_HL\_added\_traffic\_annual\_standalone (np.ndarray):} Annual incremental traffic for standalone nodes (years $\times$ nodes).
		\item \texttt{self.lowest\_HL\_added\_traffic\_annual\_colocated (np.ndarray):} Annual incremental traffic for co-located nodes (years $\times$ nodes).
	\end{itemize}
	\vspace{-0.5em}
	\paragraph{Returns}
	\begin{itemize}[noitemsep, topsep=0pt, leftmargin=*]
		\item This method does not return a value; instead, it updates the internal attributes listed above.
	\end{itemize}
}
\vspace{-0.5em}
\subsection{\texttt{\_spectrum\_assignment()} Function}

\noindent\texttt{\_spectrum\_assignment(path\_IDx, path\_type, year, \\ K\_path\_attributes\_df, 
	pure\_traffic\_to\_assign, BVT\_number, \\node\_IDx, node\_list, GSNR\_link, LSP\_array\_pair, \\Year\_FP\_pair, HL\_subnet\_links) $\rightarrow$ dict, np.ndarray, np.ndarray}

\begin{figure}
	\centering
	\includegraphics[height = 8cm]{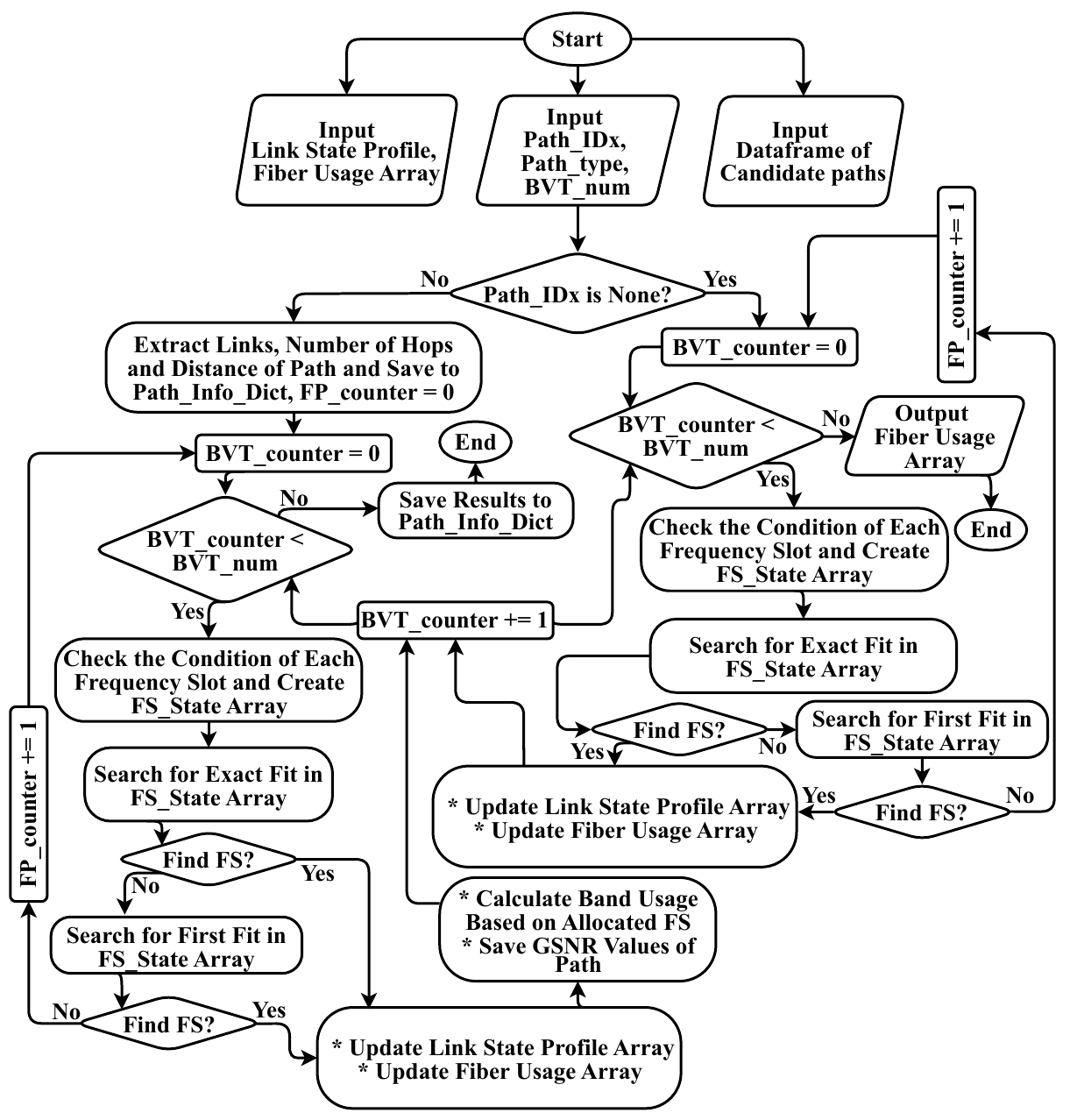}
	\caption{Flowchart of \texttt{\_spectrum\_assignment()} function.}
	\label{fig:spectrum_assignment_flowchart}
\end{figure}

\vspace{-0.6em}
\paragraph{Description}
The \texttt{\_spectrum\_assignment()} function performs spectrum and fiber pair assignment for a given lightpath in a hierarchical network. It supports primary and secondary paths for standalone HL nodes as well as co-located HL nodes. The function tracks spectrum occupancy, updates fiber usage, and computes GSNR and per-BVT costs. The flowchart of this function is shown in Fig.~\ref{fig:spectrum_assignment_flowchart}.
As illustrated in the flowchart, there exists a slight variation between the spectrum assignment procedure of the primary co-located path and that of the other paths. This distinction is represented by setting the input parameter \texttt{path\_IDx = None}. For the primary and secondary paths of standalone nodes, as well as the secondary paths of co-located nodes, the algorithm follows a unified approach.

The algorithm first extracts relevant information about each path and stores it in a dictionary named \texttt{path\_info\_storage}. It then performs spectrum assignment iteratively for a specified number of BVTs. During the assignment of each BVT, the algorithm initially examines the availability condition of every frequency slot (FS) along the path. It subsequently attempts to identify a suitable slot using the exact-fit method. If no suitable slot is found, the first-fit method is applied as a fallback.

Once an appropriate FS is located, both the link state profile and the fiber pair usage arrays are updated before proceeding to the next BVT. It is important to note that if no FS can be allocated after applying both methods (exact-fit and first-fit), the algorithm advances to the next available fiber pair.

The bitrate of each established BVT is specified using it's GSNR value, so after allocating frequency slots for all BVTs, we calculate the total bitrate of all established BVT and compare it with pure\_traffic\_to\_assign, if the total bitrate value doesn't exceed the pure\_traffic\_to\_assign, the algorithm try to deploy another BVT. Finally the results are stored in \texttt{path\_info\_storage} to be used for calculating the cost of each LAND pair, enabling the selection of the pair with the lowest overall cost.

For the primary co-located path (where \texttt{path\_IDx = None}), a few minor deviations from the general algorithm exist, as depicted in the flowchart.

This function is used internally by the \texttt{run\_planner()} method.

{\small
	\vspace{-0.5em}
	\paragraph{Parameters}
	\begin{itemize}[noitemsep, topsep=0pt, parsep=0pt, partopsep=0pt, leftmargin=*]
		
		\item \texttt{path\_IDx (int or None):} Index of the selected path in \\ \texttt{K\_path\_attributes\_df}. \texttt{None} indicates a primary path of a co-located node.
		\item \texttt{path\_type (str):} Type of path (\texttt{'primary'} or \texttt{'secondary'}).
		\item \texttt{year (int):} Planning year for multi-period simulation.
		\item \texttt{K\_path\_attributes\_df (pd.DataFrame):} DataFrame containing attributes of all K-shortest paths.
		\item \texttt{pure\_traffic\_to\_assign (float):} Pure traffic value of this node that must be routed in this year.
		\item \texttt{BVT\_number (int):} Number of BVTs to assign for this path.
		\item \texttt{node\_IDx (int):} Index of the current node being processed.
		\item \texttt{node\_list (List):} List of all node identifiers.
		\item \texttt{GSNR\_link (np.ndarray):} GSNR values per frequency slot per link.
		\item \texttt{LSP\_array\_pair (np.ndarray):} Spectrum occupancy array [FS, link, fiber pair].
		\item \texttt{Year\_FP\_pair (np.ndarray):} Annual fiber pair usage array [year, link].
		\item \texttt{HL\_subnet\_links (np.ndarray):} List of hierarchical level subnetwork link indices.
	\end{itemize}
	\vspace{-0.5em}
	\paragraph{Updates}
	\begin{itemize}[noitemsep, topsep=0pt, leftmargin=*] 
		\item
		 \texttt{LSP\_array\_pair}: Updates spectrum occupancy for allocated frequency slots and fiber pairs.
		\item
		 \texttt{Year\_FP\_pair}: Updates annual fiber pair usage for links in the assigned path.
		\item
		 \texttt{Year\_FP\_HL\_colocated}: Updated for primary path of co-located HL nodes when \texttt{path\_IDx} is \texttt{None}.
		\item
		 \texttt{BVT\_CBand\_count\_path, BVT\_superCBand\_count\_path, \\BVT\_superCLBand\_count\_path}: Counts BVTs assigned in each spectral band.
		\item 
		\texttt{self.num\_added\_license\_this\_year\_primary / \_secondary}: Stores number of added licenses (primary and secondary paths).
		\item 
		\texttt{self.HL\_BVT\_number\_all\_annual}: Annual count of deployed BVTs across all nodes.
		\item 
		\texttt{self.HL\_BVT\_number\_per\_node}: Annual count of deployed BVTs per node.
		\item 
		\texttt{self.BVT\_establishment\_info}: Information of established BVTs per node.
		\item 
		\texttt{self.num\_license\_last\_BVT\_primary}: Number of activated licenses in the last established BVT (primary path).
		\item 
		\texttt{self.num\_license\_last\_BVT\_secondary}: Number of activated licenses in the last established BVT (secondary path).
		\item 
		\texttt{self.last\_BVT\_type\_primary}: Bitrate of the last established BVT (primary) per node. 
		\item 
		\texttt{self.last\_BVT\_type\_secondary}: Bitrate of the last established BVT (secondary) per node. 
		\item 
		\texttt{self.last\_BVT\_Band\_primary}: Band in which the last BVT (primary) is established. (1: C-band, 2: SuperC-band, 3: L-band)
		\item
		\texttt{self.last\_BVT\_Band\_secondary}: Band in which the last BVT (secondary) is established. (1: C-band, 2: SuperC-band, 3: L-band)
	\end{itemize}
	\vspace{-0.5em}
	\paragraph{Returns}
	\begin{itemize}[noitemsep, topsep=0pt, leftmargin=*]
		\item If \texttt{path\_IDx} is not \texttt{None} (normal path):
		\begin{itemize}[noitemsep, topsep=0pt, leftmargin=*]
			\item \texttt{path\_info\_storage (dict):} Contains fiber usage, band counts and other cost metrics.
			\item \texttt{LSP\_array\_pair (np.ndarray):} Updated spectrum occupancy.
			\item \texttt{Year\_FP\_pair (np.ndarray):} Updated fiber usage per link.
		\end{itemize}
		\item If \texttt{path\_IDx} is \texttt{None} (co-located node):
		\begin{itemize}[noitemsep, topsep=0pt, leftmargin=*]
			\item \texttt{Year\_FP\_HL\_colocated (np.ndarray):} Updated fiber pair usage for co-located HL node (primary path).
			\item 
			\texttt{BVT\_bitrate\_storage}: Store the bitrate of established BVTs.
			\item 
			\texttt{FS\_BVT\_storage}: Store the allocated frequency slots of established BVTs.
		\end{itemize}
	\end{itemize}
}

\vspace{-0.5em}
\subsection{\texttt{\_update\_hl\_node\_degrees()} Function}

\noindent\texttt{\_update\_hl\_node\_degrees(hierarchy\_level, Year\_FP) $\rightarrow$ None}
\vspace{-1.6em}
\paragraph{Description}
The \texttt{\_update\_hl\_node\_degrees()} function computes the average node degree in each hierarchical level over time based on annual fiber-pair allocations. It tracks how the number of active fiber-pair connections evolves year by year by comparing each year's fiber-pair matrix (\texttt{Year\_FP}) with the previous year. The results are stored internally for further analysis of network connectivity dynamics.

{\small
	\vspace{-0.5em}
	\paragraph{Parameters}
	\begin{itemize}[noitemsep, topsep=0pt, parsep=0pt, partopsep=0pt, leftmargin=*]
		\item \texttt{hierarchy\_level (int):} The HL level to analyze (e.g., 4 for HL4 nodes).
		\item \texttt{Year\_FP (np.ndarray):} 2D array (years $\times$ links) indicating allocated fiber pairs per link for each year.
	\end{itemize}
	\vspace{-0.5em}
	\paragraph{Updates}
	\begin{itemize}[noitemsep, topsep=0pt, leftmargin=*]
		\item \texttt{self.degree\_number\_HLs (np.ndarray):} Stores the average node degree of HL nodes for each simulated year.
	\end{itemize}
	\vspace{-0.5em}
	\paragraph{Returns}
	\begin{itemize}[noitemsep, topsep=0pt, leftmargin=*]
		\item This method does not return a value; the results are stored in \texttt{self.degree\_number\_HLs}.
	\end{itemize}
}
\vspace{-0.5em}
\subsection{\texttt{\_calculate\_BVT\_usage()} Function}

\noindent\texttt{\_calculate\_BVT\_usage() $\rightarrow$ None}

\vspace{-0.6em}
\paragraph{Description}
The \texttt{\_calculate\_BVT\_usage()} function computes the cumulative usage of BVTs and the total 100G license counts for each year of the planning period. It aggregates deployed BVTs across all optical bands (C, SuperC, and L) and stores the results internally for reporting or visualization. The calculation assumes four 100G licenses per BVT and reflects cumulative infrastructure growth over time.

{\small
	\vspace{-0.5em}
	\paragraph{Updates}
	\begin{itemize}[noitemsep, topsep=0pt, leftmargin=*]
		\item \texttt{self.HL\_All\_100G\_lincense (np.ndarray):} Cumulative total 100G license usage across all nodes per year.
		\item \texttt{self.HL\_BVTNum\_All (np.ndarray):} Cumulative total number of BVTs (all bands combined) per year.
		\item \texttt{self.HL\_BVTNum\_CBand (np.ndarray):} Cumulative number of C-band BVTs per year.
		\item \texttt{self.HL\_BVTNum\_SuperCBand (np.ndarray):} Cumulative number of SuperC-band BVTs per year.
		\item \texttt{self.HL\_BVTNum\_LBand (np.ndarray):} Cumulative number of L-band BVTs per year.
	\end{itemize}
	\vspace{-0.5em}
	\paragraph{Returns}
	\begin{itemize}[noitemsep, topsep=0pt, leftmargin=*]
		\item This method does not return a value. Results are stored in the instance attributes listed above..
	\end{itemize}
}
\vspace{-0.5em}
\subsection{\texttt{\_save\_network\_results()} Function}

\noindent\texttt{\_save\_network\_results(hierarchy\_level,\\ minimum\_hierarchy\_level, result\_directory)}

\vspace{-0.6em}
\paragraph{Description}
The \texttt{\_save\_network\_results()} function saves detailed network planning results for a specified hierarchy level in compressed NPZ files. It generates the subgraph corresponding to the current hierarchy level, extracts relevant link indices, computes degree metrics per link and spectral band, and stores results including BVT allocations, link usage, node capacity profiles, traffic flows, and GSNR measurements.

{\small
	\vspace{-0.5em}
	\paragraph{Parameters}
	\begin{itemize}[noitemsep, topsep=0pt, parsep=0pt, partopsep=0pt, leftmargin=*]
		\item \texttt{hierarchy\_level (int):} Current hierarchy level being analyzed.
		\item \texttt{minimum\_hierarchy\_level (int):} Minimum hierarchy level considered when generating the subgraph.
		\item \texttt{result\_directory (Path):} Directory where output NPZ files will be saved.
	\end{itemize}
	\vspace{-0.5em}
	\paragraph{Saves}
	\begin{itemize}[noitemsep, topsep=0pt, leftmargin=*]
		\item \texttt{\{topology\_name\}\_HL\{hierarchy\_level\}\_bvt\_info.npz}:
		\begin{itemize}[noitemsep, leftmargin=*]
			\item
			 \texttt{HL\_All\_100G\_lincense}: Cumulative number of activated 100G licenses.
			 \item 
			 \texttt{HL\_annual\_license}: Number of activated 100G licenses per year.
			 \item 
			 \texttt{HL\_CBand\_license}: Number of activated 100G licenses per year in C-band.
			 \item
			 \texttt{HL\_SuperCBand\_license}: Number of activated 100G licenses per year in SuperC-band.
			 \item 
			 \texttt{HL\_LBand\_license}: Number of activated 100G licenses per year in L-band.
			\item
			 \texttt{HL\_BVTNum\_All}: Cumulative number of deployed BVTs.
			\item
			 \texttt{HL\_BVTNum\_CBand}: Cumulative number of deployed BVTs in the C-band.
			\item
			 \texttt{HL\_BVTNum\_SuperCBand}: Cumulative number of deployed BVTs in the SuperC-band.
			\item
			 \texttt{HL\_BVTNum\_LBand}: Cumulative number of deployed BVTs in the L-band.
			\item
			\texttt{BVT\_establishment\_info}: Information of established BVTs.
		\end{itemize}
		\item \texttt{\{topology\_name\}\_HL\{hierarchy\_level\}\_link\_info.npz}:
		\begin{itemize}[noitemsep, leftmargin=*]
			\item \texttt{HL\_links\_indices}: Indices of links in the HL subgraph.
			\item \texttt{num\_link\_CBand\_annual}: Cumulative number of FPs that used in C-band in each link.
			\item \texttt{num\_link\_SupCBand\_annual}: Cumulative number of FPs that used in SuperC-band in each link.
			\item \texttt{num\_link\_LBand\_annual}: Cumulative number of FPs that used in L-band in each link.
			\item \texttt{HL\_CDegree\_Domain}: Weighted node degree for C-band links (2 per link per endpoint).
			\item \texttt{HL\_SuperCDegree\_Domain}: Weighted node degree for SuperC-band links.
			\item \texttt{HL\_LDegree\_Domain}: Weighted node degree for L-band links.
			\item \texttt{Total\_effective\_FP\_new\_annual}: Total fiber-pair length used across all links per year (km).
			\item \texttt{HL\_FPNum}: Fiber pair usage per link per year (\texttt{Year\_FP\_new}).
			\item \texttt{HL\_FPNumCo}: Fiber pair usage per co-located HL node (just primary paths) per year (\texttt{Year\_FP\_HL\_colocated}).
			\item \texttt{degree\_number\_HLs}: Node degree per HL node per year.
			\item \texttt{traffic\_flow\_links\_array}: Added traffic flow on each link per year.
			\end{itemize}
		\item \texttt{\{topology\_name\}\_HL\{hierarchy\_level\}\_path\_GSNR\_info.npz}:
		\begin{itemize}[noitemsep, leftmargin=*]
			\item \texttt{GSNR\_all\_paths}: GSNR values for all BVTs per year.
			\item \texttt{GSNR\_primary}: GSNR values for BVTs of primary paths per year.
			\item \texttt{GSNR\_secondary}: GSNR values for BVTs of secondary paths per year.
		\end{itemize}
		\item \texttt{\{topology\_name\}\_HL\{hierarchy\_level\}\_node\_capacity\_\\
			profile\_array.npz}:
		\begin{itemize}[noitemsep, leftmargin=*]
			\item \texttt{node\_capacity\_profile\_array (np.ndarray):} Array representing the evolution of node capacity over the simulation period.  
			Each element indicates the amount of additional traffic aggregated at a node in a given year.  
			For the first year, the values correspond to the initial traffic aggregated at each node.

		\end{itemize}
		
		\item \texttt{\{topology\_name\}\_HL\{hierarchy\_level\}\_segments\_latency.npz}:
		\begin{itemize}[noitemsep, leftmargin=*]
			\item \texttt{latency}: Latency values of the primary and secondary paths for each node within this hierarchy level.
			\item \texttt{destinations}: Destination node IDs corresponding to the primary and secondary paths for each node in this hierarchy level.
		\end{itemize}
	\end{itemize}
}

\vspace{-0.7em} 
\subsection{\texttt{run\_planner()} Function}

\noindent\texttt{run\_planner(hierarchy\_level, prev\_hierarchy\_level,\\ pairs\_disjoint, kpair\_standalone, kpair\_colocated,\\ candidate\_paths\_standalone\_df,\\ candidate\_paths\_colocated\_df, GSNR\_opt\_link,\\ Nspan\_array, all\_node\_degree, P\_opt\_links, \\ minimum\_level, node\_cap\_update\_idx, result\_directory)}

\vspace{-0.6em}
\paragraph{Description}
The \texttt{run\_planner()} function executes the hierarchical optical network planning algorithm for a specified hierarchy level. It performs traffic allocation, spectrum assignment, and BVT deployment for both standalone and co-located hierarchical level (HL) nodes over multiple years. 

\vspace{0.5em}
\noindent
The function iteratively handles:
\begin{itemize}[noitemsep, topsep=0pt, parsep=0pt, partopsep=0pt, leftmargin=*]
	\item Traffic growth and residual throughput updates
	\item Spectrum assignment via \texttt{\_spectrum\_assignment()}
	\item BVT count and 100G license tracking
	\item Fiber pair and link utilization updates
	\item GSNR computation and aggregation
	\item Node capacity profile updates for each year
\end{itemize}

\begin{figure}
	\centering
	\includegraphics[height = 8.5cm]{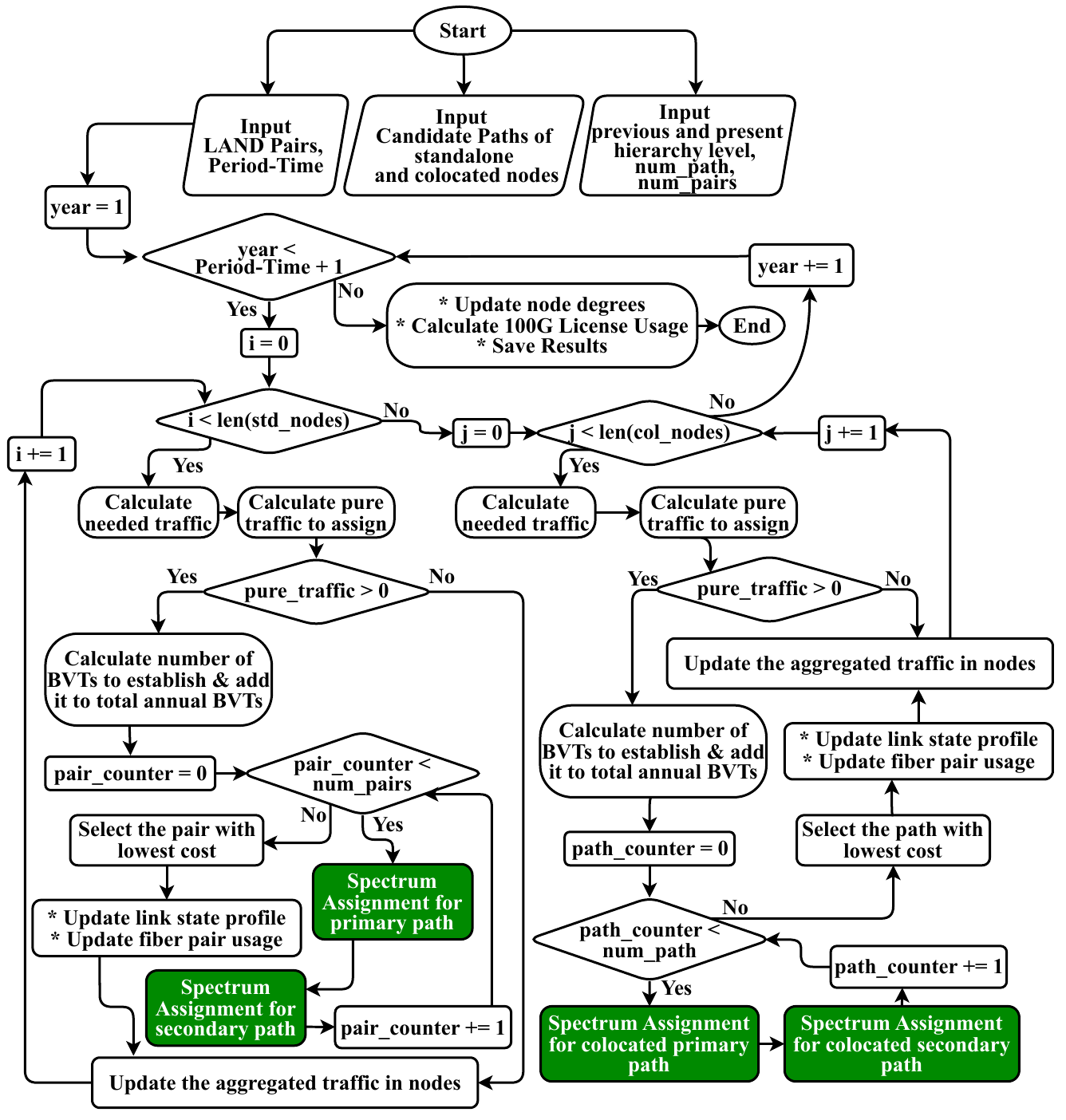}
	\caption{Flowchart of \texttt{run\_planner()} function.}
	\label{fig:run_planner_flowchart}
\end{figure}

\vspace{0.5em}
The flowchart of the \texttt{run\_planner()} function is shown in Fig.~\ref{fig:run_planner_flowchart}.  As shown in the flowchart the planner executes the network design process hierarchically and iteratively over time, progressing through each hierarchy level and year in sequence.
For each simulation year, the planning procedure begins with standalone nodes belonging to the current hierarchy level.

For each standalone node, the algorithm first extracts the required traffic demand and computes the pure throughput to assign. The planning strategy is designed such that a new BVT is only established when the capacity of all previously deployed BVTs (e.g., 400G BVTs) is fully utilized. Therefore, if the computed pure throughput remains positive, it indicates that additional BVTs must be established for that node.

To determine the number of BVTs to deploy, the algorithm divides the unallocated traffic by the largest BVT capacity (400G). For example, if the residual traffic is 350 Gbps, a single 400G BVT must be established. It is important to note that due to the multi-homed protection scheme adopted in the SixGman planning algorithm, each BVT must be deployed in both destination nodes (i.e., the primary and secondary destinations) as well as in the source node for each connection. Consequently, the total number of tracked BVTs per year corresponds to four times the calculated number of required BVTs. 

Once the number of BVTs has been determined, spectrum assignment is performed for each one. In this step, the algorithm utilizes the precomputed LAND pairs. Spectrum assignment is executed for every candidate LAND pair, and its corresponding cost is computed. The LAND pair with the minimum total cost is then selected for deployment. After the optimal pair is identified, the BVT establishment is finalized, and both the global link-state profile and the fiber pair usage arrays are updated accordingly. It is worth noting that the bitrate of each BVT is specified based on its GSNR value, for example if the target SNR of PM-64QAM and PM-32QAM are 19.73~dB and 16.85~dB respectively, and if the GSNR value of an established BVT is 18~dB, the bitrate of the established BVT is 320~Gbps due to the selection of PM-32QAM modulation. After establishing all BVTs, the algorithm calculate the total bitrate of all established BVTs, if the value doesn't exceed the pure\_traffic\_to\_assign, the algorithm tries to deploy another BVT.

After updating the aggregated node traffic, the planner proceeds to the next standalone node. Once all standalone nodes at the given hierarchy level have been processed, the same procedure is repeated for the co-located nodes. It is worth noting that in the primary paths of co-located nodes the algorithm always establish a BVT with the largest bitrate (400~Gbps) due to the lack of GSNR constraints.
When every node at the current level has been handled, the algorithm advances to the next simulation year.

At the end of the planning period, the algorithm updates the nodal degrees based on the established fiber pairs and computes the cumulative number of 100G licenses utilized. Finally, all computed results are saved, ensuring that they can be used as input for analyzing and planning the next hierarchical level.

{\small
	\vspace{-0.5em}
	\paragraph{Parameters}
	\begin{itemize}[noitemsep, topsep=0pt, parsep=0pt, partopsep=0pt, leftmargin=*]
		\item \texttt{hierarchy\_level (int):} Current HL being processed.
		\item \texttt{prev\_hierarchy\_level (int):} Previous HL for continuity and reference.
		\item \texttt{pairs\_disjoint (pd.DataFrame):} Disjoint path pairs for routing and path computation in the stanalone HL nodes.
		\item \texttt{kpair\_standalone (int):} Maximum number of LAND pairs to process for standalone nodes at the current hierarchy level.  
		\item \texttt{kpair\_colocated (int):} Maximum number of secondary paths to process for co-located nodes at the current hierarchy level.  
		\item \texttt{candidate\_paths\_standalone\_df (pd.DataFrame):} Candidate paths for standalone HL nodes.
		\item \texttt{candidate\_paths\_colocated\_df (pd.DataFrame):} Candidate paths for co-located HL nodes.
		\item \texttt{GSNR\_opt\_link (np.ndarray):} GSNR values per link in the subnetwork.
		\item 
		\texttt{Nspan\_array (np.ndarray)}: Array that contains number of spans per link in the whole network.
		\item 
		\texttt{all\_node\_degree (np.ndarray)}: Initial node degress for WSS penalty calculation.
		\item \texttt{minimum\_level (int):} The lowest hierarchy level to include in processing (e.g., 4 if analyzing levels HL4, HL3, and HL2).  
		\item \texttt{node\_cap\_update\_idx (int):} Index in the\\ \texttt{node\_capacity\_profile\_array} where the aggregated traffic for each node in the current year will be recorded.
		\item \texttt{result\_directory (Path):} Directory to save annual and summary results.
	\end{itemize}
	\vspace{-0.5em}
	\paragraph{Updates}
	The function updates multiple internal state variables for network performance tracking:
	
	\begin{itemize}[noitemsep, leftmargin=*]
		\item 
		\texttt{Residual\_Throughput\_BVT\_standalone\_HLs\_primary}: Residual capacity of the last established BVT in standalone nodes (primary paths).
		\item 
		\texttt{Residual\_Throughput\_BVT\_standalone\_HLs\_secondary}: Residual capacity of the last established BVT in standalone nodes (secondary paths).
		\item 
		\texttt{Residual\_Throughput\_BVT\_colocated\_HLs\_primary}: Residual capacity of the last established BVT in co-located nodes (primary paths).
		\item 
		\texttt{Residual\_Throughput\_BVT\_colocated\_HLs\_secondary}: Residual capacity of the last established BVT in co-located nodes (secondary paths).
		\item 
		\texttt{Year\_FP}: Fiber pair allocation per link per year.
		\item 
		\texttt{Year\_FP\_HL\_colocated}: Fiber pair allocation per co-located HL node per year.
		\item 
		\texttt{Year\_FP\_new}: Fiber pair allocations based on spectrum assignment per year.
		\item 
		\texttt{Total\_effective\_FP\_new\_annual}: Total km of fiber pair usage across all links per year.
		\item 
		\texttt{GSNR\_BVT\_all\_annual}: GSNR records of all BVTs (primary and secondary) per year.
		\item 
		\texttt{GSNR\_BVT\_primary\_annual}: GSNR records of primary BVTs per year.
		\item 
		\texttt{GSNR\_BVT\_secondary\_annual}: GSNR records of secondary BVTs per year.
		\item 
		\texttt{node\_capacity\_profile\_array}: Node capacity evolution per year.
		\item
		 \texttt{traffic\_flow\_links\_array}: Annual traffic per link.
		\item
		 \texttt{num\_link\_CBand\_annual}:  Number of fiber pairs with at least one allocated C-band FS in each link per year.
		 \item 
		 \texttt{num\_link\_SupCBand\_annual}: Number of fiber pairs with at least one allocated SuperC-band FS in each link per year.
		 \item
		 \texttt{num\_link\_LBand\_annual}: Number of fiber pairs with at least one allocated L-band FS in each link per year.
		\item 
		\texttt{num\_100G\_licence\_annual}: Annual count of activated 100G licenses.
		\item 
		\texttt{num\_100G\_licence\_CBand\_annual}: Annual count of activated 100G licenses in C-band.
		\item 
		\texttt{num\_100G\_licence\_superCBand\_annual}: Annual count of activated 100G licenses in SuperC-band.
		\item 
		\texttt{num\_100G\_licence\_LBand\_annual}: Annual count of activated 100G licenses in L-band.
		\item 
		\texttt{Residual\_Throughput\_LC\_standalone\_HLs\_primary}: Residual capacity of the last activated licenses (primary path) per standalone node.
		\item 
		\texttt{Residual\_Throughput\_LC\_standalone\_HLs\_secondary}: Residual capacity of the last activated licenses (secondary path) per standalone node.
		\item 
		\texttt{Residual\_Throughput\_LC\_colocated\_HLs\_primary}: Residual capacity of the last activated licenses (primary path) per co-located node.
		\item 
		\texttt{Residual\_Throughput\_LC\_colocated\_HLs\_secondary}: Residual capacity of the last activated licenses (secondary path) per co-located node.
		\item 
		\texttt{LAND\_Links\_Storage}: Links of selected LAND pair for each node.
		\item 
		\texttt{LSP\_array}: Link-State-Profile array that show the occupied frequency slots in different fiber pairs.
		\item 
		\texttt{LSP\_array\_Colocated}: Link-State-Profile array that show the occupied frequency slots in different fiber pairs for primary co-located paths.
		\item 
		\texttt{self.path\_latency\_storage (list):} Latency of primary and secondary paths. We consider the latency of $5\ \mu s$ per km in the path.
		\item 
		\texttt{self.destinations\_storage (list):} Destination nodes for primary and secondary paths.
	\end{itemize}
	\vspace{-0.5em}
	\paragraph{Returns}
	\begin{itemize}[noitemsep, topsep=0pt, leftmargin=*]
		\item This method does not return a value. Results are stored in the instance attributes listed above.
	\end{itemize}
}

\vspace{-1em}
\section{Post Process Module}
\label{sec:post_functions}
This section describes the functions implemented in the \texttt{analyse\_result} class, 
which is the only class included in this module.

The \texttt{analyse\_result} class provides analytical and visualization tools for 
evaluating the performance of the optical network based on the results generated by 
the Planning module. It performs post-processing of simulation outputs, computes key 
network performance indicators, and supports hierarchical-level comparisons across 
different planning stages.

\vspace{-0.5em}
\subsection{\texttt{\_load\_data()} Function}
\noindent\texttt{\_load\_data() $\rightarrow$ None}

\vspace{-0.6em}
\paragraph{Description}
The \texttt{\_load\_data()} function loads the compressed \texttt{.npz} result files 
generated by the \texttt{planning} module for each hierarchy level (HL). 
It extracts essential data related to link utilization, BVT deployment, and GSNR measurements.  
The parsed information is organized and stored internally in the class attributes 
\texttt{link\_data}, \texttt{bvt\_data}, and \texttt{GSNR\_data}, 
which are later used for performance analysis, visualization, and reporting.

{\small
	\vspace{-0.5em}
	\paragraph{Updates}
	\begin{itemize}[noitemsep, topsep=0pt, leftmargin=*]
		\item \texttt{self.link\_data (dict):} Contains link-related results for each hierarchy level, 
		including link utilization, fiber-pair usage, and spectrum occupancy.
		\item \texttt{self.bvt\_data (dict):} Stores per-level BVT deployment metrics, 
		such as BVT counts and 100G license usage.
		\item \texttt{self.GSNR\_data (dict):} Includes GSNR measurements for all analyzed paths, 
		separated by primary and secondary routes where applicable.
	\end{itemize}
	\vspace{-0.5em}
	\paragraph{Raises}
	\begin{itemize}[noitemsep, topsep=0pt, leftmargin=*]
		\item \texttt{IOError:} Raised if one or more expected result files are missing 
		or cannot be accessed.
	\end{itemize}
	\vspace{-0.5em}
	\paragraph{Returns}
	\begin{itemize}[noitemsep, topsep=0pt, leftmargin=*]
		\item All loaded data are stored internally in the class attributes listed above.
	\end{itemize}
}
\vspace{-0.5em}
\subsection{\texttt{plot\_link\_state()} Function}
\noindent\texttt{plot\_link\_state(minimum\_hierarchy\_level: int, splitter: List, save\_flag: int, save\_suffix: str = "", flag\_plot: int = 1) $\rightarrow$ Optional[np.ndarray]}

\vspace{-0.6em}
\paragraph{Description}
The \texttt{plot\_link\_state()} function visualizes or returns the evolution of link states 
(i.e., Fiber Pair (FP) numbers) across all hierarchy levels throughout the multi-year 
network planning period.  
It generates a heatmap-style plot where each row represents a network link and each column 
corresponds to a simulation year. The color scale indicates the number of active Fiber Pairs (FPs) 
assigned to each link.  
Vertical dashed lines are drawn to separate links belonging to different hierarchy level subnetworks, 
as defined by the \texttt{splitter} parameter.

{\small
	\vspace{-0.5em}
	\paragraph{Parameters}
	\begin{itemize}[noitemsep, topsep=0pt, leftmargin=*]
		\item \texttt{minimum\_hierarchy\_level (int):} The lowest hierarchy level included in the visualization.
		\item \texttt{splitter (List):} A list specifying the number of links at each hierarchy level subnetwork; 
		used to place dashed separators in the plot.
		\item \texttt{save\_flag (int):} If set to 1, the generated plot is saved to disk.
		\item \texttt{save\_suffix (str, optional):} Custom suffix appended to the saved file name. Default is an empty string.
		\item \texttt{flag\_plot (int, optional):} If 1 (default), the plot is displayed; 
		if 0, the numerical FP state matrix is returned instead.
	\end{itemize}
	\vspace{-0.5em}
	\paragraph{Returns}
	\begin{itemize}[noitemsep, topsep=0pt, leftmargin=*]
		\item \texttt{np.ndarray:} Returned only if \texttt{flag\_plot == 0}. 
		A matrix of shape $(\text{years} \times \text{total\_links})$, 
		where each element represents the number of Fiber Pairs (FPs) assigned 
		to a specific link in a given year.
	\end{itemize}
	\vspace{-0.5em}
	\paragraph{Notes}
	\begin{itemize}[noitemsep, topsep=0pt, leftmargin=*]
		\item When \texttt{save\_flag = 1}, the figure is saved in the output directory 
		with the filename pattern\\ \texttt{\{topology\_name\}\_Link\_State\{save\_suffix\}.png}.
		\item The visualization allows monitoring the temporal evolution of FP allocation 
		and link utilization across all hierarchical levels.
	\end{itemize}
}
\vspace{-0.5em}
\subsection{\texttt{plot\_FP\_usage()} Function}
\noindent\texttt{plot\_FP\_usage(save\_flag: int, save\_suffix: str = "", flag\_plot: int = 1) $\rightarrow$ None}

\vspace{-0.6em}
\paragraph{Description}
The \texttt{plot\_FP\_usage()} function visualizes the evolution of Fiber Pair (FP) usage over time 
across all hierarchy levels in the optical network.  
The generated plot provides two complementary views:
\begin{itemize}[noitemsep, topsep=0pt, leftmargin=*]
	\item The \textbf{left y-axis} (logarithmic scale) shows the cumulative FP usage in kilometers.
	\item The \textbf{right y-axis} (linear scale) shows the total number of FPs deployed in each year.
\end{itemize}
This visualization provides insight into both network growth and physical infrastructure scaling over the planning period.

{\small
	\vspace{-0.5em}
	\paragraph{Parameters}
	\begin{itemize}[noitemsep, topsep=0pt, leftmargin=*]
		\item \texttt{save\_flag (int):} If set to 1, the generated figure is saved in the result directory.
		\item \texttt{save\_suffix (str, optional):} Custom suffix appended to the saved file name. Default is an empty string.
		\item \texttt{flag\_plot (int, optional):} If 1 (default), the plot is displayed; 
		if 0, plotting is skipped (useful for automated processing).
	\end{itemize}
	\vspace{-0.5em}
	\paragraph{Returns}
	\begin{itemize}[noitemsep, topsep=0pt, leftmargin=*]
		\item This method does not return a value.
	\end{itemize}
	\vspace{-0.5em}
	\paragraph{Notes}
	\begin{itemize}[noitemsep, topsep=0pt, leftmargin=*]
		\item The cumulative FP usage represents the total deployed fiber length across all active links and years.
		\item When \texttt{save\_flag = 1}, the figure is stored as \\
		\texttt{\{topology\_name\}\_FP\_Usage\{save\_suffix\}.png} in the result directory.
	\end{itemize}
}
\vspace{-0.5em}
\subsection{\texttt{plot\_FP\_degree()} Function}
\noindent\texttt{plot\_FP\_degree(save\_flag: int, save\_suffix: str = "", flag\_plot: int = 1) $\rightarrow$ None}

\vspace{-0.6em}
\paragraph{Description}
The \texttt{plot\_FP\_degree()} function generates a dual-axis visualization showing 
the evolution of cumulative Fiber Pair (FP) usage and the band degree metrics 
for each optical spectrum band (C-band, SuperC-band, and L-band) 
over the simulation period.

The plot allows comparison between the overall infrastructure growth (in terms of 
fiber deployment) and the evolution of network connectivity across the spectral bands. The generated plot provides two complementary views:

\begin{itemize}[noitemsep, topsep=0pt, leftmargin=*]
	\item \textbf{Left y-axis (log scale):} Represents the cumulative FP usage in kilometers.
	\item \textbf{Right y-axis (linear scale):} Represents the average band degree per spectral band (C, SuperC, and L).
	\item Different colors are used to distinguish spectral bands.
\end{itemize}

{\small
	\vspace{-0.5em}
	\paragraph{Parameters}
	\begin{itemize}[noitemsep, topsep=0pt, leftmargin=*]
		\item \texttt{save\_flag (int):} If set to 1, the generated plot is saved in the result directory.
		\item \texttt{save\_suffix (str, optional):} Custom suffix to append to the saved filename. Default is an empty string.
		\item \texttt{flag\_plot (int, optional):} If 1 (default), the plot is displayed; if 0, visualization is skipped.
	\end{itemize}
	\vspace{-0.5em}
	\paragraph{Returns}
	\begin{itemize}[noitemsep, topsep=0pt, leftmargin=*]
		\item This method does not return a value.
	\end{itemize}
	\vspace{-0.5em}
	\paragraph{Notes}
	\begin{itemize}[noitemsep, topsep=0pt, leftmargin=*]
		\item FP usage represents the total deployed fiber length in kilometers, summed across all links and years.
		\item Band degree metrics indicate number of fiber-pairs with at least one frequency slot allocated in that band for each link of the network.
		\item When \texttt{save\_flag = 1}, the plot is stored in the results directory with filename pattern \\
		\texttt{\{topology\_name\}\_FP\_Degree\{save\_suffix\}.png}.
	\end{itemize}
}

\subsection{\texttt{plot\_bvt\_license()} Function}
\noindent\texttt{plot\_bvt\_license(save\_flag: int, save\_suffix: str = "", flag\_plot: int = 1) $\rightarrow$ None}

\vspace{-0.6em}
\paragraph{Description}
The \texttt{plot\_bvt\_license()} function generates a dual-axis visualization illustrating the cumulative 
usage of BVTs and the corresponding allocation of 100G licenses 
over the planning years. This function provides insight into how BVT deployment evolves 
across different spectral bands (C-band, SuperC-band, and L-band), enabling capacity planners to 
evaluate growth trends and technology adoption rates. The generated plot provides two complementary views:

\begin{itemize}[noitemsep, topsep=0pt, leftmargin=*]
	\item \textbf{Left y-axis (log scale):} Represents the cumulative number of deployed BVTs per spectral band (C, SuperC, and L).
	\item \textbf{Right y-axis (linear scale):} Represents the cumulative number of activated 100G licenses.
	\item Each spectral band is represented with distinct colors to highlight comparative deployment rates.
\end{itemize}

\paragraph{Parameters}
\begin{itemize}[noitemsep, topsep=0pt, leftmargin=*]
	\item \texttt{save\_flag (int):} If set to 1, the generated plot is saved in the result directory.
	\item \texttt{save\_suffix (str, optional):} Custom suffix appended to the saved filename. Default is an empty string.
	\item \texttt{flag\_plot (int, optional):} If 1 (default), the plot is displayed; if 0, the visualization is skipped.
\end{itemize}

\paragraph{Returns}
\begin{itemize}[noitemsep, topsep=0pt, leftmargin=*]
	\item This method does not return a value.
\end{itemize}

\paragraph{Notes}
\begin{itemize}[noitemsep, topsep=0pt, leftmargin=*]
	\item The plotted data reflects cumulative deployment across simulation years, combining all hierarchy levels.
	\item 100G license counts are derived proportionally from BVT allocations, assuming four 100G licenses per BVT.
	\item When \texttt{save\_flag = 1}, the plot is saved in the results directory under the filename pattern \\
	\texttt{\{topology\_name\}\_BVT\_License\{save\_suffix\}.png}.
\end{itemize}

\subsection{\texttt{calc\_cost()} Function}
\noindent\texttt{calc\_cost(save\_flag: int, save\_suffix: str = "", \\C\_100GL\_First: float = 1, C\_100G\_Added: float = 0.333, C\_MCS: float = 0.7, C\_RoB: float = 1.9, C\_IRU: float = 0.5) $\rightarrow$ pd.DataFrame}

\vspace{-0.6em}
\paragraph{Description}
The \texttt{calc\_cost()} function computes the capital expenditure (CAPEX) and operational expenditure (OPEX) 
for optical network deployment and expansion across multiple years. The calculation integrates hardware 
and infrastructure costs derived from the evolution of network topology, fiber-pair usage, and 
Bitrate Variable Transceiver (BVT) deployment.  

The function accounts for the cost contributions of 100G licenses, Multi-Cast Switches (MCS), ROADMs-on-Blade (RoB), 
and leased fiber-pair infrastructure (IRU), providing a comprehensive cost assessment framework.  

\paragraph{Calculation Overview}
\begin{itemize}[noitemsep, topsep=0pt, leftmargin=*]
	\item \textbf{CAPEX:} Computed based on hardware installations and upgrades (e.g., 100G licenses, ROADMs, and MCS units).
	\item \textbf{OPEX:} Derived from IRU fiber pair usage and ongoing operational costs.
	\item Both metrics are computed annually to track cost evolution over the network's planning horizon.
\end{itemize}

\paragraph{Cost Model}
In this framework, we introduce a comprehensive cost model for the optical infrastructure of metropolitan area networks (MANs) in multi-band systems. The model quantifies the total cost of ownership (TCO) at the optical layer, defined as the sum of capital expenditures (CAPEX) and operational expenditures (OPEX).

\vspace{0.5em}
\noindent\textbf{I) OPEX Component}

For a non-incumbent operator obtaining the indefeasible right of use (IRU) from an incumbent, the annual OPEX is defined as
\begin{equation}
	C_{\mathrm{OPEX}}(y)
	= 2 \times C_{\mathrm{IRU}} 
	\sum_{l=1}^{N_{\mathrm{link}}'}
	\sum_{d=1}^{D^l(y)}
	L^l \times \zeta^{l,d},
	\label{eq:OPEX}
\end{equation}
where $C_{\mathrm{IRU}}$ is the IRU cost unit [strand/km/year], $N_{\mathrm{link}}'$ is the number of links, $L^l$ is the link length, $\zeta^{l,d}$ is the utilization factor, and $D^l(y)$ denotes the number of active demands on link $l$ in year $y$.

\vspace{0.5em}
\noindent\textbf{II) CAPEX Component}

The CAPEX considers the cost of multicast switches (MCSs), ROADMs-on-a-blade (RoBs), and 100G licenses:
\begin{equation}
	C_{\mathrm{CAPEX}}(y)
	= C_{\mathrm{MCS}}(y)
	+ C_{\mathrm{RoB}}(y)
	+ C_{100\mathrm{G}}(y).
	\label{eq:CAPEX}
\end{equation}

The MCS cost is modeled as
\begin{equation}
	C_{\mathrm{MCS}}(y)
	= C_{\mathrm{MCS}}
	\sum_{n=1}^{|\mathcal{N}|}
	\Bigg\lceil
	\frac{\mathrm{H}\!\left(N_{\mathrm{port}}^{\mathrm{req}}(y,n)\right)}{16}
	\Bigg\rceil,
	\label{eq:MCS}
\end{equation}
where $\mathrm{H}(\cdot)$ is the Heaviside step function and $N_{\mathrm{port}}^{\mathrm{req}}(y,n)$ is the number of required MCS ports at node $n$ in year $y$.

The RoB and 100G line cost components are given by
\begin{align}
	C_{\mathrm{RoB}}(y)
	&= C_{\mathrm{RoB}}
	\sum_{n=1}^{|\mathcal{N}|}
	N_{\mathrm{Degree}}^{\mathrm{new}}(y,n),
	\label{eq:RoB} \\
	C_{100\mathrm{G}}(y)
	&= C_{100\mathrm{G}}
	\sum_{n=1}^{|\mathcal{N}|}
	N_{100\mathrm{G}}^{\mathrm{new}}(y,n),
	\label{eq:100G}
\end{align}
where $N_{\mathrm{Degree}}^{\mathrm{new}}(y,n)$ and $N_{100\mathrm{G}}^{\mathrm{new}}(y,n)$ represent the number of newly deployed degrees and 100G line bit rates at node $n$, respectively.

\vspace{0.5em}
\noindent\textbf{III) Optical TCO and Band Adjustment}

The overall optical TCO over $y'$ years is expressed as
\begin{equation}
	\mathrm{OTCO}(y') 
	= \sum_{y=1}^{y'}
	\big[ C_{\mathrm{CAPEX}}(y) + C_{\mathrm{OPEX}}(y) \big].
	\label{eq:OTCO}
\end{equation}

To account for equipment cost variations across spectral bands, the multi-band CAPEX adjustment is given by
\begin{equation}
	C_{\mathrm{CAPEX}}^{\mathrm{SupC/L}}(y)
	= \big(1 + \alpha \times \beta(y)\big)
	\times C_{\mathrm{CAPEX}}^{\mathrm{C\text{-}Band}}(y = 1),
	\label{eq:band}
\end{equation}
where $\alpha = \{0.1, 0.2\}$ for SupC- and L-band, respectively, and $\beta(y) = (1 - \gamma)^y$ with $\gamma = 0.1$ representing annual depreciation.

{\small
	\vspace{-0.5em}
	\paragraph{Parameters}
	\begin{itemize}[noitemsep, topsep=0pt, leftmargin=*]
		\item \texttt{save\_flag (int):} If set to 1, saves the computed cost breakdown to a CSV file.
		\item \texttt{save\_suffix (str):} Optional string appended to the output filename. Default is an empty string.
		\item \texttt{C\_100GL\_First (float):} Unit cost of the first activated 100G license [default = 1].
		\item \texttt{C\_100G\_Added (float):} Unit cost of the following activated 100G licenses after the first one [default = 0.333].
		\item \texttt{C\_MCS (float):} Unit cost coefficient for Multi-Cast Switches. Default = 0.7.
		\item \texttt{C\_RoB (float):} Unit cost coefficient for ROADMs-on-Blade. Default = 1.9.
		\item \texttt{C\_IRU (float):} Cost per kilometer for leased fiber pair (IRU). Default = 0.5.
	\end{itemize}
	\vspace{-0.5em}
	\paragraph{Returns}
	\begin{itemize}[noitemsep, topsep=0pt, leftmargin=*]
		\item \texttt{pd.DataFrame:} A structured dataframe containing annualized OPEX and CAPEX components, 
		including subcategories such as hardware cost, license cost, and IRU fiber costs.
	\end{itemize}
	\vspace{-0.5em}
	\paragraph{Notes}
	\begin{itemize}[noitemsep, topsep=0pt, leftmargin=*]
		\item The resulting cost dataframe is used for techno-economic analyses of hierarchical optical networks.
		\item When \texttt{save\_flag = 1}, results are stored as a CSV file named 
		\texttt{\{topology\_name\}\_cost\_analyse\{save\_suffix\}.csv} in the result directory.
	\end{itemize}

}

\vspace{-0.5em}
\subsection{\texttt{compute\_E2E\_path\_latency()} Function}

\noindent\texttt{compute\_E2E\_path\_latency(node, latency\_core\_array, \\ destination\_core\_array, processing\_level\_list)\\ $\rightarrow$ List[Tuple[List[int], float]]}

\vspace{-0.6em}
\paragraph{Description}
The \texttt{compute\_E2E\_path\_latency()} function recursively computes all possible end-to-end (E2E) latency paths from a given node to higher-level (top-hierarchy) destination nodes. 
It traces both primary and secondary (dual-home) connectivity paths, accumulating total propagation latency along each route.
The recursion proceeds through hierarchy levels (e.g., HL4 $\rightarrow$ HL3 $\rightarrow$ HL2 $\rightarrow$ HL1) until reaching top-level standalone HL nodes.

{\small
	\vspace{-0.5em}
	\paragraph{Parameters}
	\begin{itemize}[noitemsep, topsep=0pt, parsep=0pt, partopsep=0pt, leftmargin=*]
		\item \texttt{node (int):} Index of the starting node (e.g., a lower-level core or access node).
		\item \texttt{latency\_core\_array (np.ndarray):} 2D or ragged array storing per-path latency values (in $\mu s$) from each node to its directly connected higher-level nodes.
		\item \texttt{destination\_core\_array (np.ndarray):} 2D or ragged array storing destination node indices corresponding to \texttt{latency\_core\_array} entries.
		\item \texttt{processing\_level\_list (List[int]):} Ordered list of hierarchy levels used to determine recursion termination (e.g., \([4, 3, 2]\)).
	\end{itemize}
	\vspace{-0.5em}
	\paragraph{Returns}
	\begin{itemize}[noitemsep, topsep=0pt, leftmargin=*]
		\item \texttt{List[Tuple[List[int], float]]:} List of tuples, where each tuple contains:
		\begin{itemize}[noitemsep, leftmargin=*]
			\item \texttt{path\_list (List[int]):} Sequence of nodes along the E2E path.
			\item \texttt{total\_latency (float):} Total accumulated latency in microseconds.
		\end{itemize}
	\end{itemize}
	\vspace{-0.5em}
	\paragraph{Notes}
	\begin{itemize}[noitemsep, topsep=0pt, parsep=0pt, partopsep=0pt, leftmargin=*]
		\item Recursion terminates when the current node belongs to the top hierarchy (standalone nodes in \texttt{HL\{processing\_level\_list[-1]-1\}}).
		\item Latencies are cumulative across hops.
		\item Supports dual-homing scenarios by computing both primary and secondary paths.
		\item Latency constants (e.g., $5\ \mu s / km$  ) should be applied when constructing \texttt{latency\_core\_array} prior to calling this function.
	\end{itemize}
}

\vspace{-0.5em}
\subsection{\texttt{calc\_E2E\_latency\_Total()} Function}

\noindent\texttt{calc\_E2E\_latency\_Total(latency\_core\_array,\\ destination\_core\_array, processing\_level\_list, save\_flag, save\_suffix) $\rightarrow$ np.ndarray}

\vspace{-0.6em}
\paragraph{Description}
The \texttt{calc\_E2E\_latency\_Total()} function computes the total end-to-end (E2E) latency from the lowest hierarchy-level nodes (e.g., HL4) up to the top-level core nodes (e.g., HL1). 
It iterates over all nodes at the lowest hierarchy level and calls \texttt{compute\_E2E\_path\_latency()} recursively to determine cumulative latency of all possible paths to top-level destinations.
Latency values are derived from propagation delays stored in \texttt{latency\_core\_array} and additional per-level processing delays (default: $200 \mu s$ per hierarchy level).

{\small
	\vspace{-0.5em}
	\paragraph{Parameters}
	\begin{itemize}[noitemsep, topsep=0pt, parsep=0pt, partopsep=0pt, leftmargin=*]
		\item \texttt{latency\_core\_array (np.ndarray):} Array of per-node propagation latencies (in $\mu s$). Each element \texttt{[i]} contains a list of latencies from node \texttt{i} to its higher-level destinations.
		\item \texttt{destination\_core\_array (np.ndarray):} Array of per-node destination indices corresponding to \texttt{latency\_core\_array}.
		\item \texttt{processing\_level\_list (list[int]):} Ordered list of hierarchy levels (e.g., \([4,3,2,1]\)) defining traversal from lowest to highest hierarchy layer. The first element indicates the starting level.
		\item \texttt{save\_flag (int):} If 1, saves the computed latency results to disk as a compressed \texttt{.npz} file. If 0, results are only returned.
		\item \texttt{save\_suffix (str, optional):} Optional suffix for the output filename, e.g., \texttt{save\_suffix='\_test'} results in \\ \texttt{\{topology\_name\}\_latency\_test.npz}.
	\end{itemize}
	\vspace{-0.5em}
	\paragraph{Returns}
	\begin{itemize}[noitemsep, leftmargin=*]
		\item \texttt{np.ndarray:} A 1D array of total E2E latency values (in $\mu s$) for each lowest-level node path.
	\end{itemize}
}

\vspace{-1.5em}
\section{Simulation Setup and Results}
\label{sec:results}
In this section, we evaluate the performance of the two considered network scenarios---full hierarchical and HL3-bypassed---across a wide set of techno-economic and performance metrics. The comparison includes fiber-pair utilization, spectral band occupancy, BVT and 100G license requirements, IP router resource usage, traffic flow and congestion levels across links, quality of transmission (QoT), end-to-end latency, cost, and energy consumption. For each metric, we highlight the key observations that characterize the differences between the two scenarios and provide insights into the implications of bypassing the HL3 layer in metro-aggregation network design.

\vspace{-0.6em}
\subsection{Initial Setup}
\begin{figure}
	\centering
	\includegraphics[height = 7cm]{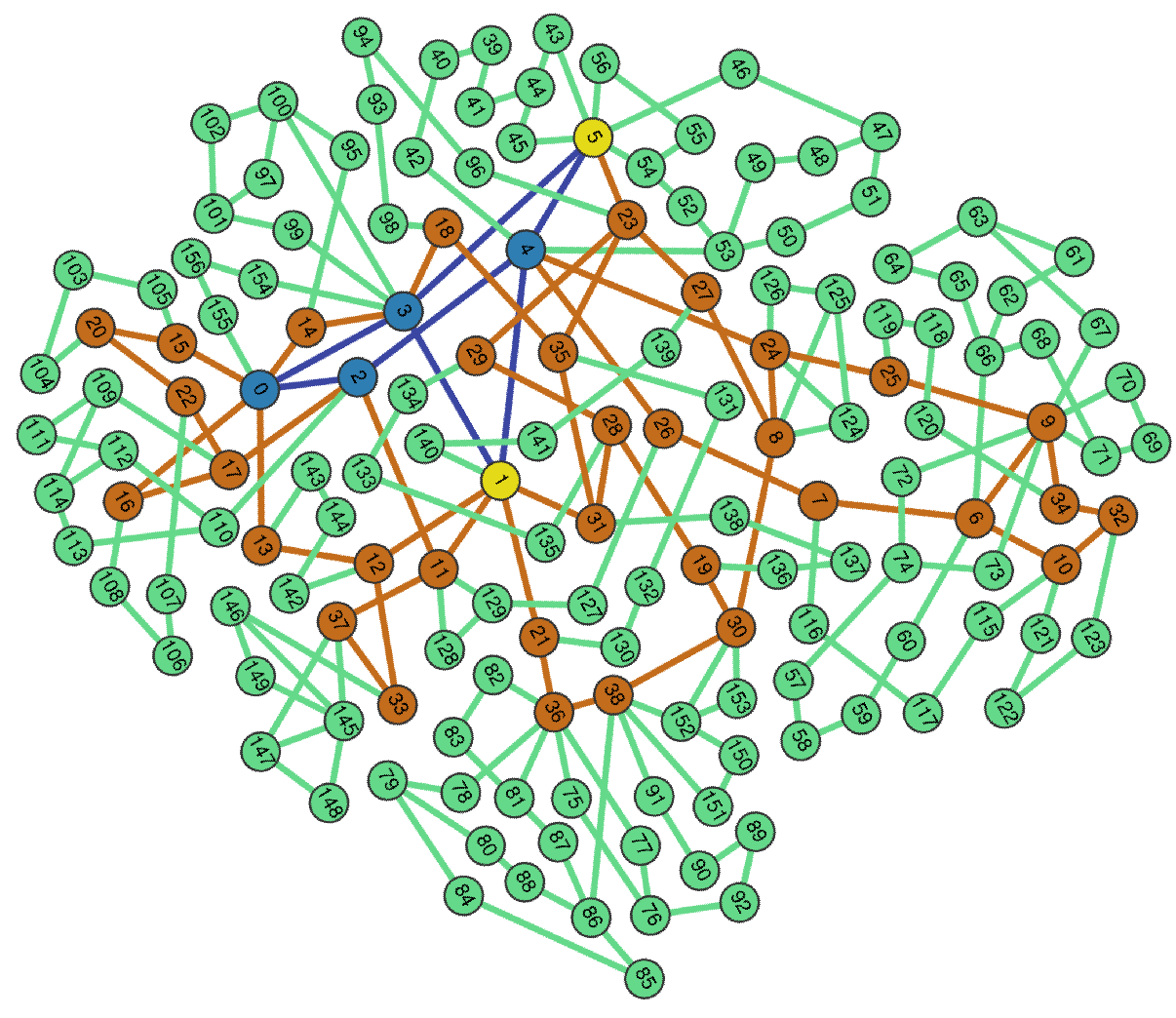}
	\caption{Telefónica metro-urban network (MAN157) topology.}
	\label{fig:man157}
\end{figure}

The performance evaluation is conducted on the Telefónica metro-urban MAN157 network, shown in Fig.~\ref{fig:man157}. The topology comprises 157 optical nodes interconnected by 220 fiber links and organized into four hierarchical layers (HL1--HL4) spanning three hierarchical domains. The network includes two HL1 nodes (yellow circles), four HL2 nodes (blue circles), thirty-three HL3 nodes (brown circles), and 118 HL4 nodes (green circles). Each HL3 site is co-located with an HL4 node; each HL2 site hosts co-located HL4 and HL3 nodes; and each HL1 site is co-located with HL4, HL3, and HL2 nodes.

Two architectural configurations are analyzed. In the full hierarchical scenario, traffic is generated at all HL4 nodes---both standalone and co-located---using 100 Monte Carlo iterations. Each iteration yields initial HL4 demands uniformly distributed between 20 and 200~Gbps (average of 100~Gbps), with a projected 40\% annual growth rate. Traffic is transported following the conventional\\ HL4~$\rightarrow$~HL3~$\rightarrow$~HL2~$\rightarrow$~HL1 sequence. In the first stage, optical signals originating at the HL4 nodes are forwarded to HL3 standalone or co-located nodes. After aggregation and electrical processing at the HL3 routers, the traffic is converted back to the optical domain and carried toward the HL2 layer. Finally, HL2 routers perform another aggregation step and forward the traffic optically to the HL1 nodes. HL1 routers perform the last aggregation step in the network. Dual-homing protection is applied throughout this process: each demand is routed over two link- and node-disjoint paths, and the carried traffic is evenly split between the primary and secondary routes.

\begin{figure*}
	\centering
	\includegraphics[width = \textwidth]{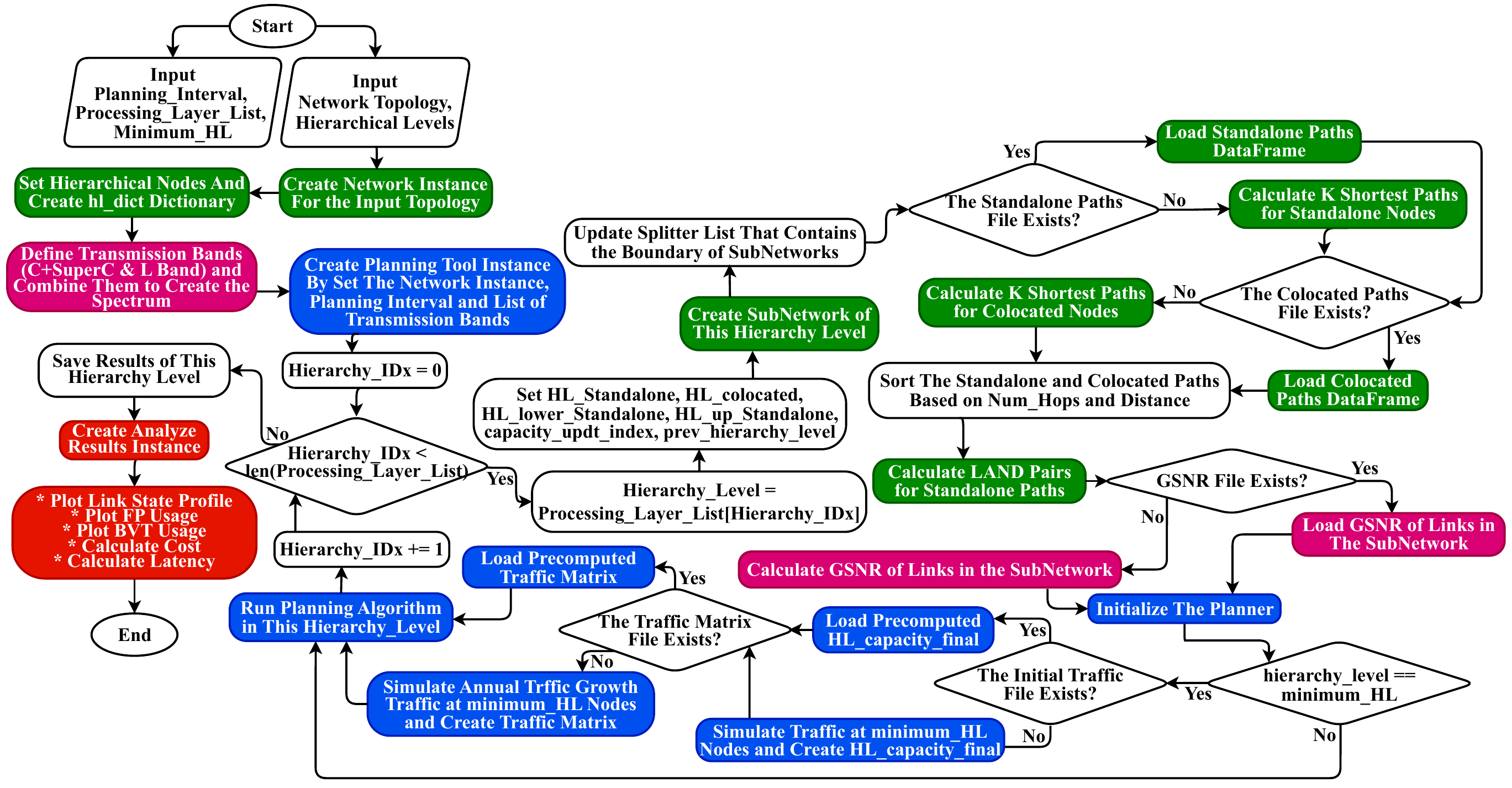}
	\caption{Overview of the network-planning procedure.}
	\label{fig:SixGman_scenarios}
\end{figure*}

In the HL3-bypassed scenario, the electrical aggregation stage at HL3 nodes is removed. Traffic generated at HL4 sites is forwarded directly to HL2 nodes, while HL3 nodes function solely as optical bypass elements, with no OEO conversion applied. The same Monte Carlo-generated traffic sets and identical physical-layer parameters are used for both network scenarios to ensure a fair and consistent comparison. Figure~\ref{fig:SixGman_scenarios} illustrates how the two configurations are defined within the open-source planning tool (SixGman) introduced in this study. In this flowchart, green blocks correspond to \texttt{network} module, pink blocks indicate the \texttt{band} module, blue blocks represent the \texttt{planning} module, and red blocks denote the \texttt{post\_process} module of SixGman.

All simulations employ a coherent transponder model configured with bit rates between 64 and 400~Gbps, a 64~GBaud symbol rate, a roll-off factor of 0.1, and 75~GHz channel spacing, while allowing FEC overhead values in the 20\%--35\% range. The GSNR thresholds associated with the considered modulation formats are 4.32~dB for PM-BPSK, 7.33~dB for PM-QPSK, 11.40~dB for PM-8QAM, 13.90~dB for PM-16QAM, 16.85~dB for PM-32QAM, and 19.73~dB for PM-64QAM, corresponding to modulation cardinalities $m=2,4,6,8,10,$ and $12$, respectively. These threshold values are taken from the analytical model reported in~\citep{arpanaei2024enabling}. The optical spectrum extends over the combined C, Super-C, and L bands, offering 12~THz of usable bandwidth corresponding to 160 channels at 75~GHz spacing.

Quality-of-transmission evaluations account for several practical impairments. The filter-penalty term $\mathrm{SNR}_{\mathrm{Pen_{filter}}}$ ranges from 0.3~dB to 7~dB, following the hop-count and degree-dependent values reported in Fig.~4(a) of~\citep{sequeira2018impact}. An additional 1~dB aging margin is included, and a 36~dB back-to-back transponder SNR is assumed to incorporate connector losses, polarization-dependent effects, and MUX/DEMUX insertion loss.
	
	\subsection{Performance Comparison of Two Network Scenarios}
	
	\vspace{1em}
	\subsubsection{Fiber Pair Usage}
\begin{figure}
		\centering
		\includegraphics[height = 3.5cm]{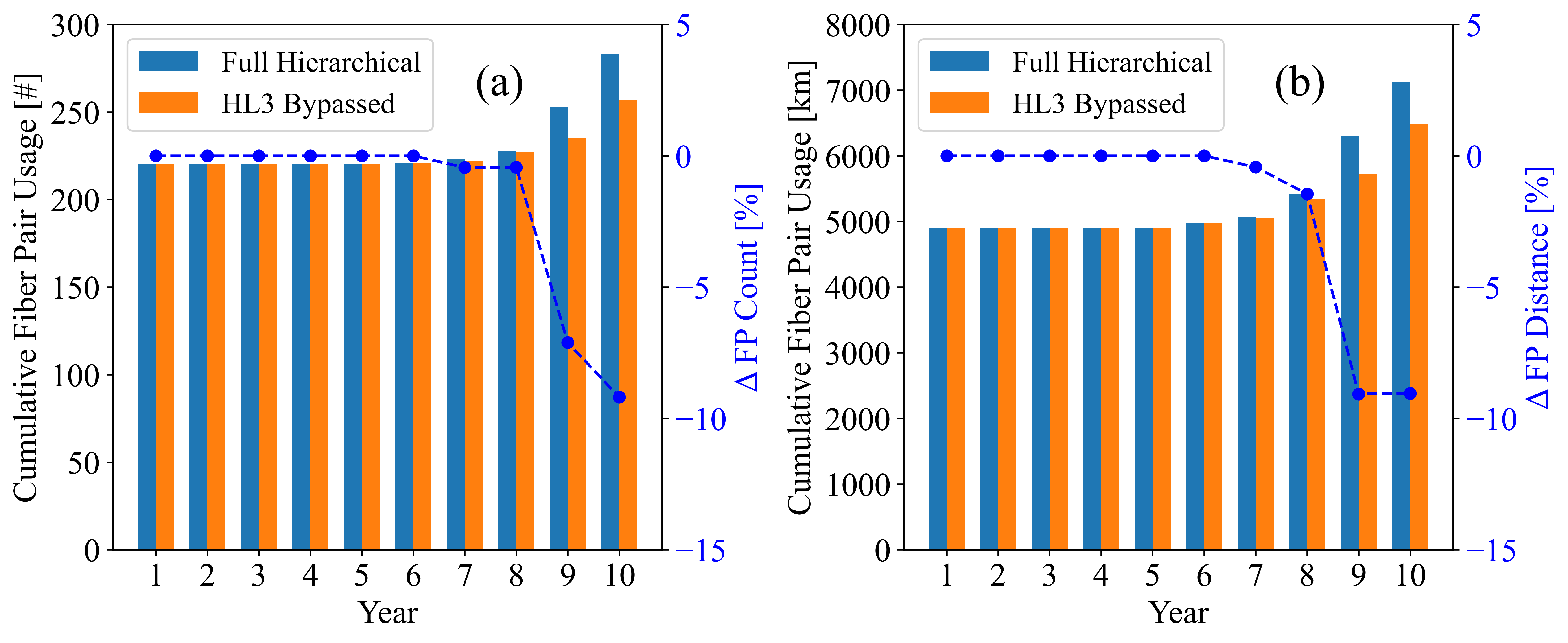}
		\caption{Cumulative (a) fiber pair count, (b) fiber pairs usage in km.} 
		\label{fig:cum_fp_usage}
	\end{figure}

	Figure~\ref{fig:cum_fp_usage} illustrates the cumulative fiber pair (FP) usage for different links across multiple years under two scenarios: (i) full hierarchical and (ii) HL3-bypassed. The relative difference used throughout this study is calculated as: 
	
	{\small
		\[
		\text{Relative Difference} = \frac{\text{HL3-Bypassed} - \text{Full Hierarchical}}{\text{Full Hierarchical}} \times 100
		\]
	}
	
	Figure~\ref{fig:cum_fp_usage}(a) presents the cumulative FP counts in both scenarios. The FP usage remains constant during Years~1--5. In Year~6, both scenarios experience an identical increase; therefore, the relative difference remains zero. From Year~7 onward, the FP usage in the full hierarchical scenario grows more rapidly, resulting in lower FP counts for the HL3-bypassed case. This trend continues through Years~9--10.
	
	Figure~\ref{fig:cum_fp_usage}(b) shows the cumulative total FP distance over time. The trend largely follows that of FP count. In Years~9--10, the growth in FP distance remains lower for the HL3-bypassed case, leading to negative relative differences.
	
	Overall, bypassing HL3 nodes reduces FP usage in terms of both count and distance over the study period, indicating a more cost-efficient architecture compared to the full hierarchical scenario.
	
\begin{figure}
		\centering
		\includegraphics[height = 6cm]{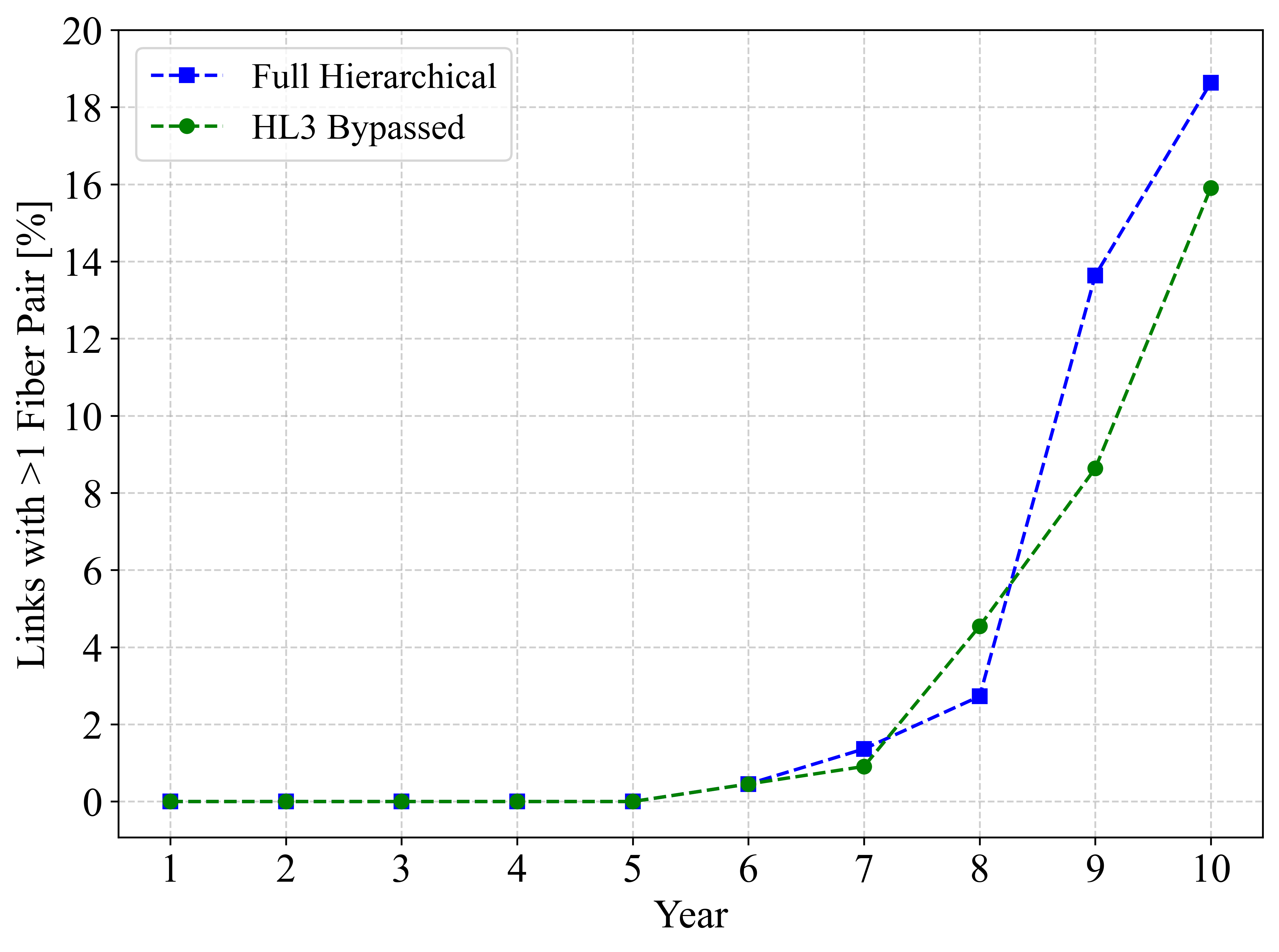}
		\caption{Percentage of links with more than one fiber pair.}
		\label{fig:>1_FP_links}
	\end{figure}

	Figure~\ref{fig:>1_FP_links} presents the percentage of links requiring more than one fiber pair (FP) across different years under both scenarios. During Years~1--5, no links exceed a single FP in either case. In Year~6, both scenarios show a similar increase in multi-FP usage. In Year~7, the growth is more pronounced in the full hierarchical scenario. In Year~8, a sharp rise occurs in the HL3-bypassed scenario, leading to a higher percentage than the full hierarchical case. However, in Years~9--10, the full hierarchical scenario again shows greater multi-FP utilization.
	
	\begin{figure}
		\centering
		\includegraphics[height = 4cm]{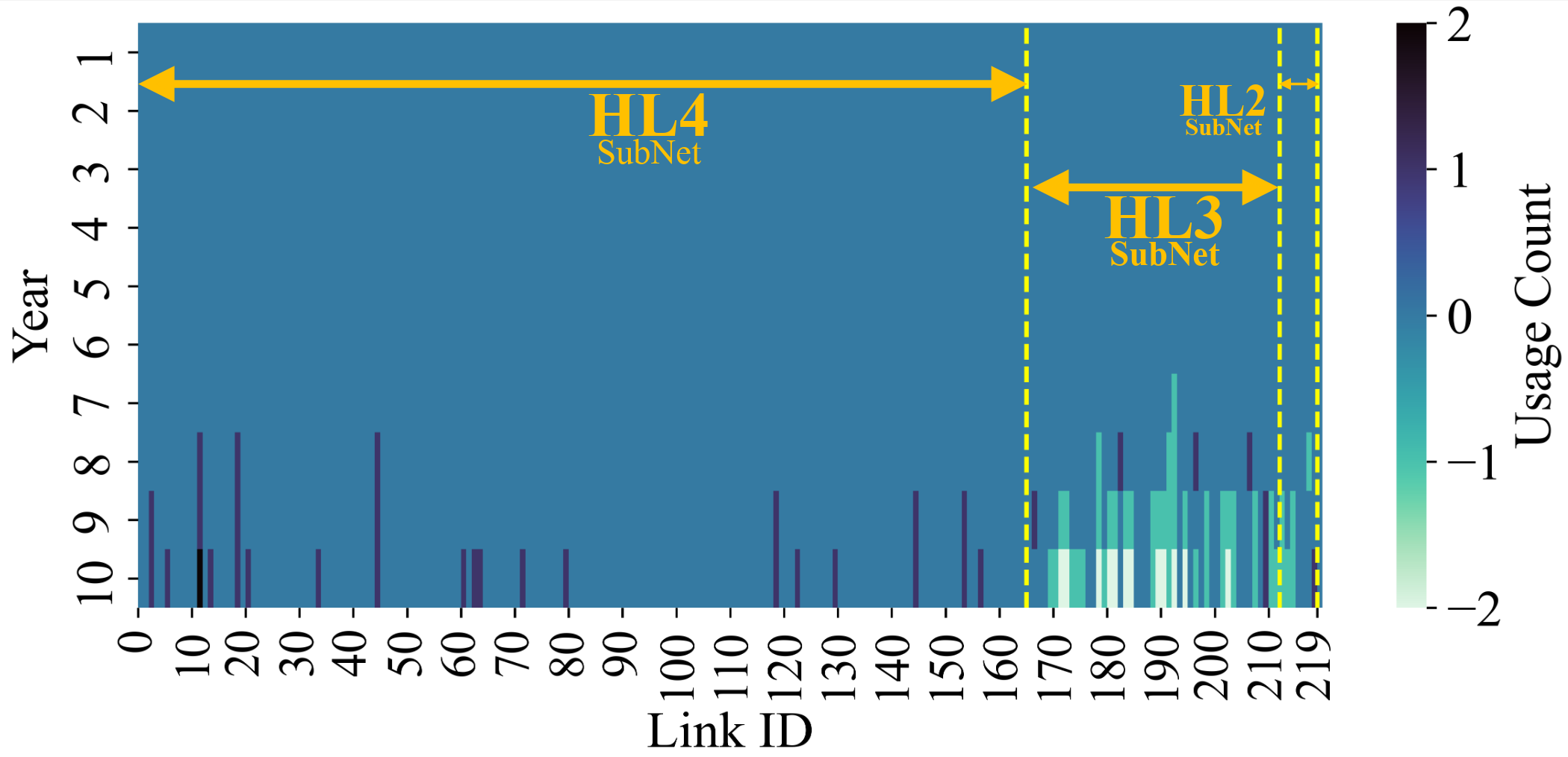}
		\caption{Fiber pair usage difference (HL3-Bypassed - Full Hierarchical) heatmap of network links over the planning horizon.}
		\label{fig:FP_Usage_Heatmap}
	\end{figure}
	
	On average, 3.68\% of links require more than one FP in the full hierarchical scenario, compared to 3.04\% in the HL3-bypassed scenario. These results align with those in Figure~\ref{fig:cum_fp_usage}, demonstrating that bypassing HL3 nodes generally reduces overall FP usage.
	
	

    Figure~\ref{fig:FP_Usage_Heatmap} illustrates the difference in FP count per link over time between the HL3-bypassed and full hierarchical scenarios in different sub-networks. In the HL4 subnet, bypassing HL3 nodes results in higher FP usage on several links in the later years, and no link shows a reduction in FP count under this configuration. Conversely, in the HL3 subnet, most links experience a reduction in FP usage when HL3 nodes are bypassed.

    While Figures~\ref{fig:cum_fp_usage} and~\ref{fig:>1_FP_links} confirm that bypassing HL3 nodes generally decreases total FP usage, Figure~\ref{fig:FP_Usage_Heatmap} highlights how this reduction is spatially distributed across the network. Specifically, shifting traffic away from HL3 nodes lowers demand in the HL3 subnet while increasing FP pressure in the HL4 subnet.
    
    This behavior arises from two opposing effects introduced by bypassing HL3 nodes. On one hand, traffic originating from HL4 nodes must be routed directly toward HL2 nodes without aggregation at HL3 nodes. This may lead to less efficient spectrum utilization for low-capacity optical signals, which in turn increases the number of fiber pairs required on some links, particularly within the HL4 subnet. On the other hand, bypassing HL3 nodes eliminates certain intermediate lightpaths that would otherwise exist in the HL3 subnet in the full hierarchical architecture.
    
    Under the considered dual-homed protection scheme, traffic from HL4 nodes toward HL2 destinations typically requires disjoint lightpaths. In the full hierarchical scenario, these paths traverse HL3 nodes, creating additional lightpaths within the HL3 subnet. When HL3 nodes are bypassed, the continuation of the HL4 subnet paths toward HL2 nodes removes the need for these intermediate HL3 lightpaths. As a result, the number of lightpaths—and consequently FP usage—decreases across many links in the HL3 subnet.
    
    In the MAN157 topology, although the HL4 subnetwork is significantly larger, comprising 166 links, the HL3 subnet contains only 47 links. Nevertheless, the number of links in the HL4 subnetwork that actually experience an FP increase is relatively limited. As illustrated in Figure~\ref{fig:FP_Usage_Heatmap}, only 19 links in the HL4 subnet show an increase in FP usage, whereas a larger number of links in the HL3 subnet experience reductions in FP count when HL3 nodes are bypassed. Therefore, the reduction in FP usage within the HL3 subnet outweighs the increase observed in the HL4 subnet, leading to an overall decrease in network-wide FP usage. Consequently, the aggregate reduction across the HL3 subnet dominates the localized increases observed in the HL4 subnet, resulting in a net decrease in total FP usage across the network. 
	
	\subsubsection{Band Usage}
	
	\begin{figure*}
		\centering
		\includegraphics[width = \textwidth]{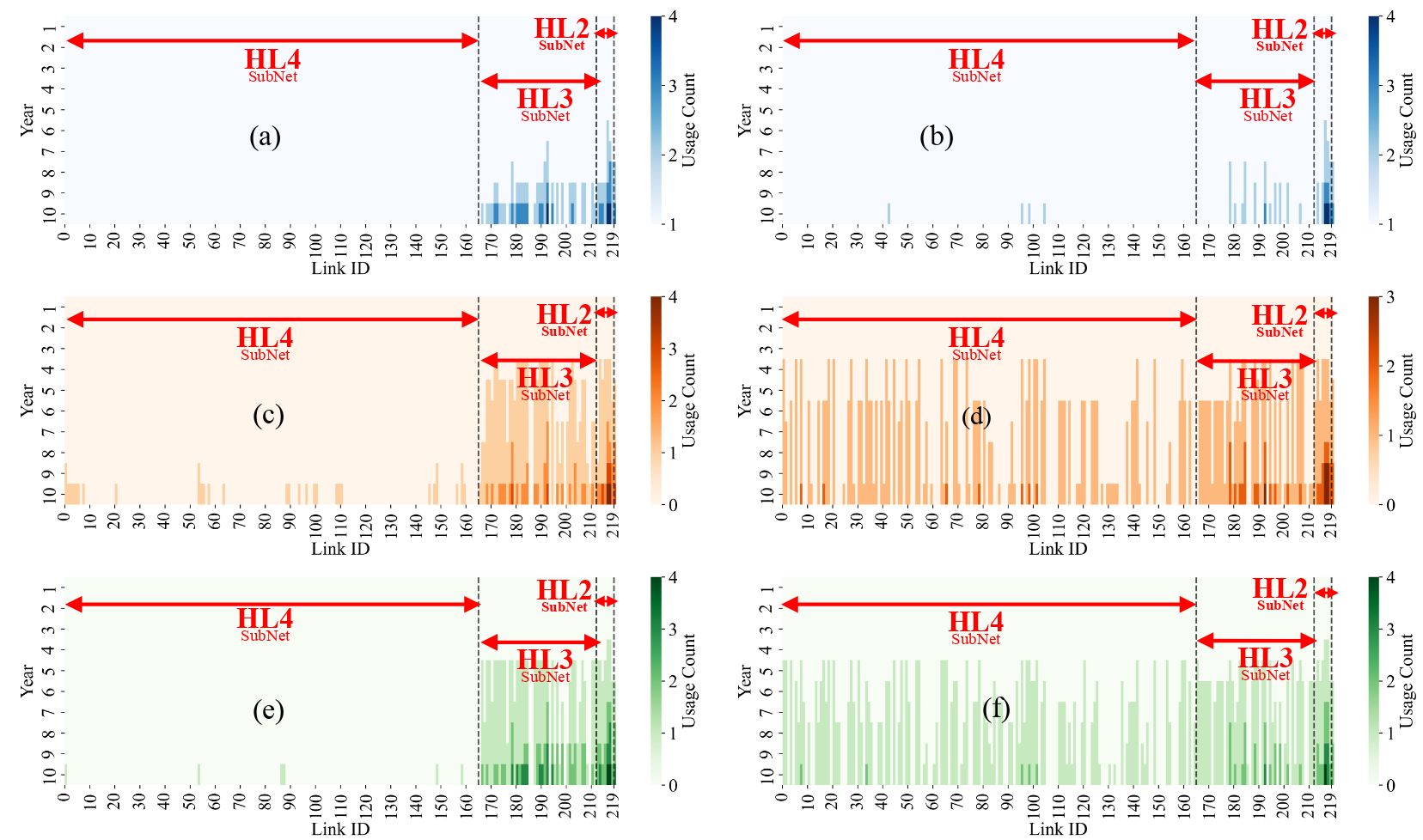}
		\caption{Band usage heatmaps of network links over the planning horizon for (a) C-band in Full Hierarchical, (b) C-band in HL3-Bypassed, (c) SuperC-band in Full Hierarchical, (d) SuperC-band in HL3-Bypassed, (e) L-band in Full Hierarchical and (f) L-band in HL3-Bypassed.}
		\label{fig:band_usage}
	\end{figure*}
	
	Figure~\ref{fig:band_usage} presents the annual band usage per link for the C-band, SuperC-band, and L-band under both network scenarios. Figures~\ref{fig:band_usage}(a--b) correspond to the C-band, (c--d) to the SuperC-band, and (e--f) to the L-band, with odd-indexed subfigures representing the full hierarchical scenario and even-indexed subfigures representing the HL3-bypassed scenario.
	
	Band usage for each link in a given year is defined as the number of fiber pairs in which at least one frequency slot (FS) is utilized within a given band. For example, if in Year~7 a link uses three FSs on the first fiber pair and one FS on the second, the band degree is considered 2. The maximum possible band degree for any link equals the maximum number of fiber pairs (20 in this study).
	
	As shown in Figures~\ref{fig:band_usage}(a), (c), and (e), in the full hierarchical scenario, no link within the HL4 subnet uses more than one fiber pair in the C-band. Conversely, Figures~\ref{fig:band_usage}(b), (d), and (f) show that in the HL3-bypassed scenario, several HL4 links exceed one fiber pair in the C-band. Since the planner allocates spectrum sequentially (C-band~$\rightarrow$~SuperC-band~$\rightarrow$~L-band) within each fiber pair before assigning new pairs, exceeding the C-band indicates full utilization of the lower bands on prior fiber pairs.
	
	These results suggest that bypassing HL3 nodes leads to higher congestion and greater band utilization in the HL4 subnet, while reducing usage in the HL3 subnet. Band usage in the HL2 subnet remains relatively stable across both scenarios. Overall, bypassing HL3 nodes shifts network load downward, increasing pressure on HL4 links while alleviating it on HL3 nodes.
	
	\begin{figure}
		\centering
		\includegraphics[height = 3.5cm]{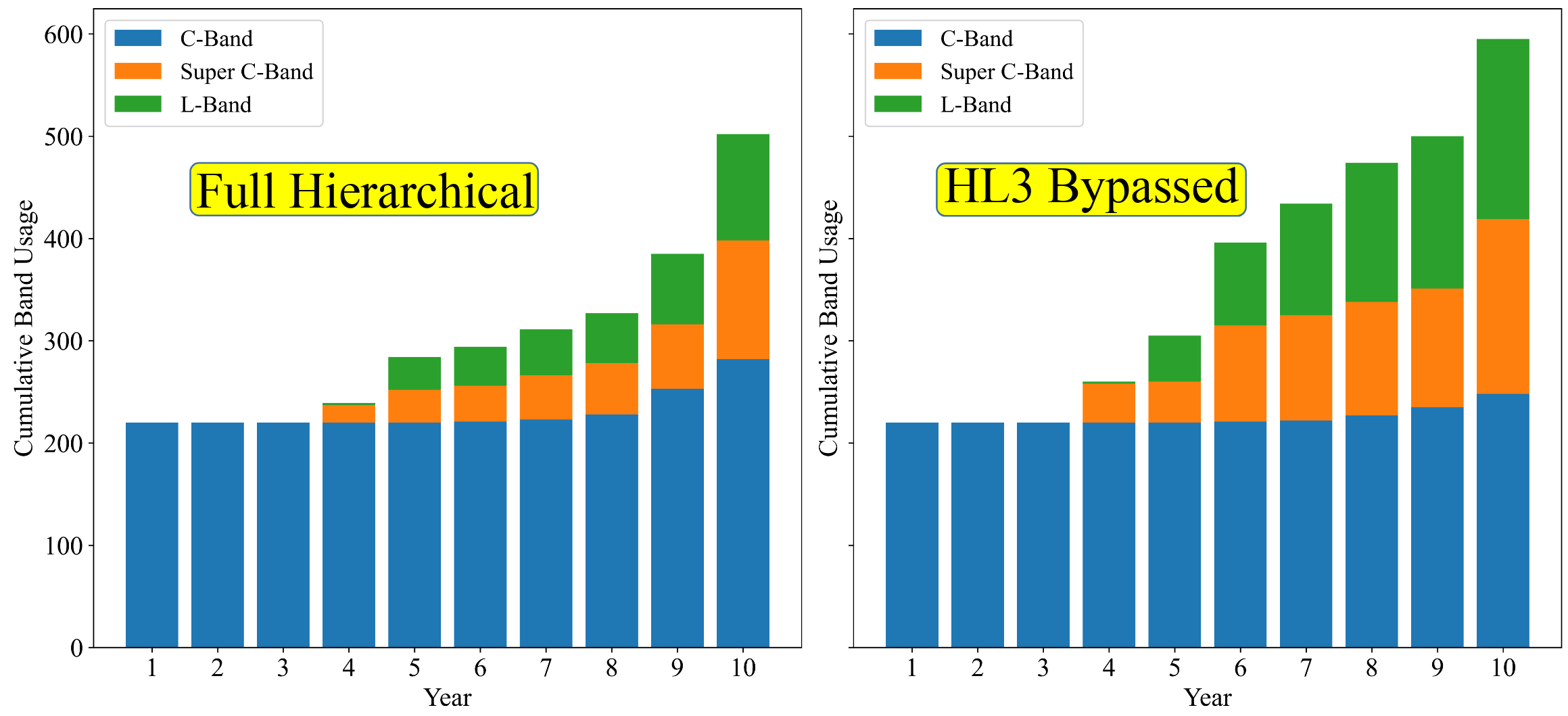}
		\caption{Cumulative band degree for (a) Full Hierarchical, (b) HL3-Bypassed scenario.}
		\label{fig:cumulative_band_degree}
	\end{figure}
	
	Figure~\ref{fig:cumulative_band_degree} presents the cumulative band degree over time for both network scenarios: (a) full hierarchical and (b) HL3-bypassed. In the full hierarchical scenario, the SuperC-band begins to be utilized in Year~4, followed by the L-band in the same year. In the HL3-bypassed scenario, SuperC-band and L-band usage also start in Year~4 but at a higher cumulative degree.
	
	These observations indicate that bypassing HL3 nodes have no impact on the initial activation of higher bands (SuperC and L) compared to the full hierarchical structure. However, once these bands begin to be used, their cumulative utilization increases more rapidly in the HL3-bypassed scenario. This suggests that HL3 bypassing promotes full exploitation of the C-band before transitioning to higher bands, ultimately leading to greater overall usage of the SuperC and L bands in later years.
	
	\subsubsection{BVT \& 100G License Usage}
	
	\begin{figure}
		\centering
		\includegraphics[height = 6.5cm]{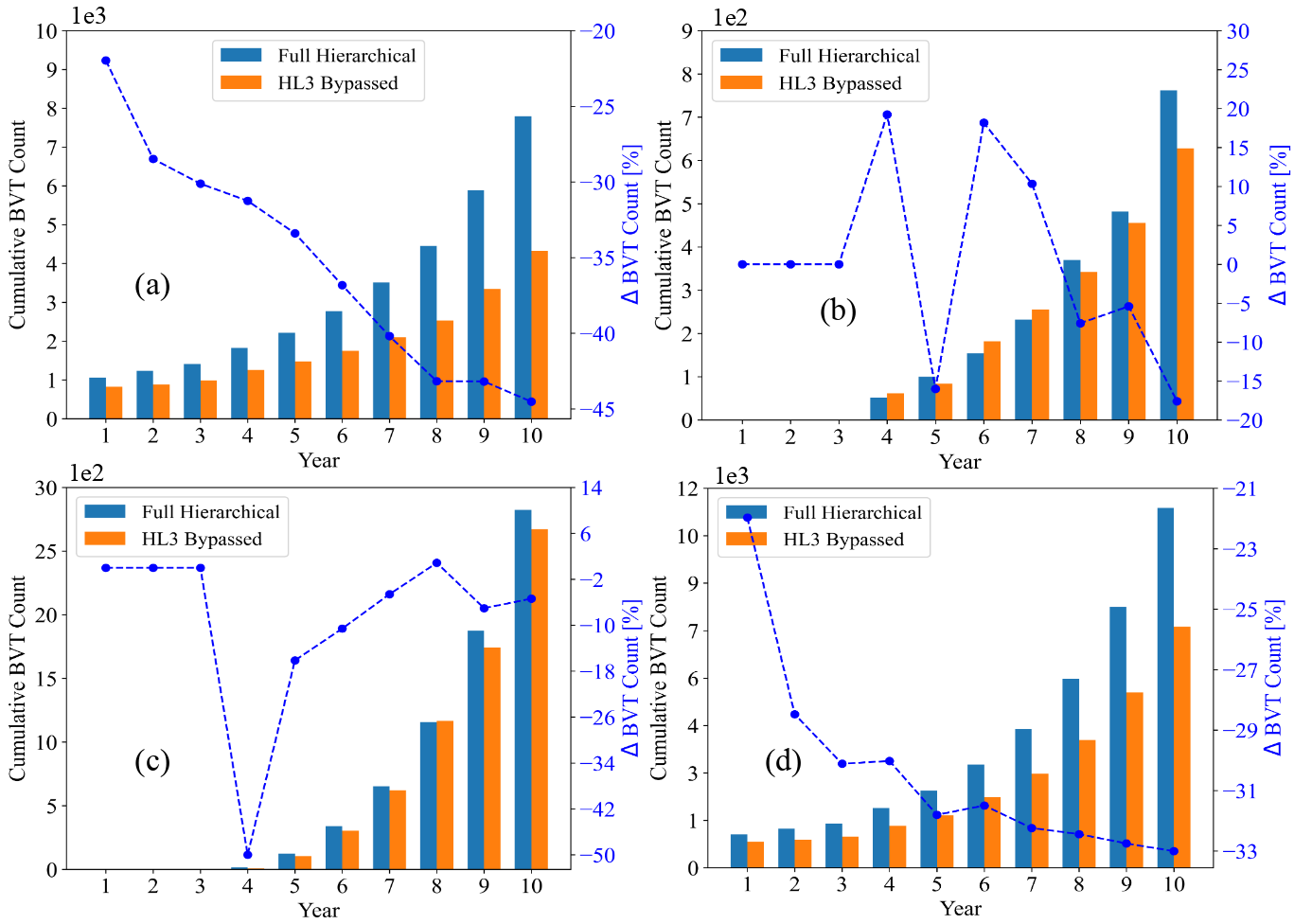}
		\caption{Cumulative (a) C-band BVT usage, (b) SuperC-band BVT usage, (c) L-band BVT usage, and (d) Total BVT usage.}
		\label{fig:bvt_usage}
	\end{figure}

	Figure~\ref{fig:bvt_usage} summarizes the number of bitrate-variable transponders (BVTs) established each year under the two network scenarios.
	
	Figure~\ref{fig:bvt_usage}(a) presents the BVTs deployed in the C-band. From the beginning of the planning horizon, the HL3-bypassed scenario consistently requires fewer BVTs to support the same traffic load. Moreover, the relative difference between the two scenarios increases over time, indicating that bypassing HL3 nodes leads to progressively more efficient BVT utilization. By Year~10, the HL3-bypassed scenario deploys approximately 44.5\% fewer BVTs in the C-band compared to the full hierarchical scenario.
	
	Figures~\ref{fig:bvt_usage}(b) and~\ref{fig:bvt_usage}(c) show the BVTs established in the SuperC-band and L-band, respectively. No BVTs are deployed in the SuperC-band in either scenario before Year~4. In the L-band, both scenarios show no activity from Years~1--4.
	By Year~10, BVT usage in both SuperC- and L-bands remains lower in the HL3-bypassed scenario.
	
	Figure~\ref{fig:bvt_usage}(d) reports the total number of BVTs established across all bands. Over the 10-year planning period, the HL3-bypassed scenario achieves an overall 33\% reduction in total BVT usage compared to the full hierarchical scenario. This demonstrates that bypassing HL3 nodes significantly improves transponder efficiency across the network.
	
	\begin{figure}
		\centering
		\includegraphics[height = 5.8cm]{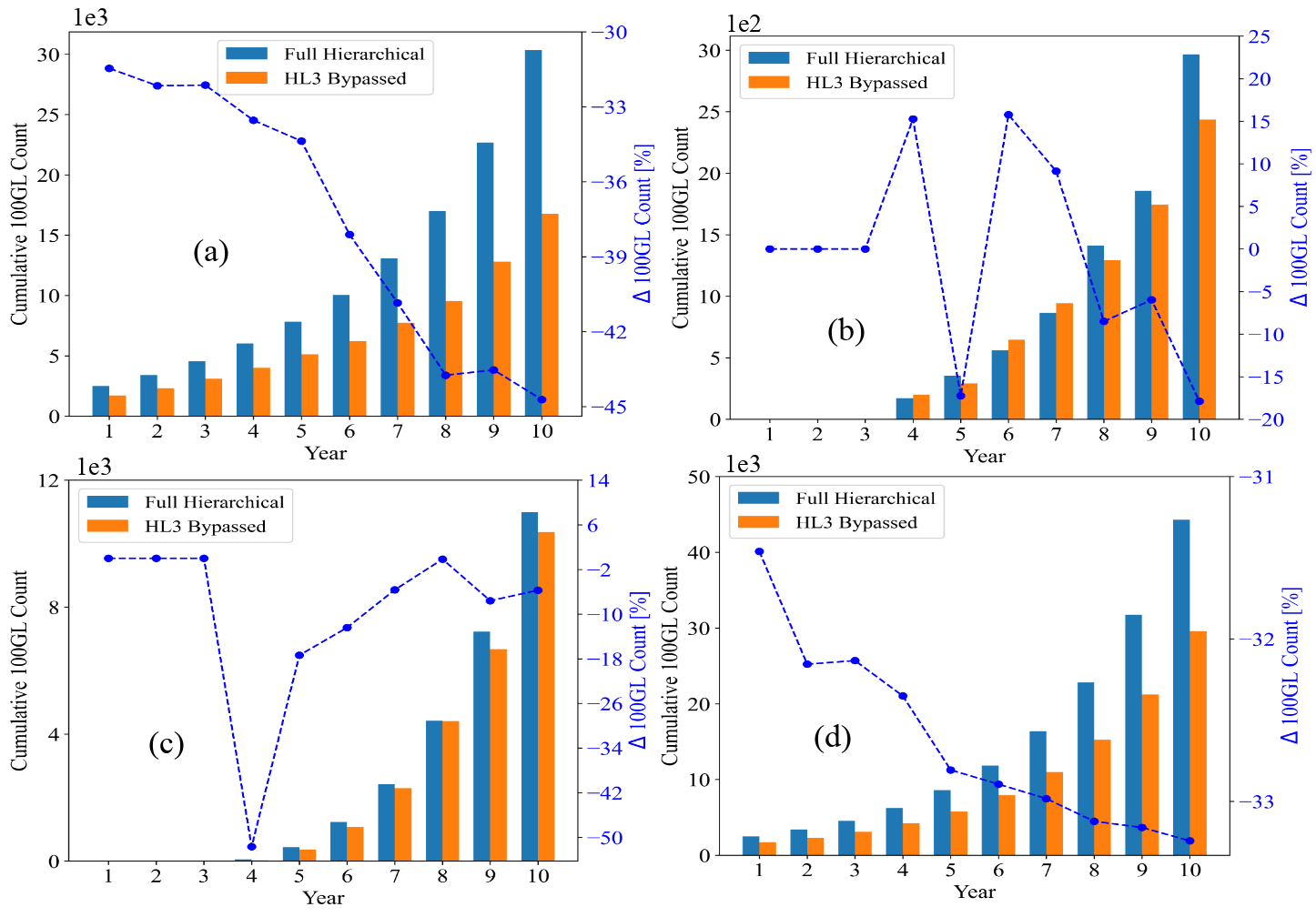}
		\caption{Cumulative (a) C-band 100G license usage, (b) SuperC-band 100G license usage, (c) L-band 100G license usage, and (d) Total 100G license usage.}
		\label{fig:100G_licenses}
	\end{figure}
	
	Figure~\ref{fig:100G_licenses} presents the number of activated 100G licenses per year for both network scenarios. Each 400G BVT contains four 100G licenses, which are activated sequentially as capacity is consumed. After all four licenses of a BVT are fully utilized, the planner deploys a new BVT, whose licenses are then activated gradually as traffic increases.
	
	As shown in Figure~\ref{fig:100G_licenses}(a), the number of 100G licenses activated in the C-band is consistently lower in the HL3-bypassed scenario for all years of the planning interval. However, Figures~\ref{fig:100G_licenses}(b) and~\ref{fig:100G_licenses}(c) reveal that toward the end of the planning period, the HL3-bypassed scenario activates more 100G licenses in the SuperC- and L-bands than the full hierarchical scenario in several years. This behavior aligns with earlier observations: bypassing HL3 nodes delays the initial use of higher bands, but once the SuperC- and L-bands begin to be used, their activation levels grow more rapidly than in the full hierarchical case.
	
	Figure~\ref{fig:100G_licenses}(d) shows the total number of activated 100G licenses across all bands. Overall, the HL3-bypassed scenario results in an approximately 33\% reduction in total licenses by Year~10. The relative-difference trend indicates that the greatest efficiency gains occur in the final years. This demonstrates that bypassing HL3 nodes significantly reduce license activation across the network.
	
	\subsubsection{IP Router Usage}
	\begin{figure}
		\centering
		\includegraphics[height = 6.3cm]{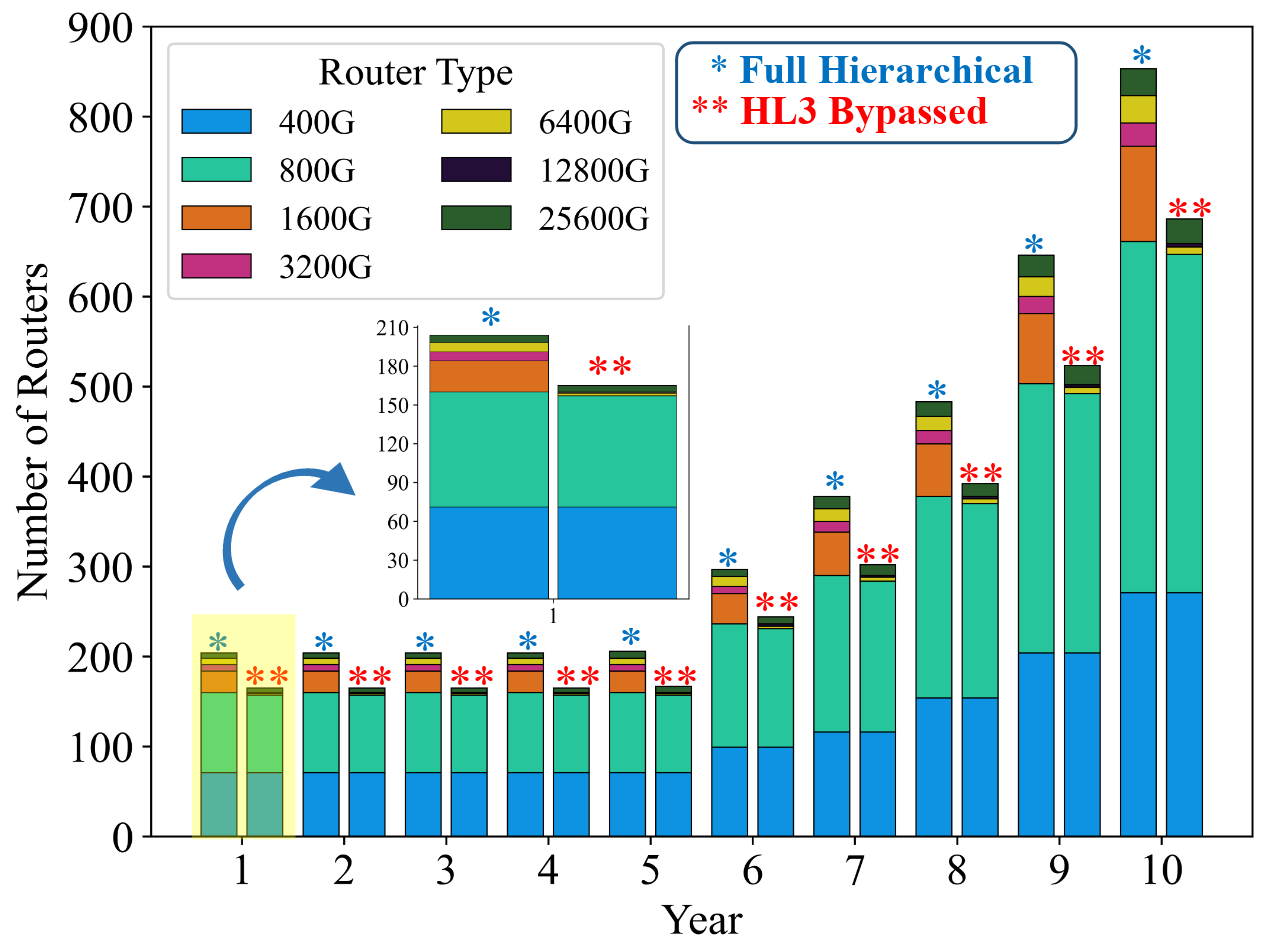}
		\caption{Cumulative number of IP routers.}
		\label{fig:ip_routers_usage}
	\end{figure}
	
	Figure~\ref{fig:ip_routers_usage} illustrates the cumulative number of IP routers deployed across all network nodes in each year for the two scenarios. The plot distinguishes router types using different colors. Seven base router capacities are considered in this study: 400~G, 800~G, 1600~G, 3200~G, 6400~G, 12800~G, and 25600~G. For each node, the traffic in Year~5 is used to select the base router capacity. for example, If the Year-5 traffic is 450~G, an 800~G router is selected as the base router; additional routers of the same type are deployed as traffic at that node grows. If a node's Year-5 traffic exceeds 25600~G, the largest router (25600~G) is used as the base.
	
	Each year contains two bar groups: the left bar (marked ``*'') corresponds to the full hierarchical scenario, and the right bar (marked ``**'') corresponds to the HL3-bypassed scenario. As shown, the number of 400~G routers is identical in both scenarios for all years, indicating that deployments of the smallest-capacity routers are unaffected by HL3 bypassing.
	
	In contrast, the deployment of higher-capacity routers differs significantly between the two architectures. When HL3 nodes are bypassed, no traffic aggregation occurs at HL3; traffic remains in the optical domain and is forwarded directly to higher-level nodes. Consequently, the traffic that would normally concentrate at HL3 nodes shifts toward HL2 nodes, altering the traffic distribution and resulting router capacities. In the full hierarchical scenario, no node requires a 12800~G router as its base router. However, under the HL3-bypassed scenario, the modified traffic distribution leads some nodes to deploy 12800~G base routers, while the number of 6400~G and 25600~G routers decreases.
	
	These differences in router-type selection directly affect the total number of routers deployed and the associated CAPEX. It is important to note that the total traffic volume in HL1 and HL2 nodes remains identical in both scenarios, as the initial traffic values and annual growth rates are the same; only the distribution of traffic among nodes changes due to the bypassing of HL3 nodes.
	
	\subsubsection{Cost Analysis}
	
	In this section, the cost performance of the two network scenarios is evaluated. The analysis is organized into three components: optical-layer cost, electrical (IP-layer) cost, and total network cost. Table~\ref{tab:cost_units} summarizes the base cost units assigned to the various network elements considered in this study. The Optical CAPEX and OPEX cost units is taken from \citep{arpanaei2024enabling}, and the Electrical CAPEX cost units is taken from \citep{hernandez2020comprehensive}. In this study, the reference cost unit is set to 15{,}000~Euro, corresponding to the price of the first activated 100G license in a BVT. It is worth noting that when a new BVT is established, for the first activated 100G license the cost of unity and for the second, third and the fourth activated 100G license, the cost of 0.33 is considered.
	
	\begin{table}[ht]
		{\fontsize{7}{8}\selectfont
		\centering
		\caption{\textbf{Estimated CAPEX and OPEX Values Based on the Cost Unit [CU]}} 
		\label{tab:cost_units}
		\renewcommand{\arraystretch}{1.1}
		\setlength{\tabcolsep}{6pt}
		\begin{tabular}{l >{\centering\arraybackslash}p{3cm} l}
			\toprule
			\textbf{Cost Type} & \textbf{Equipment} & \textbf{Cost [CU]} \\
			\midrule
			
			\multirow{6}{*}{Optical CAPEX} 
			& 100G          & $C_{\mathrm{TRx}} = 1$ \\
			& 200G          & $C_{\mathrm{TRx}} = 1.33$ \\
			& 300G          & $C_{\mathrm{TRx}} = 1.66$ \\
			& 400G          & $C_{\mathrm{TRx}} = 2$ \\
			& $1 \times 20$ CDC-RoB & $C_{\mathrm{RoB}} = 1.9$ \\
			& $8 \times 16$ MCS     & $C_{\mathrm{MCS}} = 0.7$ \\
			
			\midrule
			
			OPEX 
			& IRU/lease \\[-3pt]
			& {\footnotesize [strand/km/year]} 
			& $C_{\mathrm{IRU}} = 0.5$ \\
			
			\midrule
			
			\multirow{6}{*}{Electrical CAPEX} 
			& 400G IP Router & $C_{\mathrm{IP}} = 1.6$  \\
			& 800G IP Router & $C_{\mathrm{IP}} = 3.2$  \\
			& 1600G IP Router & $C_{\mathrm{IP}} = 6.4$  \\
			& 3200G IP Router & $C_{\mathrm{IP}} = 12.8$  \\
			& 6400G IP Router & $C_{\mathrm{IP}} = 25.6$  \\
			& 12800G IP Router & $C_{\mathrm{IP}} = 51.2$  \\
			& 25600G IP Router & $C_{\mathrm{IP}} = 102.4$  \\
			
			\bottomrule
		\end{tabular}
	}
	\end{table}
	
\begin{figure}
		\centering
		\includegraphics[height = 4.2cm]{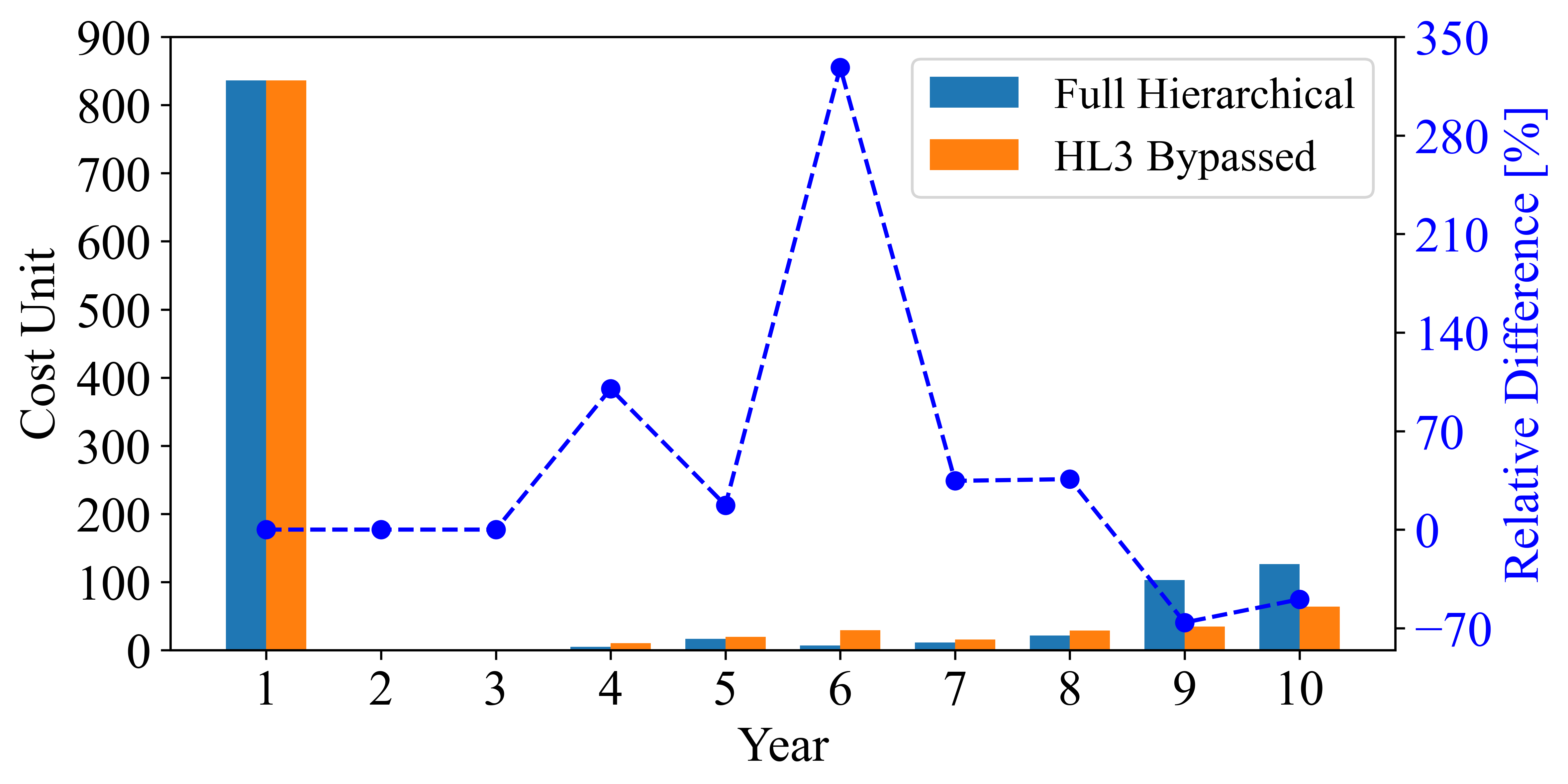}
		\caption{Annual CAPEX of ROADM on the Blade (RoB).}
		\label{fig:rob_cost}
	\end{figure}

	Figure~\ref{fig:rob_cost} shows the annual cost associated with ROADM-on-the-Blade (RoB) deployments for both network scenarios. In Year~1, the number of ROADMs deployed is identical in the two scenarios. During Years~2--3, no new fiber pairs are added to the network; consequently, no additional ROADMs are required, and the RoB cost remains zero for both cases.
	
	In Year~4, new fiber pairs are introduced in both scenarios, leading to the deployment of new ROADMs. However, the number of ROADMs established in the HL3-bypassed scenario is higher than in the full hierarchical scenario, indicating that although both architectures begin using additional fiber pairs at the same time, the HL3-bypassed scenario requires more extensive FP expansion at this stage.
	
	A same behavior emerges in Years~5--8, where the HL3-bypassed scenario requires significantly more new fiber pairs and therefore more ROADM installations than the full hierarchical scenario. In Years~9 and~10, this trend reverses, with the full hierarchical scenario exhibiting higher new fiber-pair deployment and correspondingly greater RoB cost.
	
	Overall, the RoB cost profile directly mirrors the pattern of new fiber-pair establishment in each scenario, reflecting how differences in network-layer architecture influence optical-layer expansion requirements.
	
\begin{figure}
		\centering
		\includegraphics[height = 4.2cm]{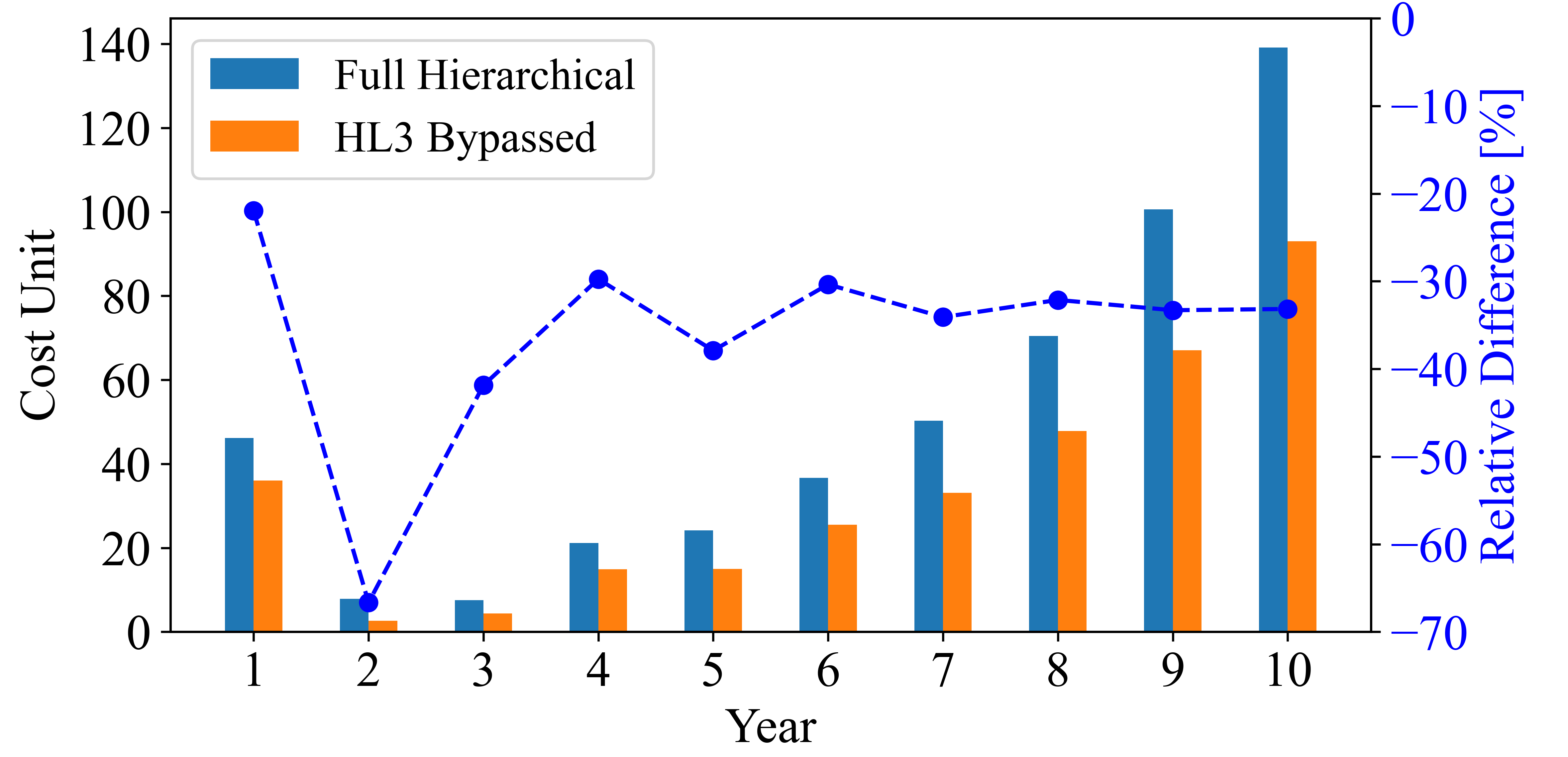}
		\caption{Annual CAPEX of Multi-Cast Switch (MCS).}
		\label{fig:mcs_cost}
	\end{figure}

	Figure~\ref{fig:mcs_cost} presents the annual cost of the multicast switch (MCS) for both network scenarios. As shown, the MCS cost is consistently lower in the HL3-bypassed scenario across all years. This reduction is directly attributed to the smaller number of BVTs deployed each year when HL3 nodes are bypassed. Since each BVT requires an associated MCS port, fewer BVT installations naturally result in lower overall MCS cost in the HL3-bypassed architecture compared to the full hierarchical scenario.
	
\begin{figure}
		\centering
		\includegraphics[height = 4.2cm]{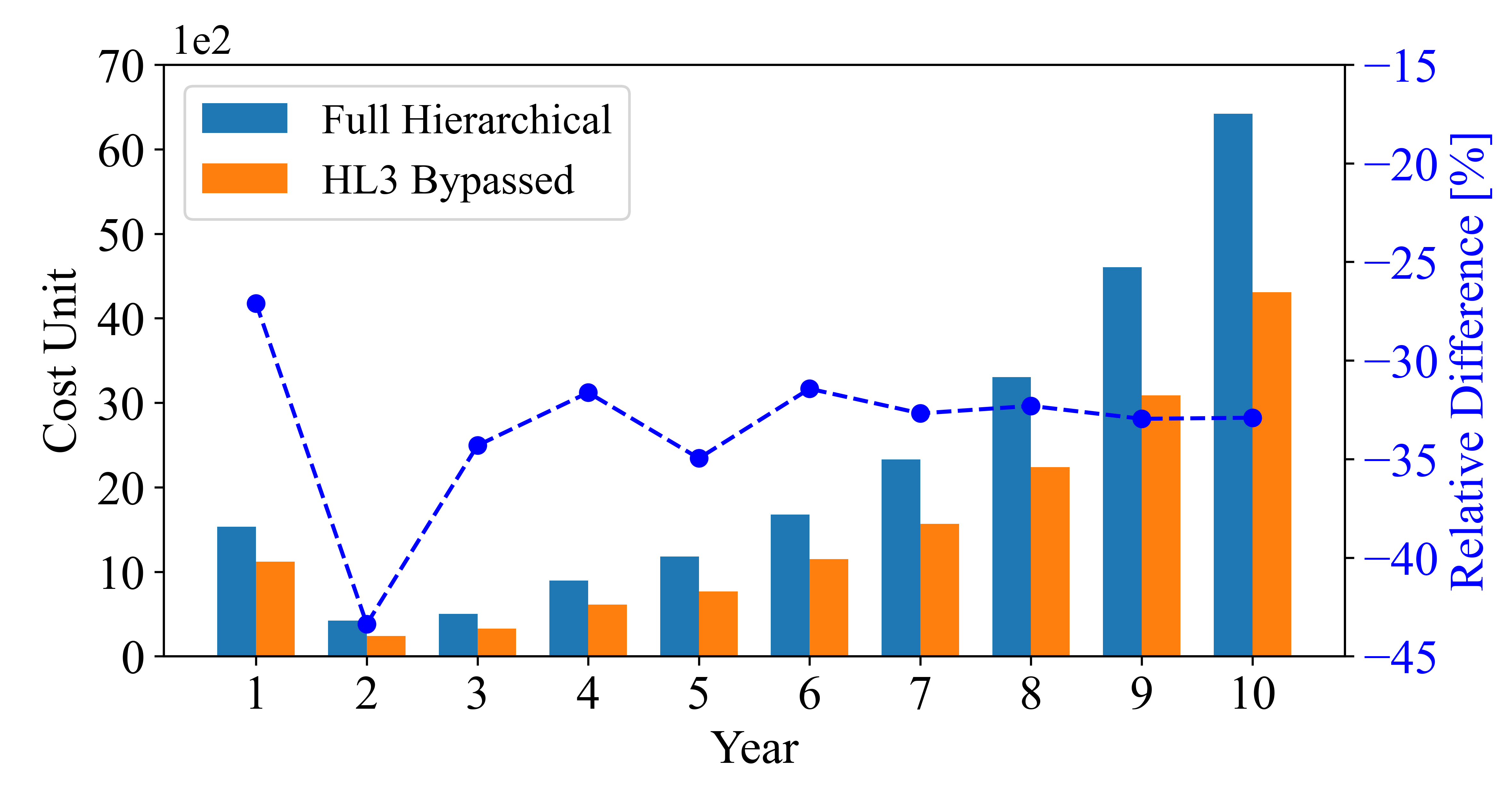}
		\caption{Annual CAPEX of 100G Licenses (100GL).}
		\label{fig:license_cost}
	\end{figure}

	Figure~\ref{fig:license_cost} illustrates the annual cost associated with 100G license activation across the entire network. As shown, the number of activated licenses and therefore the cost is consistently higher in the full hierarchical scenario compared to the HL3-bypassed scenario. As discussed earlier, the gap between the two scenarios decreases toward the end of the planning period, indicating that 100G license activation accelerates in the HL3-bypassed scenario in the later years.
	
\begin{figure*}
		\centering
		\includegraphics[width =\textwidth]{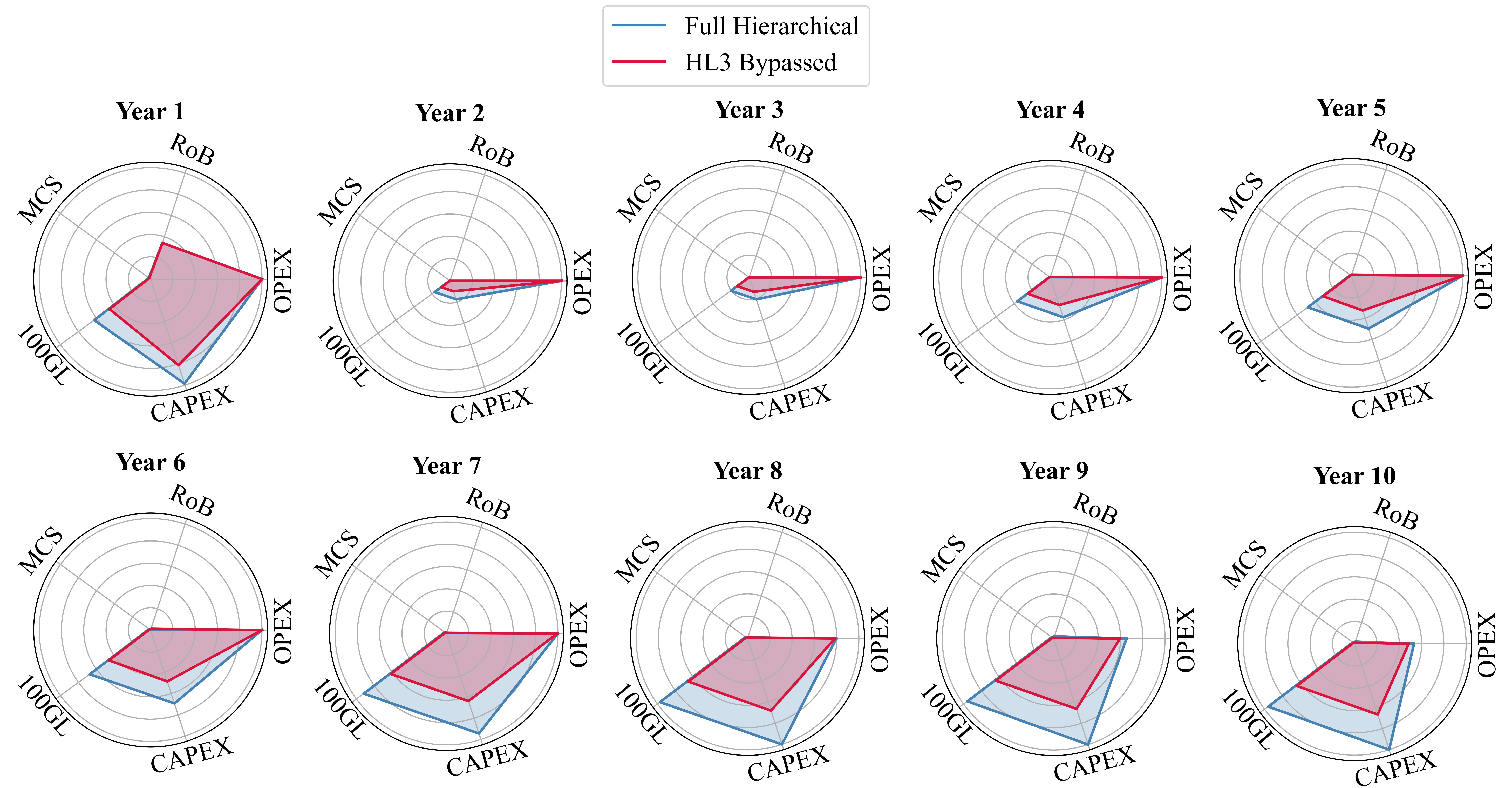}
		\caption{Optical cost structure per year.}
		\label{fig:spider_cost}
	\end{figure*}

	Figure~\ref{fig:spider_cost} presents a spider plot comparing the cost structure of the two network scenarios. As shown, the CAPEX components for the full hierarchical scenario are consistently higher than those of the HL3-bypassed scenario across all cost categories. The figure also highlights that the contribution of multicast switches and ROADM-on-the-Blade (RoB) to total CAPEX is relatively small compared to the contribution of 100G license activation. This emphasizes that 100G license costs are the dominant driver of optical-layer CAPEX in both architectures. The OPEX component, previously analyzed, is included here for completeness and to provide a comprehensive comparison of the overall cost structure between the two scenarios.

\begin{figure}
		\centering
		\includegraphics[height = 6.5cm]{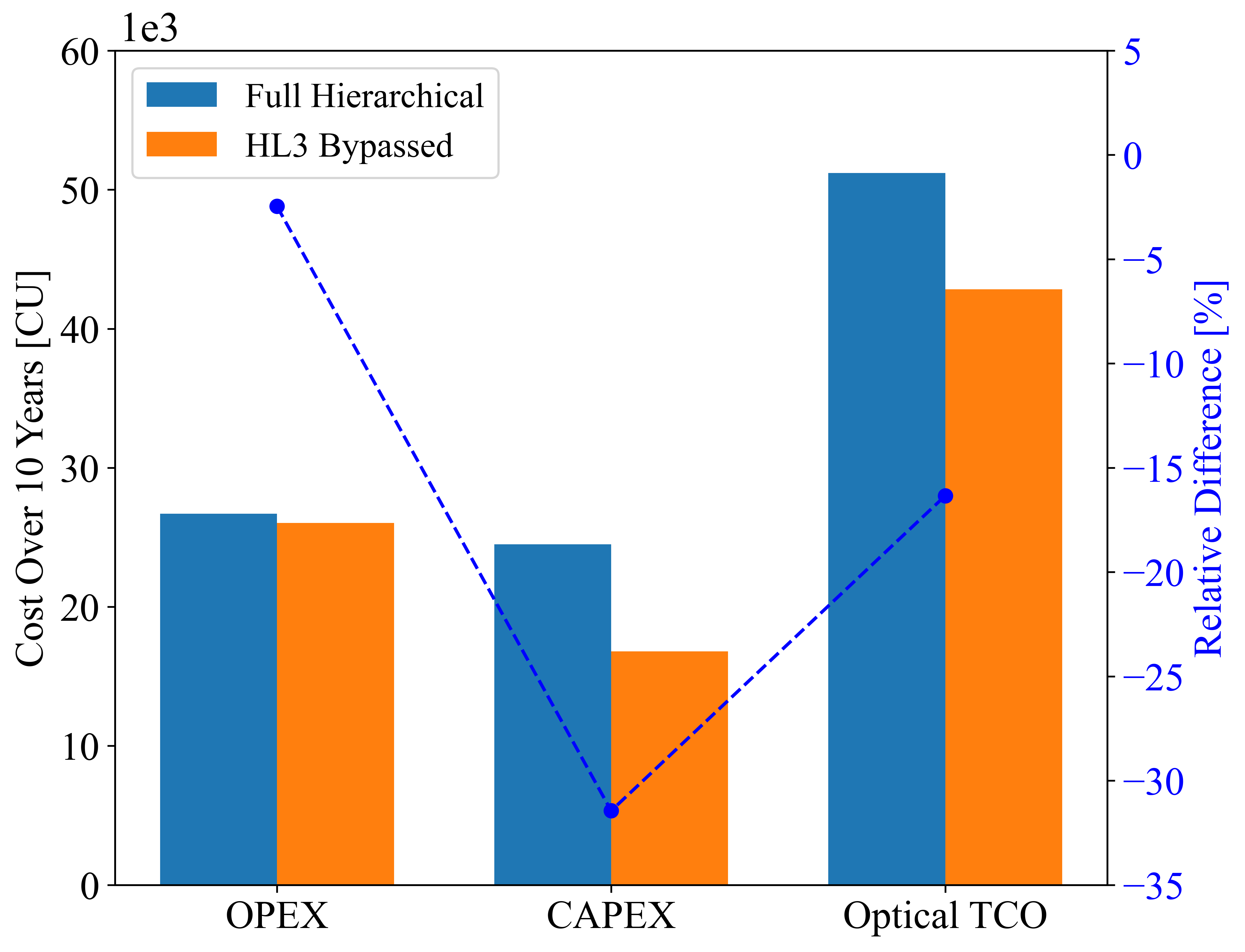}
		\caption{Total CAPEX, OPEX and Optical TCO over 10 years.}
		\label{fig:total_opt_cost}
	\end{figure}

	Figure~\ref{fig:total_opt_cost} illustrates the total CAPEX, OPEX, and optical Total Cost of Ownership (TCO) accumulated over the 10-year planning horizon for both network scenarios. As shown, the OPEX of the HL3-bypassed scenario is approximately 2.5\% lower than that of the full hierarchical scenario, indicating that the overall fiber-pair usage remains relatively similar in both architectures; hence, the OPEX difference is modest.
	
	In contrast, the CAPEX of the HL3-bypassed scenario is approximately 33.5\% lower than the full hierarchical scenario. Consequently, the optical TCO in the HL3-bypassed architecture is about 16.3\% lower compared to the full hierarchical design.
	
	Overall, these results demonstrate that bypassing HL3 nodes leads to reductions in both CAPEX and OPEX, and therefore yields a substantially lower optical TCO. This indicates that the HL3-bypassed architecture can provide notable cost efficiencies over long-term network operation.
	
\begin{figure}
		\centering
		\includegraphics[height = 3.5cm]{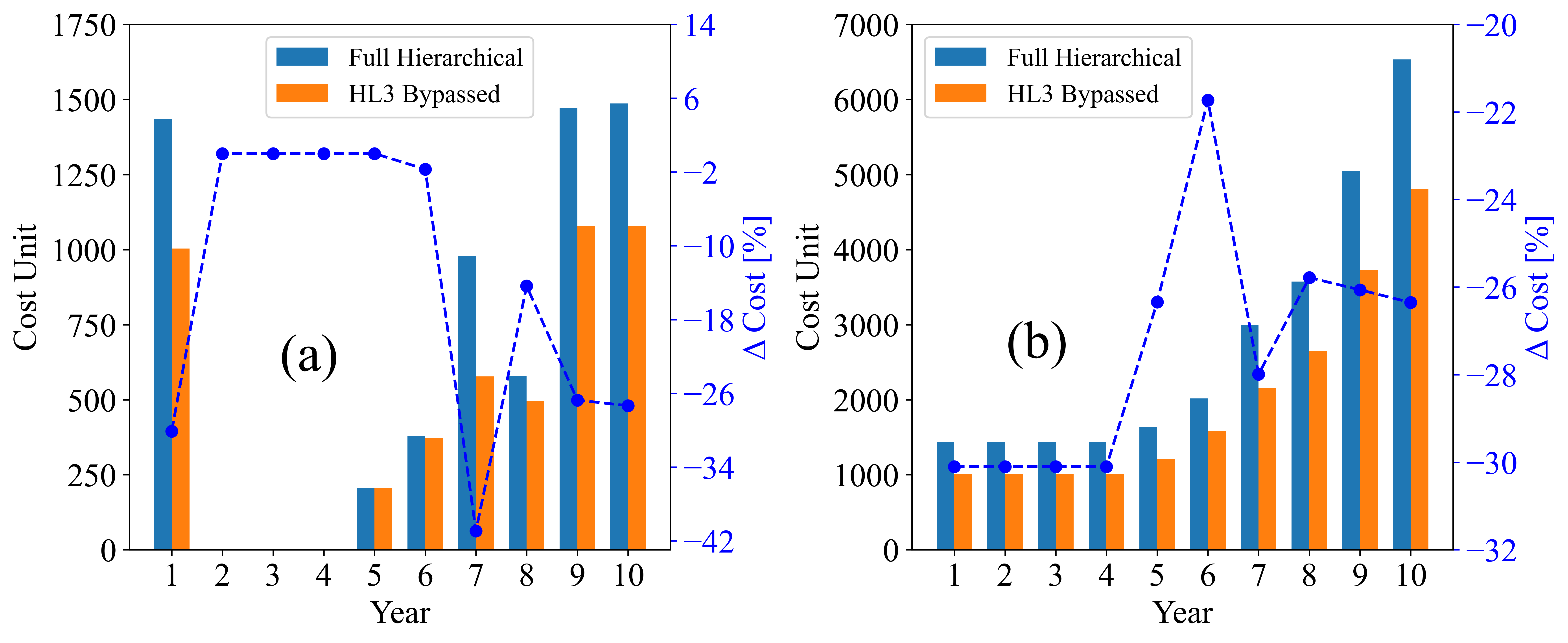}
		\caption{(a) Annual, (b) Cumulative Cost of IP Routers (electrical cost).}
		\label{fig:ip_cost}
	\end{figure}

	Figure~\ref{fig:ip_cost}(a) presents the annual IP router cost (i.e., the electrical-layer cost) for both network scenarios. As shown, in Year~1 the total cost of IP routers deployed across the network in the full hierarchical scenario is approximately 30.1\% higher than in the HL3-bypassed scenario. This difference arises from the elimination of IP routers at HL3 nodes in the bypassed architecture.
	
	From Years~2 to~4, although each node experiences a 40\% annual traffic growth, no additional IP routers are required in either scenario. In Year~5, the cost of IP router deployment becomes identical in both scenarios. This is expected because, as previously explained, the “base router” at each node is selected according to the node's traffic load in Year~5; thus, both scenarios deploy routers sized to accommodate the full Year-5 demand.
	
	From Years~6 to~10, new IP routers are added as traffic continues to grow. Across all these years, the HL3-bypassed scenario consistently incurs lower annual router deployment costs. This reflects the changed traffic distribution resulting from bypassing HL3 nodes, which reduces the need for high-capacity routers at intermediate aggregation points.
	
	Figure~\ref{fig:ip_cost}(b) shows the cumulative IP router cost over the planning interval. By the end of Year~10, the HL3-bypassed scenario exhibits a cumulative electrical cost that is approximately 26\% lower than that of the full hierarchical scenario. This confirms that bypassing HL3 nodes fundamentally alters the traffic distribution across the network, resulting in different base-router selections and fewer total router installations which ultimately yielding significantly lower electrical-layer CAPEX.

\begin{figure}
		\centering
		\includegraphics[height = 6.5cm]{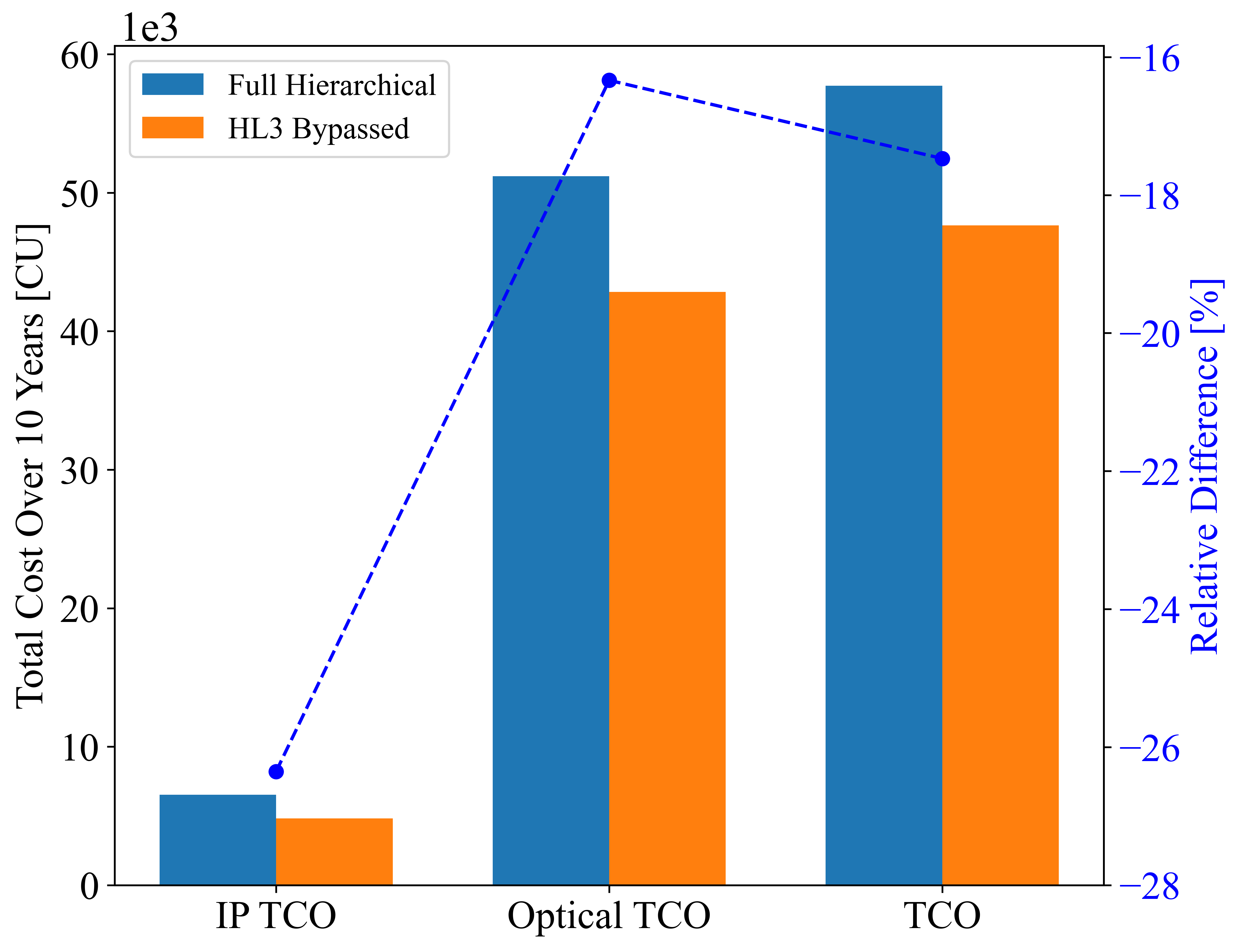}
		\caption{IP TCO, Optical TCO and TCO over 10 years.}
		\label{fig:total_tco}
	\end{figure}

	Figure~\ref{fig:total_tco} presents the Optical TCO, Electrical(IP) TCO, and the combined Optical~+~Electrical Total Cost of Ownership (TCO) for both network scenarios over the 10-year planning interval. As shown, the Electrical TCO which corresponds to the cumulative IP router cost is approximately 26.3\% lower in the HL3-bypassed scenario compared to the full hierarchical architecture.
	
	Similarly, the Optical TCO in the HL3-bypassed scenario is about 16.3\% lower than in the full hierarchical scenario, reflecting the reduced need for 100G license activations when HL3 nodes are removed from the optical switching hierarchy.
	
	When combining both layers, the resulting Total Cost of Ownership (TCO) in the HL3-bypassed scenario becomes approximately 17.5\% lower than that of the full hierarchical scenario.
	
	Overall, this figure demonstrates that bypassing HL3 nodes leads to substantial cost reductions in both the electrical and optical layers and consequently yields a significantly lower total network TCO.

	\subsubsection{Energy Consumption}
	
	In this section, the two scenarios are compared in terms of energy consumption. The analysis is divided into three components: the energy consumption of optical components (BVTs), the energy consumption of electrical components (IP routers), and the total energy consumption for both scenarios. Table~\ref{tab:power_values} lists the power consumption values of the various network elements considered in this study.
	
	\begin{table}[ht]
		{\fontsize{7}{8}\selectfont
			\centering
			\caption{\textbf{Power Consumption of different network elements in terms of Watt}} 
			\label{tab:power_values}
			\renewcommand{\arraystretch}{1.1}
			\setlength{\tabcolsep}{6pt}
			
			\begin{tabular}{l >{\centering\arraybackslash}p{3cm} l}
				\toprule
				\textbf{Equipment Type} & \textbf{Equipment} & \textbf{Power (Watt)} \\
				\midrule
				
				Optical Equipments
				& 400G BVT          & $P_{\mathrm{BVT}} = 160$ \\
				
				\midrule
				
				\multirow{6}{*}{Electrical Equipments} 
				& 400G IP Router & $P_{\mathrm{IP}} = 2000$  \\
				& 800G IP Router & $P_{\mathrm{IP}} = 2500$  \\
				& 1600G IP Router & $P_{\mathrm{IP}} = 3100$  \\
				& 3200G IP Router & $P_{\mathrm{IP}} = 3900$  \\
				& 6400G IP Router & $P_{\mathrm{IP}} = 4900$  \\
				& 12800G IP Router & $P_{\mathrm{IP}} = 6100$  \\
				& 25600G IP Router & $P_{\mathrm{IP}} = 7600$  \\
				
				\bottomrule
			\end{tabular}
		}
	\end{table}

\begin{figure}
		\centering
		\includegraphics[height = 6cm]{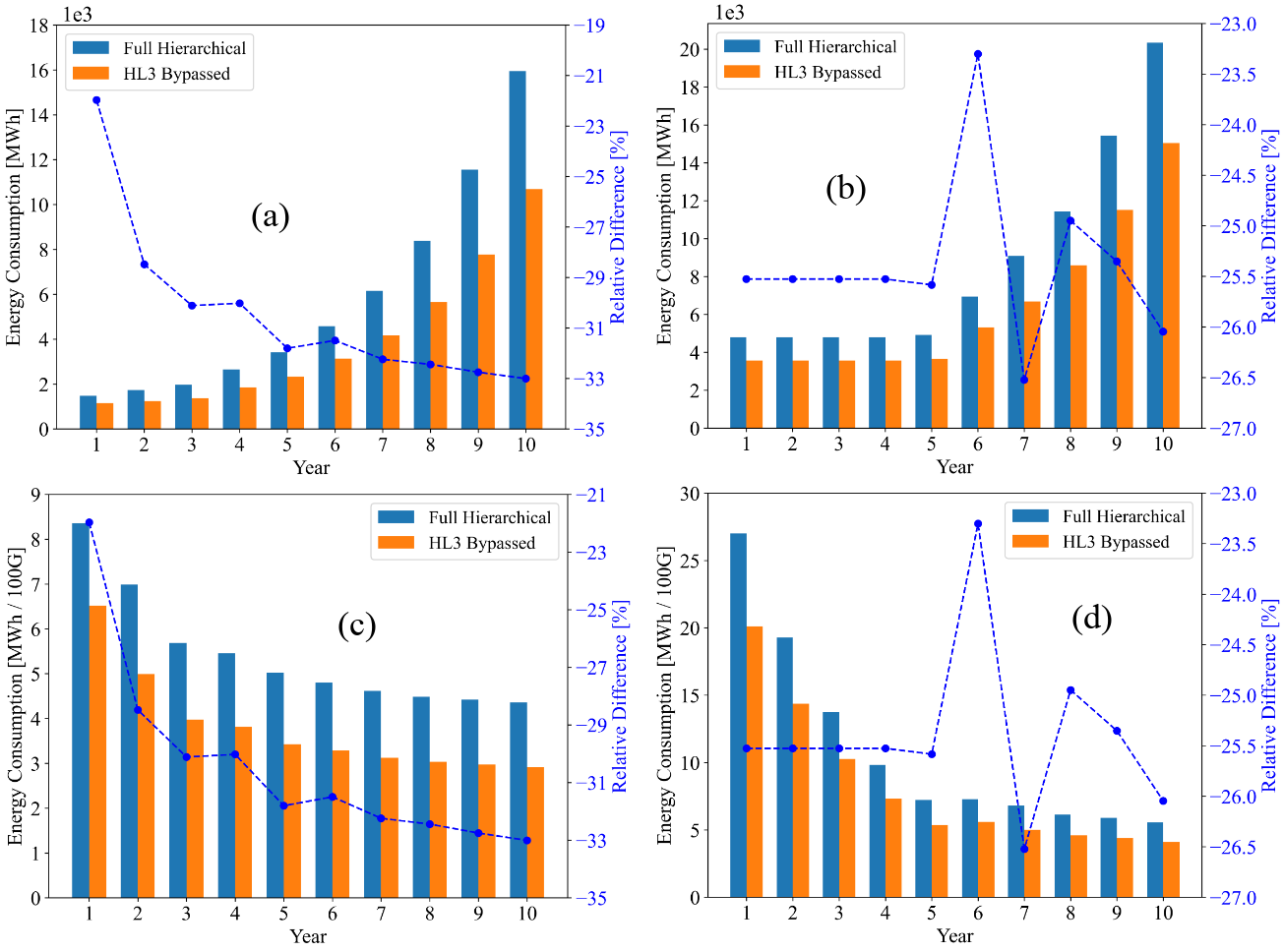}
		\caption{Cumulative (a) optical, (b) electrical energy consumption. Cumulative (c) optical, (d) electrical energy consumption per 100G traffic.}
		\label{fig:energy_components}
	\end{figure}

	Figure~\ref{fig:energy_components} illustrates the cumulative optical and electrical energy consumption (in MWh) for both network scenarios over the 10-year planning horizon.
	
	Figure~\ref{fig:energy_components}(a) presents the cumulative optical energy consumption, which is derived from the energy usage of BVTs. As shown, the HL3-bypassed scenario consistently consumes less optical energy compared to the full hierarchical architecture. This behavior reflects the lower number of BVTs deployed in the HL3-bypassed scenario. In the first year, the optical energy consumption is approximately 22\% lower in the HL3-bypassed scenario. As the planning period progresses, the relative difference increases due to the faster growth of BVT deployment in the full hierarchical scenario. By the final year, the optical energy consumption in the HL3-bypassed architecture is approximately 33\% lower.
	
	Figure~\ref{fig:energy_components}(b) shows the cumulative electrical energy consumption, corresponding to the energy usage of IP routers. In the first year, electrical energy consumption is about 25.5\% lower in the HL3-bypassed scenario. Because no new IP routers are deployed during Years~2--4 in either scenario, the electrical energy consumption remains unchanged during that period. By the end of the planning horizon, the HL3-bypassed scenario maintains an electrical energy consumption that is roughly 26\% lower than that of the full hierarchical scenario.
	
	Figure~\ref{fig:energy_components}(c) presents the normalized optical energy consumption, defined as the energy consumed per 100G of traffic carried in each year. Although the total optical energy consumption increases over time due to higher traffic demand, the normalized optical energy decreases for both scenarios. Because the total annual traffic is identical in both architectures, the relative difference between scenarios mirrors the difference observed in total optical energy consumption.
	
	Figure~\ref{fig:energy_components}(d) illustrates the normalized electrical energy consumption, defined as the electrical energy consumed per 100G of traffic. Similar to the optical case, the normalized electrical energy decreases over time in both scenarios despite the increasing total electrical energy consumption, reflecting improved energy efficiency as traffic scales.
	
	Overall, bypassing HL3 nodes leads to lower total optical and electrical energy consumption, and it also results in lower normalized optical and electrical energy consumption compared to the full hierarchical scenario.
	
\begin{figure}
		\centering
		\includegraphics[height = 3cm]{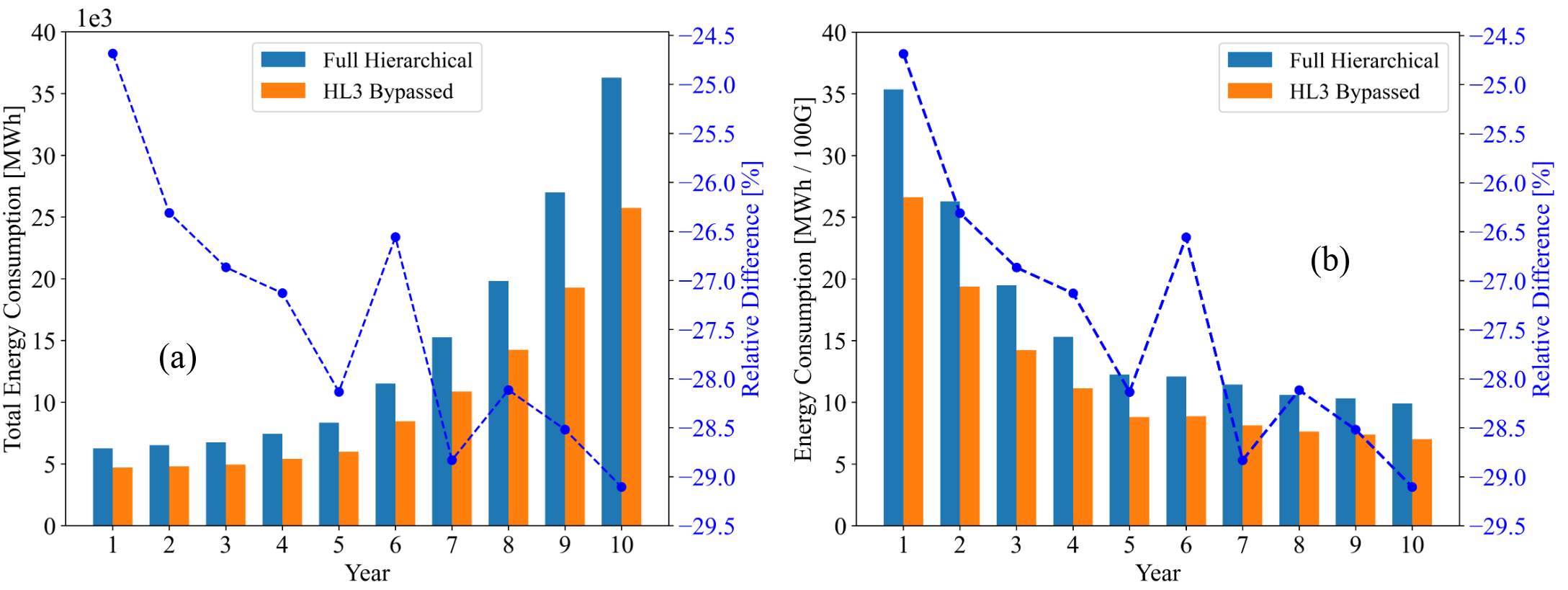}
		\caption{Cumulative total (a) energy consumption, (b) energy consumption per 100G traffic.}
		\label{fig:energy_total}
	\end{figure}

	Figure~\ref{fig:energy_total} presents the cumulative total energy consumption including both optical and electrical components for the two network scenarios over the 10-year planning horizon.
	
	Figure~\ref{fig:energy_total}(a) shows the cumulative total energy consumption in MWh. As illustrated, the HL3-bypassed scenario consistently requires less total energy in every year of the planning interval. Across the entire 10-year period, the overall energy consumption in the HL3-bypassed scenario is approximately 29\% lower compared to the full hierarchical architecture.
	
	Figure~\ref{fig:energy_total}(b) shows the normalized total energy consumption, defined as the total energy consumed per 100G of traffic carried in each year. Similar to the total energy results, the normalized energy consumption in the HL3-bypassed scenario is consistently lower than in the full hierarchical scenario, demonstrating improved energy efficiency when HL3 nodes are bypassed.
	
\begin{figure}
		\centering
		\includegraphics[height = 6.5cm]{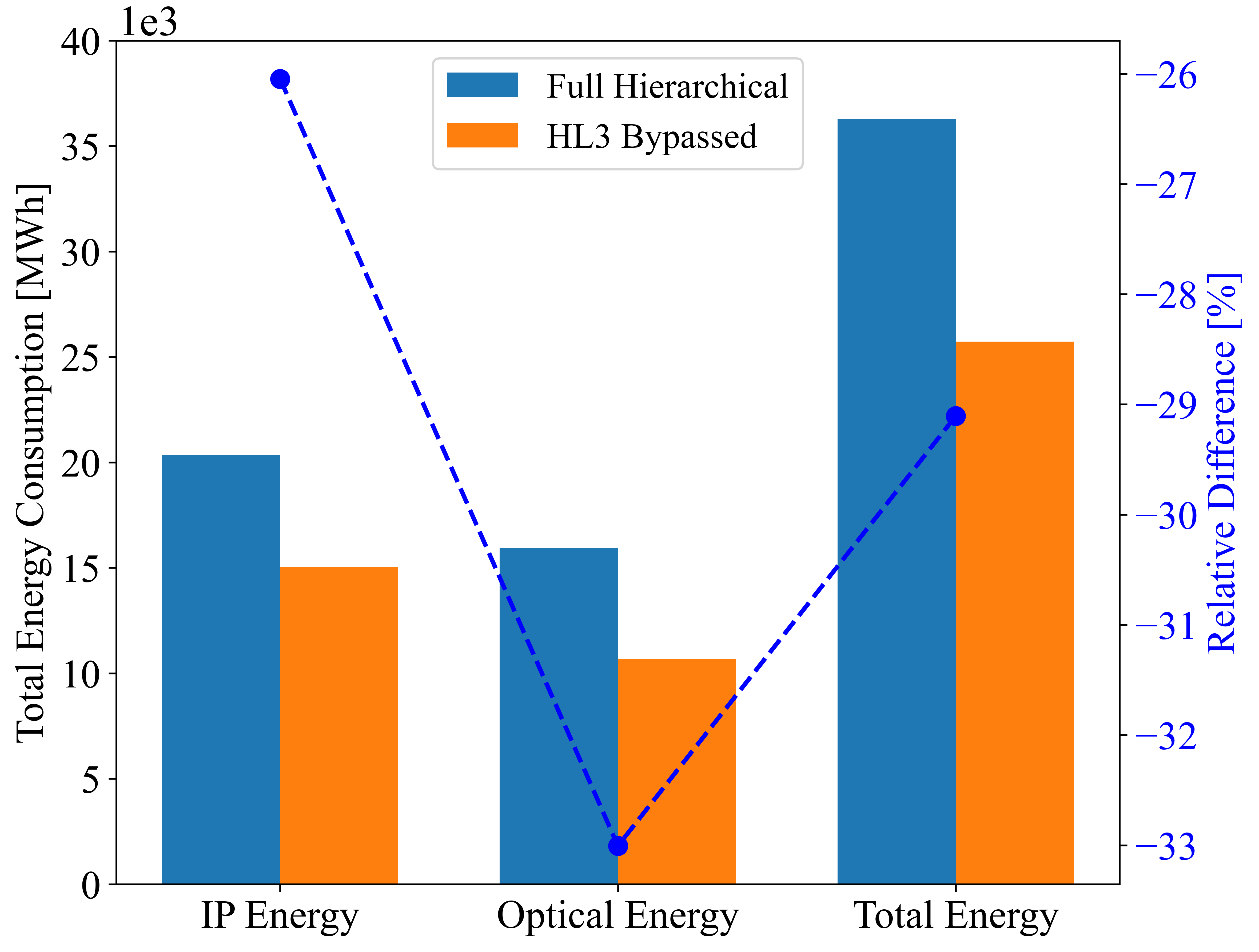}
		\caption{IP, optical and total energy consumption over 10 years.}
		\label{fig:energy_total_10yr}
	\end{figure}

	Figure~\ref{fig:energy_total_10yr} illustrates the overall electrical, optical, and total energy consumption accumulated over the entire 10-year planning interval for both network scenarios. As shown, the optical layer dominates the total energy consumption, contributing significantly more than the electrical (IP) layer in both scenarios.
	
	Bypassing HL3 nodes yields substantial energy savings across all components. Specifically, the HL3-bypassed scenario achieves approximately a 26\% reduction in electrical (IP) energy, a 33\% reduction in optical energy, and an overall 29.1\% reduction in total energy consumption compared to the full hierarchical architecture.
	
	These results demonstrate that bypassing HL3 nodes is not only cost-efficient but also environmentally advantageous, as it significantly reduces the network's energy footprint and thus lowers its associated carbon emissions.
	
	\subsubsection{Traffic Flow on Links}
	
\begin{figure*}
		\centering
		\includegraphics[width = \textwidth]{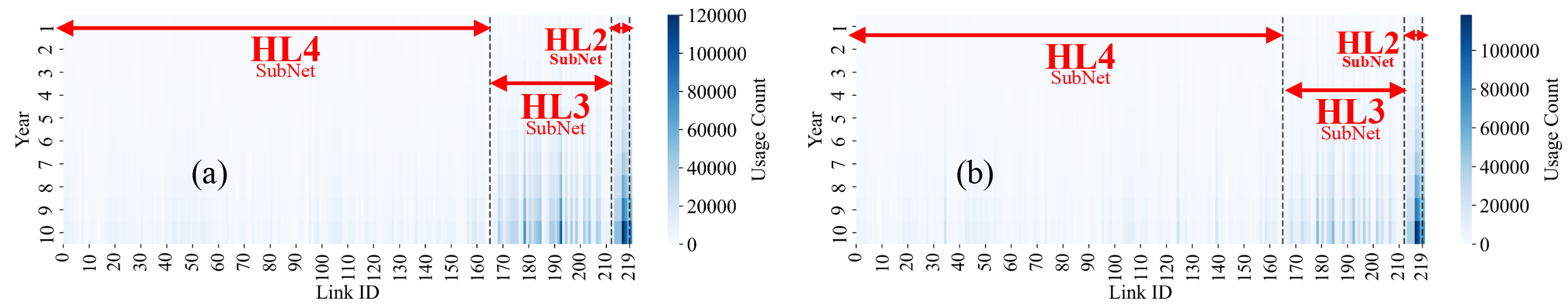}
		\caption{Traffic flow heatmaps of network links over the planning horizon for (a) full hierarchical and (b) HL3-bypassed scenario.}
		\label{fig:traffic_heatmaps}
	\end{figure*}

	Figure~\ref{fig:traffic_heatmaps} presents traffic flow heatmaps of network links over the 10-year planning interval for both network scenarios. Figure~\ref{fig:traffic_heatmaps}(a) corresponds to the full hierarchical architecture, while Figure~\ref{fig:traffic_heatmaps}(b) shows the HL3-bypassed scenario.
	
	As illustrated, bypassing HL3 nodes results in consistently lower traffic levels in both the HL4 and HL3 subnetworks, reflecting the redistribution of traffic when aggregation at HL3 nodes is removed. In contrast, traffic levels in the HL2 subnetwork remain largely unchanged between the two scenarios, indicating that HL2 is less sensitive to the architectural modification.
	
	In both cases, the heatmaps clearly show an increasing traffic trend over time, driven by the annual traffic growth applied across the planning interval.

\begin{figure*}
		\centering
		\includegraphics[width = \textwidth]{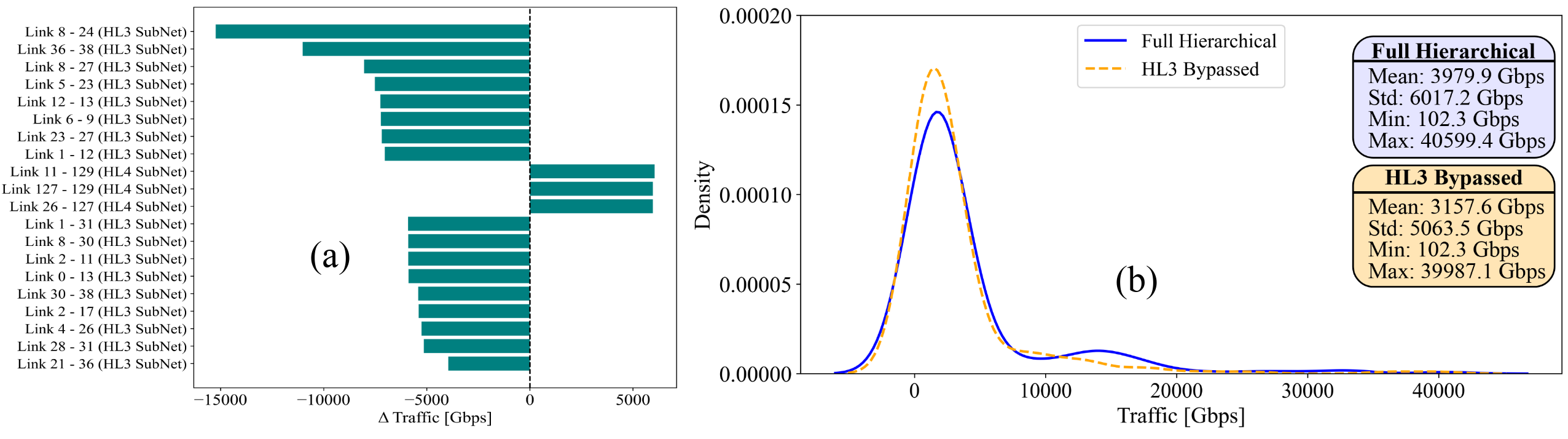}
		\caption{(a) Top links with the highest traffic change, (b) probability density function (PDF) of average links traffic.}
		\label{fig:traffic_flow_analysis}
	\end{figure*}

	Figure~\ref{fig:traffic_flow_analysis} presents the average traffic flow across all network links for the two considered network scenarios. 
		
	Figure~\ref{fig:traffic_flow_analysis}(a) highlights the top 20 links with the largest average traffic changes resulting from bypassing HL3 nodes. The most affected link is Link~8--24 in the HL3 subnetwork, which experiences a reduction of approximately 15.2~Tbps. Notably, the five links with the largest traffic decreases are all located in the HL3 subnetwork, indicating that this layer is the most impacted by the HL3 bypass. Although most of the top 20 links show a reduction in traffic, three links that belong to the HL4 subnetwork experience increase. Considering the average traffic across all links and all years, the HL4 subnetwork decreases from 4721.1~Gbps to 3692.3~Gbps, the HL3 subnetwork decreases from 1661.4~Gbps to 1512.3~Gbps, and the HL2 subnetwork decreases from 1970.2~Gbps to 1523.3~Gbps. These results confirm that bypassing HL3 nodes reduces the average traffic load across all subnetworks.
	
	Figure~\ref{fig:traffic_flow_analysis}(b) depicts the probability density function (PDF) of average link traffic. In the HL3-bypassed scenario, the distribution exhibits a higher peak and a lighter tail, indicating that heavily loaded links become less common and the traffic distribution becomes more concentrated. The tail approaches zero earlier, reflecting a reduction in extreme traffic values. Furthermore, the variance of link traffic is lower in the HL3-bypassed case, demonstrating a more balanced traffic distribution across the network. While the minimum link traffic remains unchanged, the maximum link traffic decreases when HL3 nodes are bypassed, further illustrating the trend toward a more uniform traffic pattern.
	
	\subsubsection{Latency}
	
\begin{figure}
		\centering
		\includegraphics[width = 8.5cm]{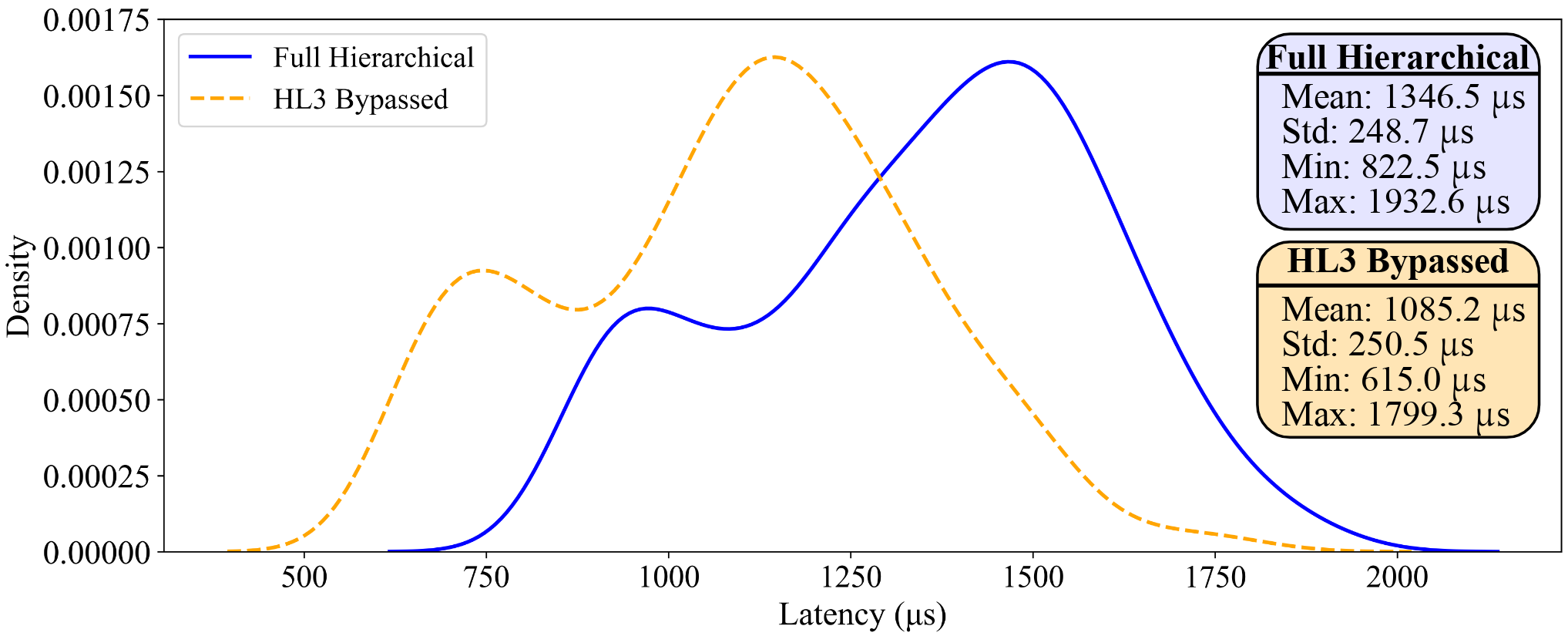}
		\caption{Probability density function (PDF) of end-to-end latency.}
		\label{fig:latency_pdf}
	\end{figure}

	Figure~\ref{fig:latency_pdf} illustrates the probability density function (PDF) of end-to-end latency for the two network scenarios. The end-to-end latency is computed by considering an optical propagation delay of 5$\mu$s per km and an electrical aggregation delay of 200$\mu$s for each aggregation stage. When HL3 nodes are bypassed, the 200$\mu$s aggregation delay associated with HL3 processing is eliminated. However, bypassing HL3 nodes may also lead to longer physical paths, which can increase propagation delay.
	
	As shown in the figure, the peak of the latency PDF in the HL3-bypassed scenario occurs at a lower latency value compared to the full hierarchical scenario. This indicates that bypassing HL3 nodes leads to a reduction in average end-to-end latency. Numerically, the average latency in the HL3-bypassed architecture is 1085.2$\mu$s, while the full hierarchical scenario exhibits an average latency of 1346.5$\mu$s, representing a 19.4\% reduction in latency when HL3 nodes are removed.
	
	The standard deviation of latency remains nearly the same for both scenarios, indicating that bypassing HL3 nodes does not significantly change the distribution of physical path lengths. Thus, the primary contributor to latency reduction is the removal of the 200$\mu$s electrical aggregation delay at HL3 nodes rather than changes in optical propagation distance.
	
	\subsubsection{Quality of Transmission (QoT)}
	
\begin{figure}
		\centering
		\includegraphics[height = 5.8cm]{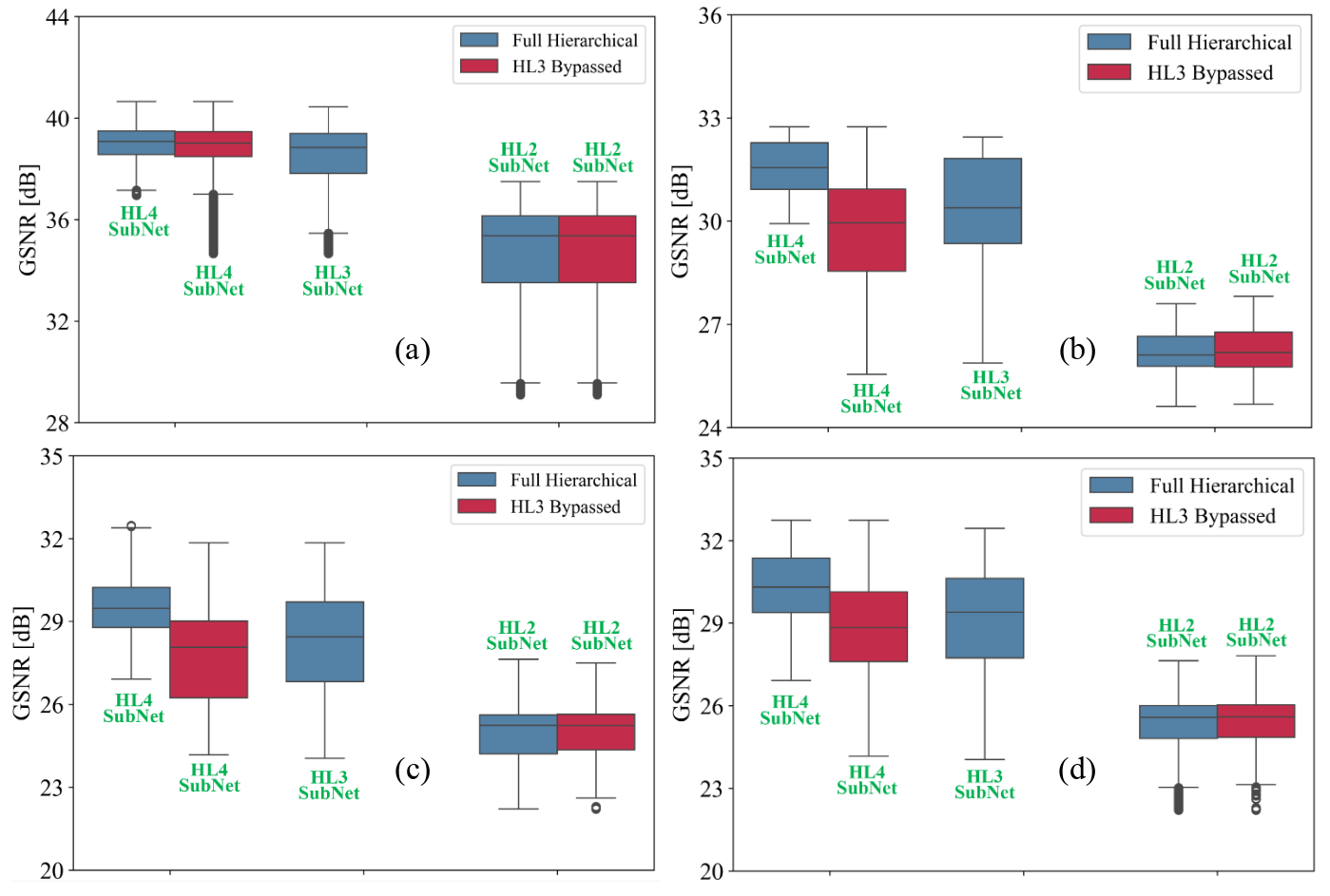}
		\caption{GSNR boxplot of (a) network links, (b) primary paths, (c) secondary paths, (d) all paths (primary + secondary) in different subnetworks.}
		\label{fig:gsnr_analysis}
	\end{figure}

	Figure~\ref{fig:gsnr_analysis} shows the GSNR characteristics of links and optical paths in the two network scenarios. In the full hierarchical scenario, routing is performed in three sequential stages: HL4--to--HL3, HL3--to--HL2, and HL2--to--HL1. When HL3 nodes are bypassed, HL3 nodes cannot be the destination of HL4 traffic, so they are treated as part of the HL4 group. A new subnetwork is therefore constructed to route traffic directly from HL4 nodes to HL2 nodes (either standalone or co-located with HL1 nodes), and the traffic that originally terminated at HL3 nodes simply passes through them toward HL2 nodes.
	
	Figure~\ref{fig:gsnr_analysis}(a) shows the GSNR of optical links in the different subnetworks for the two scenarios. As shown, the GSNR of HL2-subnetwork links is identical in both scenarios, since these links are not affected by bypassing HL3 nodes. In the HL3-bypassed scenario, the HL3-subnetwork links are included inside the HL4 subnetwork, so the GSNR values of HL4 links differ between the two scenarios due to the addition of HL3-subnetwork links into the HL4 group.
	
	Figure~\ref{fig:gsnr_analysis}(b) shows the GSNR of primary optical paths in the two scenarios. As shown, bypassing HL3 nodes leads to lower path GSNR values in the HL4 subnetwork in the HL3-bypassed scenario. This is because, in the full hierarchical scenario, the optical signal is converted to electrical and regenerated at HL3 nodes before reaching the HL2 layer, while in the HL3-bypassed scenario this regeneration step does not occur, resulting in lower signal quality at the HL2 destinations. For the final routing stage from HL2 nodes to HL1 nodes, the GSNR values are almost identical in both scenarios.
	
	Figure~\ref{fig:gsnr_analysis}(c) shows the GSNR of secondary (protection) paths. Similar to the primary paths, bypassing HL3 nodes results in lower GSNR at the HL2 destinations. This again comes from the absence of electrical regeneration in HL3 nodes, which normally improves the received signal quality.
	
	Figure~\ref{fig:gsnr_analysis}(d) shows the GSNR of all optical paths (primary and secondary) in the two network scenarios. The overall GSNR behavior follows the same trend observed in Figures~\ref{fig:gsnr_analysis}(b) and~\ref{fig:gsnr_analysis}(c). Therefore, bypassing HL3 nodes results in lower received GSNR at the destination nodes. This implies that node-bypassing designs such as bypassing HL3 nodes should be selected cautiously in applications with strict QoT requirements, since removing regeneration stages may reduce the attainable signal quality.

\section{Conclusion}
\label{sec:conclusion}
This paper introduced \textit{SixGman}, a fully open-source, multi-layer, multi-band, and multi-hierarchical optical network planning framework tailored for real-world metropolitan area networks. The tool provides comprehensive support for MAN-scale network design by integrating physical-layer modeling, routing and resource allocation, cost computation, and energy assessment, while enabling flexible architectural evaluations such as hierarchical-layer bypassing. Using the Telefónica MAN157 topology as a case study, we evaluated two representative architectures, a conventional full-hierarchical design and a scenario in which HL3 electrical aggregation is bypassed to assess their performance across a wide range of operational and techno-economic metrics.

The results demonstrate that bypassing HL3 nodes yields substantial improvements across multiple dimensions. In the optical layer, HL3 bypassing reduces fiber-pair consumption, spectrum occupancy, and 100G license activation, leading to an 16.3\% reduction in optical CAPEX. In the electrical layer, the elimination of HL3 IP routers results in a 26.3\% reduction in electrical CAPEX, significantly lowering overall infrastructure cost. The total cost of ownership (TCO) decreases by approximately 17.5\%, reflecting both reduced CAPEX and nearly 2.5\% lower OPEX. Energy efficiency also improves notably: optical and electrical components together exhibit a 29.1\% reduction in cumulative energy consumption, with optical devices contributing most to the savings. Traffic-flow analysis reveals that bypassing HL3 produces a more balanced and less congested distribution across metro links, while the end-to-end latency improves by 19.4\%, primarily due to the removal of HL3 aggregation delays. Finally, QoT analysis shows a modest degradation in GSNR for bypassed paths, an outcome that highlights the need for careful trade-off evaluation when eliminating regeneration stages.

Overall, this study confirms that hierarchical-layer\\ simplification specifically HL3 bypass can yield significant gains in cost, energy, and latency performance, making it a compelling direction for next-generation metro architectures targeted for the 6G era. Moreover, by providing \textit{SixGman} as an open-source framework, this work enables transparent, repeatable, and extensible research for the broader optical-networking community.

Future research will extend this work in several important directions. First, incorporating network slicing and Quality-of-Service (QoS) constraints will allow the evaluation of differentiated service classes across hierarchical MAN structures. Second, integrating AI-driven network planning, including machine-learning-based traffic prediction and reinforcement-learning optimization for resource allocation, can further enhance the planning intelligence of \textit{SixGman}. Third, evaluating datacenter placement strategies within hierarchical MANs, particularly for distributed AI/edge computing will create richer interdependencies between optical topology design and computational resource distribution. Finally, future versions of \textit{SixGman} will explore autonomous multi-band and hierarchical orchestration, enabling the tool to serve as a foundation for self-optimizing transport networks in the 6G era.

\printcredits

\section*{Declaration of competing interest}
The authors declare that they have no known competing financial interests or personal relationships that could have appeared to
influence the work reported in this paper.

\section*{Acknowledgment}
The authors from UC3M would like to acknowledge the support of the FLEX-SCALE project (Grant No. 101096909) and PROTEUS-6G project (Grant No. 101139134) are funded by the European Union’s HORIZON-JU-SNS-2024 program. Additionally, this work was supported by the TUCAN6-CM project (Grant No. TEC-2024/COM-460), funded by the Community of Madrid (ORDER 5696/2024).

\section*{Data availability}
Data will be made available on request. The SixGman is distributed as a Python-based open-source software package on GitHub  (\href{http://github.com/OS-ONDT/SixGman}{github.com/OS-ONDT/SixGman}).



\bibliographystyle{elsarticle-num}
 \bibliography{sixdman-refs}

\end{document}